\documentclass[12pt]{article}
\usepackage[dvips]{epsfig}

   \def\listoftables{\section*{List of Tables\@mkboth 
          {LIST OF TABLES}{LIST OF TABLES}}\@starttoc{lot}} 
   \def\listoffigures{\section*{List of Figures\@mkboth 
          {LIST OF FIGURES}{LIST OF FIGURES}}\@starttoc{lof}}

   \catcode`\@=11 
      \def\fnum@figure{\small {\bf Figure \thefigure}} 
      \def\fnum@table{\small {\bf Table \thetable}} 
   \catcode`\@=12 
   \makeatletter 
   \@addtoreset{equation}{section} 
   \makeatother 
    
\begin{document} 
   \normalsize 
   \noindent 
\begin{flushright}December 1994\\PITHA 63/94\end{flushright}
\section*{   \begin{center} 
Constructing solutions of Hamilton--Jacobi equations for 2 D
fields with one component by means of B\"acklund transformations
\end{center}} 
\begin{center} 
      {\large 
Wulf B\"ottger\footnote{e--mail: wulf@thphys.physik.rwth--aachen.de}, 
Henning Wissowski\footnote{e--mail: henning@acds16.physik.rwth--aachen.de}, 
Hans A.\ Kastrup\footnote{e--mail: kastrup@thphys.physik.rwth--aachen.de}\\ 
 Institut f\"ur Theoretische Physik E, RWTH Aachen,\\ 
 D--52056 Aachen, Fed.\ Rep.\ Germany} 
\end{center} 
\noindent 
\begin{abstract}\footnotesize\noindent\bf 
The Hamilton--Jacobi formalism generalized to 2--dimensional field theories 
according to Lepage's canonical framework is applied to several relativistic
real scalar fields, e.g.\ massless and massive Klein--Gordon, 
Sinh \& Sine--Gordon, Liouville and $\phi^4$ theories.
The relations between the Euler--Lagrange and the Hamilton--Jacobi equations  
are discussed in DeDonder and Weyl's and the corresponding 
wave fronts are calculated in Carath\'eodory's formulation. 
\\
Unlike mechanics we have to impose certain integrability conditions on 
the velocity fields to guarantee the transversality relations and 
especially the dynamical equivalence between Hamilton--Jacobi wave fronts 
and families of extremals embedded therein.
\\ 
B\"acklund Transformations play a crucial role in solving the resulting 
system of coupled nonlinear PDEs. 
\end{abstract}
\normalsize 
\section{Introduction}
Varying a relativistically invariant action integral leads to covariant 
Euler-Lagrange equations. However, if one wants to
reformulate the theory in terms of the conventional canonical Hamiltonian
framework one has to break the manifest covariance by distinguishing a time
variable and regarding the other "spatial" coordinates as "indices"
representing an infinite number of degrees of freedom. The method is widely
known from elementary particle physics, canonical gravity and other
field theories. This approach, however, can obscure a part of the rich
geometrical structure contained in a generally covariant framework, at least
on the classical level \cite{ka}: 
\\[2mm] 
Utilizing Cartan's theory of alternating forms Lepage and others showed that 
a large variety of algebraically inequivalent covariant Hamiltonian 
formulations, including Hamil\-ton--\-Ja\-co\-bi equations, exists, e.g.\ 
that of DeDonder \& Weyl \cite{DeDonder}\hspace{0.5mm}\cite{Weyl} and that of 
Cara\-th\'{e}o\-dory \cite{Cara}, where only the latter one provides a 
Hamil\-ton--Ja\-cobi equation the associated wave fronts of which  have the 
same nice transversality properties with respect to the extremals as one has 
in mechanics. Reformulations and special examples of this covariant 
Hamiltonian approach in terms of multisymplectic frameworks may be found in 
\cite{ka}\cite{go1}\cite{go2}\cite{ki}. 
\\[2mm]
In mechanics one can construct solutions of the canonical eqs.\ of motion if
one has an appropriate solution of the corresponding Hamilton--Jacobi
equation, the solutions of which describe wave fronts which are transversal 
to a "field" of extremals and which contain the same dynamical information as 
the extremals themselves. 
\\[2mm] 
For field theories this is no longer true \cite{ka}: Solutions of the
Hamilton-Jacobi equation (HJE) associated with one of the canonical 
frameworks mentioned above provide "velocity" fields which in general do not
obey the necessary integrability conditions (IC). The latter have to be 
postulated separately and give rise to an additional set of partial 
differential equations. However, if these equations and the associated 
Hamilton--Jacobi equation are satisfied, then combined they contain the same 
dynamical content as the Euler-Lagrange equations \cite{ka}. Thus, it is 
possible to construct solutions of the Euler-Lagrange field equations (ELE) 
by solving a Hamilton-Jacobi equation and a set of integrability conditions. 
\\[2mm]
The aims of the present paper are the following ones: 
\begin{enumerate} 
\item
All the different canonical formalisms just mentioned coincide if the field 
has just one real component \cite{ka}, especially for fields in 2 spacetime 
or Euclidean dimensions. In this case one can use the more simple DeDonder 
\& Weyl HJE in order to solve it simultaneously with one integrability 
condition. Here we do not construct completely new solutions of the ELE but 
find one-parameter extensions of a given solution $\varphi_0 $.
The method is applied to the following one-component models in 2 dimensions:
massive and massless Klein--Gordon, Sinh-, Sine--Gordon, Liouville and 
$\phi^4$
theory. The solutions of the HJE plus IC are constructed by expanding the
solutions of these equations in powers of the field variables. This leads to 
a hierarchy of nonlinear PDEs that can be transformed into linear PDEs with
nonconstant coefficients. By applying integrable B\"acklund transformations
these PDEs can be reduced further to free (massive or massless) Klein-Gordon
equations. Remarkably, it is only necessary to solve just  two linear PDEs in
order to obtain the general solution
for {\em every} order of the hierarchy!
\item The solutions of the DeDonder \& Weyl HJE do not have the appropriate
transversality properties required to construct the wave fronts associated
with a given one--parameter set of extremals. Those have to be given in terms 
of the solutions of the more complicated HJE of Carath\'{e}odory. As we are
dealing with one-component fields here only, the two canonical frameworks are
equivalent and one can construct the 1--dimensional Carath\'{e}odory 
wave fronts from the solutions of the DeDonder \& Weyl HJE. This procedure 
will be outlined and applied to given solutions (e.g.\ solitons) of models 
mentioned above. 
\end{enumerate}
The paper is organized as follows:
\\[3mm]
In chapter 2 we very briefly summarize Lepage's reformulation of canonical 
mechanics.
\\[1mm]
In chapter 3 we apply this framework to field theories with one component in 
two dimensions and present the HJEs of DeDonder \& Weyl and of 
Carath\'eodory, respectively. 
\\[1mm]
In chapter 4 we study the hierarchy of equations derived within the 
Hamilton--Jacobi framework of DeDonder \& Weyl and the associated 
integrability condition for the velocity fields by expanding the wave fronts 
in powers of the field variable $\varphi$ in the neighbourhood of 
$\varphi_0$.
\\[1mm]
In chapter 5 we determine the B\"acklund transformations which reduce
this hierarchy of linear PDEs with nonconstant coefficients to free field 
equations.
\\[1mm]
In chapter 6 we apply the general results to several well known models 
and point out some relations to stability problems of solitons.
\\[1mm]
In chapter 7 we discuss the relations between the hierarchies of PDEs derived 
from expanding the HJE and the IC and a corresponding expansion of the 
Euler--Lagrange equations. Solutions of the equations of motions are 
determined from those of the HJE and the IC perturbatively. The kink 
solution of the Sine--Gordon equations is treated in considerable detail 
here.
\\[1mm]
In chapter 8 we calculate the wave fronts by means of Carath\'eodory's 
formulation, especially for the kink solution just mentioned.
\section{Lepage's Canonical Formulation of Mechanics}
We very briefly recall Lepage's main idea of introducing the canonical 
formalism in mechanics for one configuration variable $q$. The general case
is discussed in \cite{ka}.
\\[2mm] 
The starting point of the canonical theory for Lagrangian canonical systems 
is an action functional. In mechanics it is given by a Lagrangian 1--form 
$\omega$ integrated along a path ${\cal C}\!:=\!\{t,q\!=\!q(t)\}$ in the 
extended configuration space ${\cal M}_{1+1}:=\{(t,q)\}$: 
\begin{equation}
{\cal A}[{\cal C}]=\int_{\cal C}\omega= \int_{\cal C} L(t,q,\dot{q})\,
\mbox{d}t\,.
\end{equation}
As to the variational principle it is preferable to consider the generalized 
velocity $v$ as an independent variable, which coincides with $\dot{q}$  
on the extremals ${\cal C} \! = \! {\cal C}_0 \!:=\! \left\{(t,q \!=\! 
q_0(t))\right\}$. Normally this extension in the number of variables is 
performed by using Lagrangian multipliers. Lepage's reformulation of the 
variational principle is similar in spirit \cite{lep}.
\\   
The initial canonical Lagrangian form $\omega\!=\! L(t,q,\dot{q})\,\mbox{d}t$ 
is extended by the product of a Lagrangian multiplier $h(t,q,v)$ and the 
Pfaffian form $\varrho\! = \!\mbox{d}q-v\,\mbox{d}t$ vanishing on the tangent 
vectors of the extremals, which ensures the identification of $v(t)$ with 
$\dot{q}(t)$ on the solutions of the equation of motion. 
Then the action integral ${\cal A}[{\cal C}]$ over the path ${\cal C}\!:=\!
\{(t,q(t))\}$ can be modified:
\begin{equation} 
{\cal A}[{\cal C}]\,\,\rightarrow\,\,\tilde{\cal A}[{\cal C}]=\int_{\cal C}
\Omega=\int_{\cal C}\left[ L(t,q,v)\,\mbox{d}t+h(t,q,v)\,\varrho\right]\,,
\label{e1}
\end{equation}
without changing the Euler--Lagrange equations and their solutions 
${\cal C}\!=\!{\cal C}_0$.
\\[2mm]
The form $\varrho$ generates an ideal $I[\varrho]$ in the algebra $\Lambda$ 
of forms on the extended configuration space ${\cal M}_{1+1}:=\{(t,q)\}$: if 
$\alpha\in\Lambda$ and $\beta\in I[\varrho]$, then $\alpha\wedge\beta$ is 
also an element of the ideal $I[\varrho]$. 
\\
The Lagrangian multiplier $h(t,q,v)$ can be fixed by varying the action 
integral (\ref{e1}) with respect to $q,v$ independently. This leads to the 
standard definition of the canonical momentum: $h \! \stackrel{!}{=} \!
\partial_v L\! =: \!p$.
\\[2mm]
We obtain the same results by requiring $\mbox{d}\Omega$ to be an element of 
the ideal $I[\varrho]$:
\begin{equation}
\mbox{d}\Omega=(\partial_v L-h)\,\mbox{d}v
\wedge{\mbox{d}}t+(\mbox{d}h-\partial_q L\,\mbox{d}t)\wedge\varrho=
\underbrace{(\partial_v L-h)}_{\textstyle =0!}{ 
\mbox{d}}v\wedge{\mbox{d}}t+0\left(\mbox{mod}I[\varrho]\right)\,.
\vspace{-2mm} 
\end{equation}
Hence $\mbox{d}\Omega$ is a closed two form on families of extremals 
covering the extended configuration space ${\cal M}_{1+1}\!=\!\{(t,q)\}$ -- 
or correspondingly  -- a (Lagrangian) submanifold ${\cal Q}:=\{t,q,p \! = \! 
\psi(t,q)|(t,q)\in M\}$ of the extended phase space ${\cal P}_{2+1}:=\{
(t,q,p)\}$. 
\\[2mm]
Following Poincar\'e's lemma $\Omega$ is locally (at least) exact 
$\Omega\!=\!\mbox{d}S(t,q)$.
\\[2mm]
The Legendre transformation $L \rightarrow H$, $v \rightarrow p$ can 
be implemented as a change of basis in the cotangent bundle ${\cal T}^{
\ast}({\cal M}_{1+1})$, $\varrho\rightarrow \mbox{d}q$, $\mbox{d}t 
\rightarrow \mbox{d}t$: 
\begin{equation}
\Omega=L\,\mbox{d}t+p\,\varrho=L\,\mbox{d}t+p\,(\mbox{d}q-v\,\mbox{d}t)=-(pv
-L)\mbox{d}t+p\,\mbox{d}q=-H\,\mbox{d}t+p\,\mbox{d}q\,. 
\end{equation}
$H$ denotes the usual Hamiltonian $H\!=\!pv\!-\!L\!=\!H(p,q,t)$.
\\[2mm]
The existence of a potential $S(t,q)$ for the basic differential form 
$\Omega\!=\!\mbox{d}S$ yields the familiar Hamilton--Jacobi equation for 
$S(t,q)$ and the corresponding condition for the momentum: 
\begin{equation}
\Omega=-H\left(t,q,p=\psi(t,q)\right){\mbox d}t+\psi(t,q)\, 
\mbox{d}q\stackrel{\textstyle !}{=}\mbox{d}S(t,q)= 
\partial_t S(t,q)\,\mbox{d}t+\partial_q S(t,q)\,\mbox{d} 
q\,. 
\end{equation}
Comparing the coefficients of $\mbox{d}t$, $\mbox{d}q$ yields:
\begin{equation}
\partial_t S(t,q)+H\left(t,q,p=\psi(t,q)\right)=0\,,\;\; p=\psi(t,q)=
\partial_q S(t,q)\,. \vspace{3mm}
\end{equation} 
The extremals can be determined, if a complete integral of the 
Hamilton--Jacobi equation is found. Dealing with one independent 
variable $q$ this integral depends on one constant $c_0$. 
The solution of the equation of motion can be calculated by solving the 
algebraic relation $\partial_{c_0}S(t,q,c_0)\!=\!c_1$ with a second constant 
$c_1$.
\\[2mm]
In mechanics there exists a special relation between the wavefront 
$S(t,q)\!=\!\sigma\!=\!const,\, \sigma \in {\cal R}$ and the extremals $q\!=
\!q_0(t)$ called {\em ``transversality"}: the union of their respective 
tangent spaces $span\{e_t\! = \!\partial_t + \dot{q}(t)\partial_q\}$ 
and $span\{w\! = \!p\partial_t+H\partial_q\}$, span the whole tangent space 
${\cal T}_P ({\cal M}_{1+1})$ at any point $P\in{\cal M}_{1+1}$, if 
the Lagrangian $L$ does not vanish, because of     
$$
\det{(e_t,w)}\! = \!\det\left|\left(\begin{array}{lr}1 & p \\ v & H 
\end{array}\right)\right|=\left(H-\dot{q}p\right)=-L\,.
\vspace{3mm}
$$
This concept of the Hamilton--Jacobi framework developed for mechanics can 
easily be generalized to field theories. We confine 
our discussion to those theories depending on {\em one} real scalar field 
$\varphi\! = \!\varphi(z,\bar{z})$ in a $1\!+\!1$ dimensions. Here $z,\bar{z}
$ play the role of lightcone variables.
For details see 
\cite{ka}\hspace{0.5mm}\cite{DeDonder}\hspace{0.5mm}\cite{Weyl}. 
\section{The Hamilton--Jacobi theories of DeDonder \& $\,\,$ Weyl and of 
Carath\'eodory} 
As in mechanics a canonical theory for fields is based on an action 
functional ${\cal A}[\Sigma]$ defined on a two dimensional surface $\Sigma 
\! := \! \{(z,\bar{z},\varphi(z,\bar{z}))\}$ in the extended configuration 
space ${\cal M}_{2+1}\! :=\!\{(z,\bar{z},\varphi)\}$: 
\begin{equation}
{\cal A}[\Sigma]=\int_{\Sigma}\omega=\int_{\Sigma}{\cal L}\left(z,\bar{z}, 
\varphi(z,\bar{z}),v=\partial_z \varphi,\bar{v}=\partial_{\bar{z}}\varphi
\right)\,{\mbox{d}} z \wedge\mbox{d}\bar{z}\,. 
\end{equation}
Only on the extremals $\varphi\! = \!\varphi_0(z,\bar{z})$ the 
generalized velocities $v, \bar{v}$ coincide with the derivatives of the 
fields: $v\!=\!\partial_z \varphi_0, \bar{v}\!=\!\partial_{\bar{z}}
\varphi_0$.
\\[2mm]
The canonical 2--form $\omega\! = \!{\cal L}\,{ \mbox{d}}z\wedge\mbox{d}
\bar{z}$ is extended by means of two Lagrangian parameters $h(z,\bar{z},
\varphi)$, $\bar{h}(z,\bar{z},\varphi)$ and a 1--form $\varrho\!=\! \mbox{d}
\varphi-v\,\mbox{d}z-\bar{v}\,\mbox{d}\bar{z}$ that vanishes on the  
2--dimensional extremals $\varphi\! = \!\varphi_0(z,\bar{z})$: 
\begin{equation}
\Omega={\cal L}\,\mbox{d}z\wedge\mbox{d}\bar{z}+\bar{h}\,\mbox{d}z\wedge
\varrho+h\,\varrho\wedge\mbox{d}\bar{z}\,.\label{basic}
\end{equation}
The Lagrangian multipliers $h,\bar{h}$ are determined by requiring 
$\mbox{d}\Omega \in I[\varrho]$:
\begin{equation}
\mbox{d}\Omega=
(\partial_v{\cal L}-h)\mbox{d}v\wedge\mbox{d}z\wedge\mbox{d}\bar{z}+
(\partial_{\bar{v}}{\cal L}-\bar{h})\mbox{d}\bar{v}\wedge\mbox{d}z\wedge
\mbox{d}\bar{z}+0\left(\mbox{mod}I[\varrho]\right)\stackrel{!}{=}0\left(
\mbox{mod}I[\varrho]\right)\,,
\end{equation}
yielding  
\begin{equation} 
h \stackrel{!}{=} \partial_v {\cal L}=:p\,,\quad \bar{h} \stackrel{!}{=} 
\partial_{\bar{v}} {\cal L}=:\bar{p}\,.
\end{equation}
The Legendre transformation ${\cal L} \rightarrow {\cal H}$, $\{v,\bar{v}\} 
\rightarrow \{p,\bar{p}\}$ can be implemented as a change of the basis 
in the cotangent bundle ${\cal T}^{\ast}({\cal M}_{2+1})$, $\varrho
\rightarrow \mbox{d}\varphi$, $\mbox{d}z \rightarrow \mbox{d}z$, 
$\mbox{d}\bar{z} \rightarrow \mbox{d}\bar{z}$:
\begin{equation} 
\Omega=-{\cal H}\,\mbox{d}z\wedge\mbox{d}\bar{z}+\bar{p}\,\mbox{d}z\wedge
\mbox{d}\varphi+p\,\mbox{d}\varphi\wedge\mbox{d}\bar{z}\label{Omega}\quad
\mbox{with}\quad {\cal H}:=pv+\bar{p}\bar{v}-{\cal L}\,.
\label{ham}\end{equation}
Because $\mbox{d}\Omega\!=\!0\left(\mbox{mod}I[\varrho]\right)$ it is 
locally exact $\Omega\!=\! \mbox{d} {\cal S}$, ${\cal S}\in{\cal T}^{
\ast}({\cal M}_{2+1})$ on families of extremals.
However, contrary to mechanics $\Omega$ as an exact 2--form can be 
represented in different ways by means of a Pfaffian form ${\cal S}$.
In the case of DeDonder \& Weyl \cite{DeDonder}\hspace{0.5mm}\cite{Weyl}:
\begin{equation}
{\cal S}_{DW}=S(z,\bar{z},\varphi)\,\mbox{d}z-\bar{S}(z,\bar{z},\varphi)\,
\mbox{d}\bar{z}
\end{equation}
and in the case of Carath\'eodory \cite{Cara}:
\begin{equation}
{\cal S}_C=S^{z}(z,\bar{z},\varphi)\,\mbox{d}S^{\bar{z}}(z,\bar{z},\varphi)
\,.
\end{equation}
Comparing the exterior derivatives of these expressions with equation 
(\ref{Omega})
\begin{equation}
\Omega=-{\cal H}\,\mbox{d}z\wedge\mbox{d}\bar{z}+\bar{p}\,\mbox{d}z\wedge
\mbox{d}\varphi+p\,\mbox{d}\varphi\wedge\mbox{d}\bar{z}\stackrel{!}{=}\,
\mbox{d}\bar{S}\wedge\mbox{d}z-\mbox{d}S\wedge\mbox{d}\bar{z}\,,
\end{equation}
we obtain the Hamilton--Jacobi equations and the transversality
conditions for a one component field theory in DeDonder and 
Weyl's formulation:
\begin{equation}
\partial_z S+\partial_{\bar{z}} \bar{S}=-{\cal H}\,,\quad p=\partial_{
\varphi}S\, ,\quad \bar{p}=\partial_{\varphi}\bar{S}\label{trans}
\end{equation}
and in Carath\'eodory's case:
\begin{equation}
\Omega=-{\cal H}\,\mbox{d}z\wedge\mbox{d}\bar{z}+\bar{p}\,\mbox{d}z\wedge
\mbox{d}\varphi+p\,\mbox{d}\varphi\wedge\mbox{d}\bar{z}\stackrel{!}{=}
\mbox{d}S^z\wedge\,\mbox{d}S^{\bar{z}}\,,\label{basca}
\end{equation}
we get:
\begin{equation} 
\partial_zS^{z}\partial_{\bar{z}} S^{\bar{z}} - \partial_{z}S^{\bar{z}} 
\partial_{\bar{z}} S^z=-{\cal H}\,, \quad
\begin{array}{c}p=\partial_{\bar{ z}} S^{\bar{z}}\partial_{\varphi}S^z-  
\partial_{\bar{z}} S^{z}\partial_{\varphi}S^{\bar{z}}\\
\bar{p}=\partial_{z} S^{z}\partial_{\varphi}S^{\bar{z}}- 
\partial_{z} S^{\bar{z}}\partial_{\varphi}S^{z}\label{trans2}
\end{array}\,.\label{carat}\label{CHJ}
\end{equation}
The two theories here are equivalent because a $n$--form in a space of 
$n\!+\!1$ variables has always rank $n$ \cite{choq}. Due to this algebraic 
equivalence of covariant canonical formulations for one component field 
theories we may choose DeDonder and Weyl's description to embed the 
extremals of interest in a system of solutions of the Hamilton--Jacobi 
equation. We use Carath\'eodory's theory for the explicit calculation of the 
wave fronts $S^z=const.,S^{\bar{z}}=const.$:
\\
In two dimensional field theories involving one field variable 
the basic two form $\Omega$ has always the rank two, i.e.\ it can be 
constructed from two independent one 
forms by linear combination of exterior products. Because $\Omega$ is closed 
its rank is equal to its class, that gives the codimension of the integral 
submanifold determined by $\Omega$. To calculate this integral manifold --- 
the wavefronts in our case --- we can use a corollary of Frobenius'
integrability theorem \cite{ka}: there exist two functions $S^z(z,\bar{z},
\varphi)$, $S^{\bar{z}}(z,\bar{z},\varphi)$ such a manner that the exterior 
product of their differentials equals $\Omega$. The corresponding one 
dimensional wave fronts are just the submanifolds determined by the 
conditions $S^z(z,\bar{z},\varphi)\!=\!const.$, $S^{\bar{z}}(z,\bar{z},
\varphi)\!=\!const.$.
\\[2mm] 
Thus the wave fronts are equipotential surfaces of the solutions of the 
Hamilton--Jacobi equation formulated in Carath\'eodory's framework. For 
simplicity we first solve the HJE of DeDonder \& Weyl and the associated IC
and afterwards we return to Carath\'eodory's equation in order to obtain an 
explicit expression for the wave fronts. 
\\[2mm]
In mechanics for one variable $q$ it is possible to construct wave fronts for 
$1$--parametric families of extremals that cover a certain region of the 
extended configuration space ${\cal M}_{1+1}$ {\em and} vice versa. 
Provided a solution $S(t,q)$ of the Hamilton--Jacobi equation (HJE) is given, 
the corresponding velocity field, the so--called "slope function": 
\begin{equation} 
\Phi(t,q) = \partial_p H(t,q,p\! = \!\partial_q S(t,q)) 
\end{equation} 
determines the corresponding 1--parametric extremals by means of the ordinary
first order differential equation: $\dot{q}(t)\! = \!\Phi(t,q(t))$. 
\\[2mm] 
In general this is {\em not} true for field theories; the ability
to embed extremals $\varphi_0(z,\bar{z})$ in a given wave front can be 
maintained only if the slope functions (velocity fields) $v\!=\!\Phi(\varphi,
z,\bar{z})$, $\bar{v}\!=\!\bar{\Phi}(\varphi,z,\bar{z})$ obtained from the 
inverse Legendre transformation 
\begin{eqnarray}
\partial_z\varphi(z,\bar{z})\!&\!=\!&\!v\left(p=\partial_{\varphi}S,\bar{p}=
\partial_{\varphi}\bar{S},z,\bar{z},\varphi\right)=\Phi(\varphi,
z,\bar{z})\,,\nonumber\\
\partial_{\bar{z}}\varphi (z,\bar{z})\!&\!=\!&\!\bar{v}\left(p=\partial_{
\varphi}S,\bar{p}=\partial_{\varphi}\bar{S},z,\bar{z},\varphi\right)=\bar{
\Phi}(\varphi,z,\bar{z})\label{slope}
\end{eqnarray}
satisfy the integrability condition $\partial_z\partial_{\bar{z}}{
\varphi}(z,\bar{z})\!\stackrel{!}{=}\!\partial_{\bar{z}}\partial_z{
\varphi}(z,\bar{z})$, which results in the requirement  
\begin{equation} 
\fbox{$\displaystyle\frac{\mbox{d}}{\mbox{d}{\bar{z}}}\Phi(z,{\bar{z}},
\varphi(z,{\bar{z}})):=\partial_{\bar{z}}\Phi+\bar{\Phi}\cdot\partial_{
\varphi} \Phi\,=\,\frac{\mbox{d}}{\mbox{d}z}\bar{\Phi}(z,{\bar{z}},\varphi(z,
{\bar{z}})):=\partial_z\bar{\Phi}+ \Phi\cdot\partial_{\varphi}\bar{\Phi} 
$}\label{g2} 
\end{equation}
on $\Phi$ and $\bar{\Phi}$.
\\[2mm]
Provided the Hamilton--Jacobi equation and the integrability condition are 
fulfilled, the Euler--Lagrange equation is satisfied automatically. This 
can be seen as follows:
\\[2mm]
Differentiating the Hamilton--Jacobi equation (\ref{trans}) 
with respect to the field variable $\varphi$ one obtains:
\begin{equation}
\partial_z p+\partial_{\bar{z}}\bar{p}=-\partial_{\varphi}{\cal H}-\partial_p
{\cal H}\partial_{\varphi}p-\partial_{\bar{p}}{\cal H}\partial_{\varphi}
\bar{p}\,.
\end{equation}
Because $\partial_{\varphi}{\cal H}\!=\!-\partial_{\varphi}{\cal L}$,
$\partial_{p}{\cal H}\!=\!\Phi$ and $\partial_{\bar{p}}{\cal H}
\!=\!\bar{\Phi}$ (see \cite{ka}) we get: 
\begin{equation}
\frac{\mbox{d}}{\mbox{d}z}p+\frac{\mbox{d}}{\mbox{d}\bar{z}}\bar{p}=
\frac{\mbox{d}}{\mbox{d}z}\frac{\partial{\cal L}}{\partial v}+
\frac{\mbox{d}}{\mbox{d}\bar{z}}\frac{\partial{\cal L}}{\partial \bar{v}}=
\partial_{\varphi}{\cal L}\,.\label{EL}
\end{equation}
The momenta $p(z,\bar{z},\varphi)$ and $\bar{p}(z,\bar{z},\varphi)$ are 
defined on the extended configuration space ${\cal M}_{2+1}$ and are 
considered to be associated with a family of extremals $\varphi\!=\!\tilde{
\varphi}(z,\bar{z},u)$ parametrized by $u$. The 
Euler--Lagrange equation has to be fulfilled for every single extremal 
$\varphi\!=\!\tilde{\varphi}(z,\bar{z},u\!=\!\mbox{const.})$. Hence if we 
insert $p$ and $\bar{p}$ into this equation of motion, we have to be aware of 
their implicit dependence of $z,\bar{z}$ via the field variable $\varphi$. 
Taking the defining equations (\ref{slope}) for the slope functions into 
account, the total derivatives in the PDEs (\ref{g2}) and (\ref{EL}) are 
defined as: 
\begin{equation}
\frac{\mbox{d}}{\mbox{d}z}:=\left.\frac{\partial}{\partial z}\right|_{\bar{z}
,\varphi=\mbox{const}}+\left.\Phi\frac{\partial}{\partial \varphi}\right|_{
\bar{z},z=\mbox{const}}\,,\quad
\frac{\mbox{d}}{\mbox{d}\bar{z}}:=\left.\frac{\partial}{\partial \bar{z}}
\right|_{z,\varphi=\mbox{const}}+\bar{\Phi}\left.\frac{\partial}{\partial 
\varphi}\right|_{z,\bar{z}=\mbox{const}}\,.
\end{equation}
Therefore the total derivatives are nothing but derivatives with respect to 
the independent variables $z,\bar{z}$ regarding $u$ to be a constant.
\\[2mm]
Due to the extend in which the integrability condition is taken into account 
there exist two different methods of using a Hamilton--Jacobi theory: 
the weak and the strong embedding of extremals in families of wave fronts.
\\[2mm]
1.) {\em Weak embedding:}
\\[1mm]
This method is used to embed a given single extremal 
$\hat{\varphi}_0(z,\bar{z})$ in families of wave fronts. In order to obtain
a weak embedding it is sufficient to take only the Hamilton--Jacobi equation 
and the transversality conditions (\ref{slope}) on 
$\hat{\varphi}_0(z,\bar{z})$ 
into account. In this case usually one chooses a linear ansatz in the field 
variable for one of the functions $S$ or $\bar{S}$. However, this approach in
general will not provide new extremals, because the IC are fulfilled on the 
given extremal only. For details to this subject see \cite{ri}.
\\[2mm] 
2.) {\em Strong embedding:}
\\[1mm]
Here one requires the IC not only to hold on the given extremal but in a 
whole neighbourhood of it. If this is the case then one is able to generate 
new extremals from a given one by integrating the slope functions 
(\ref{slope}).
\\[2mm]
In the following we study the strong embedding of a single given extremal 
into a family of wave fronts.
\\[2mm]
Like in mechanics there exists a transversality relation between extremals 
$\varphi(z,\bar{z})$ and wave fronts  $S^z(z,\bar{z},\varphi) \!=\! const.$, 
$S^{\bar{z}}(z,\bar{z},\varphi) \!=\! const.$, which holds, if in every 
point $P\in{\cal M}_{1+2}$ the basis of tangent space ${\cal T}{ \cal M}_{1+
2}$ is given by a union of the basis of the 2--dimensional tangent 
space of the extremals $e_z \!=\!\partial_z +v\partial_{\varphi}$, $e_{\bar{
z}} \!=\! \partial_{\bar{z}}+\bar{v}\partial_{\varphi}$ and a basis vector 
$w \!=\! p\partial_z +{\bar{p}}\partial_{\bar{z}} + {\cal H}\partial_{
\varphi}$ of the  $1$--dimensional tangent space of the wave fronts, i.e.\ 
iff the Lagrangian density ${\cal L}$ does not vanish:
\begin{equation}
\det{(e_z,e_{\bar{z}},w)}\! = \!\det\left|\left(\begin{array}{
ccc}1&0&p \\ 0&1&\bar{p}\\ v&\bar{v}&{\cal H}\end{array}\right)\right|=
\left({\cal H}-pv-\bar{p}\bar{v}\right)=-{\cal L}\neq 0\,.\label{transvers}
\end{equation}
Notice that in theories with more than one real field $(d\!\ge\!2)$ both 
${\cal L}$ and ${\cal H}$ have to be nonvanishing quantities to guarantee 
the transversality relation \cite{ka}.
\section{Hamilton--Jacobi theory for one real field in DeDonder and Weyl's 
formulation}
We here restrict ourselves to Lagrangian densities of the following type: 
${\cal L}\!=\! \partial_z\varphi
\partial_{\bar{z}}\varphi\!-\!V(\varphi)$. The potential $V(\varphi)$ is an 
analytic function. Here we have the canonical 
momenta $p\!=\!\bar{v}$, $\bar{p}\!=\!v$, the Hamiltonian density ${\cal H}
\!=\!p\bar{p}+V$ and the slope functions $\Phi\!=\!\partial_{\varphi}
\bar{S}$, $\bar{\Phi}\!=\!\partial_{\varphi}S$. We have the (DeDonder \& 
Weyl)
Hamilton--Jacobi equation
\begin{equation}
\partial_z S+\partial_{\bar{z}} \bar{S} = \partial_{\varphi} S\, \partial_{
\varphi} \bar{S}+V(\varphi)\label{HJE}
\end{equation}
and the related integrability condition 
\begin{equation}
\partial_z\partial_{\varphi}S-\partial_{\bar{z}}\partial_{\varphi}\bar{S} 
=\partial_{\varphi} S\, \partial_{\varphi}^2\bar{S}-
\partial_{\varphi} \bar{S}\, \partial_{\varphi}^2 S \,.\label{ICC}
\end{equation}
Knowing solutions $S$ and $\bar{S}$ of the equations (\ref{HJE}) and 
(\ref{ICC}) a family of embedded extremals $\varphi=\tilde{\varphi}(z,
\bar{z})$ is determined by a system of first order PDEs:
\begin{eqnarray}
\partial_z\tilde{\varphi}(z,\bar{z})\!&\!=\!&\!\Phi=\partial_{\varphi}\bar{S
}(z,\bar{z},\varphi=\tilde{\varphi})\label{qq1}\,, \,\\ 
\partial_{\bar{z}}\tilde{\varphi}(z,\bar{z})\!&\!=\!&\!\bar{\Phi}=\partial_{
\varphi}S(z,\bar{z},\varphi=\tilde{\varphi})\,.\label{qq}
\end{eqnarray}
A solution is obtained by expanding $S(z,\bar{z},\varphi)$ and 
$\bar{S}(z,\bar{z},\varphi)$ in powers of the difference $y\!=\! \varphi -
\varphi_0$ between $\varphi$ and a known extremal $\varphi_0(z,\bar{z})$: 
\begin{equation}
S(z,\bar{z},\varphi)= \sum^{\infty}_{n=0} \frac{1}{n!} A_n(z,\bar{z}) y^n\,
, \quad \bar{S}(z,\bar{z},\varphi)= \sum^{\infty}_{n=0} \frac{1}{n!}
\bar{A}_n(z,\bar{z}) y^n\,.\label{Reihe}
\end{equation}
This method of expanding about a given solution of the 
equations of motion is commonly employed e.g.\ with stability investigations 
or determining (quantum) fluctuations around c--number fields in 
selfinteracting theories \cite{ja}\hspace{0.5mm}\cite{mak}.
\\[2mm]
Naturally $\varphi_0$ has to satisfy the transversality relations (\ref{qq}), 
that fix only the functions $A_1\! = \! \partial_{\bar{z}} \varphi_0$ and 
$\bar{A}_1\! = \! \partial_z \varphi_0 $, without influencing the remaining 
coefficients. Inserting the expressions (\ref{Reihe}) into the HJE 
(\ref{HJE})
and expanding the potential $V$ in powers of $y$ we see that the equation is 
automatically fulfilled up to the order $y^1$, whereas the IC (\ref{ICC})
is fulfilled on the extremals up to the order $y^0$.
\\[2mm] 
$A_0$ and $\bar{A}_0$ are only affected by the HJE of zeroth order in $y$: 
\begin{equation}
\partial_z A_0+\partial_{\bar{z}} \bar{A}_0 = \left.{\cal L}\right|_{\varphi=
\varphi_0} =  \partial_z\varphi_0\partial_{\bar{z}}\varphi_0- V(\varphi_0).
\label{0}
\end{equation}
The general solution of the homogeneous equation $\partial_z A^{hom}_0\!+\!
\partial_{\bar{z}} \bar{A}^{hom}_0\!=\!0$ is given --- at least locally --- 
by $A^{hom}_0\!=\! \partial_{\bar{z}}\chi_0(z,\bar{z})$ and $\bar{A}^{hom}_0
\!= \!- \partial_z\chi_0(z,\bar{z})$. One special solution of the 
inhomogeneous PDE (\ref{0}) can be obtained by integration:
$\bar{A}^{inh}_0\!=\!\int\!\left.{\cal L}\right|_{\varphi\!=\!\varphi_0} 
\,d\bar{z}/2$ and $A^{inh}_0\!=\!\int\!\left.{\cal L}\right|_{\varphi\!
=\!\varphi_0}\,dz/2$. This yields the general solution (locally):
\begin{eqnarray}
A_0(z,\bar{z}) &=& \frac{1}{2}\int\left[ 
\partial_z\varphi_0\partial_{\bar{z}}\varphi_0- V(\varphi_0)\right]dz+
\partial_{\bar{z}}\chi_0(z,\bar{z})\label{A0}\,,
\\
\bar{A}_0(z,\bar{z}) &=&\frac{1}{2}\int\left[\partial_z\varphi_0\partial_{
\bar{z}}\varphi_0- V(\varphi_0)\right]d\bar{z}-\partial_z\chi_0(z,\bar{z})
\label{anull}\,.
\end{eqnarray}
In order to determine the coefficients $A_2$ and $\bar{A}_2$, we insert 
$A_1\! = \! \partial_{\bar{z}} \varphi_0$ and $\bar{A}_1\! = \! 
\partial_z \varphi_0 $ into the IC of order $y$:
\begin{equation}
\partial_zA_2-A_{3}\partial_z\varphi_0+\bar{A}_{1}A_{3}+ \bar{A}_{2}A_{2}
=\partial_{\bar{z}}\bar{A}_2-\bar{A}_{3}\partial_{\bar{z}}\varphi_0+
A_{1}\bar{A}_{3}+A_{2}\bar{A}_{2}\,.\label{1}
\end{equation}
Since $y$ is a function of $z,\bar{z}$, its derivative with respect to $z$ or 
$\bar{z}$ yields $\partial_zy\!=\!-\partial_z\varphi_0$ and $\partial_{\bar{z
}}y\!=\!-\partial_{\bar{z}}\varphi_0$ respectively.
Thus we infer from equation (\ref{1}) that $\partial_z A_2 \! = \! \partial_{
\bar{z}}\bar{A}_2$ which permits to express these two functions at least 
locally by one generating potential function: $A_2 \! = \! \partial_{\bar{z}}
\zeta_2(z,\bar{z})$ and $\bar{A}_2 \! = \! \partial_z \zeta_2(z,\bar{z})$. 
$\zeta_2$ has to fulfil the PDE
\begin{equation}
\partial_{\bar{z}}\partial_z\zeta_2 (z,\bar{z})+\partial_z\zeta_2 (z,\bar{z})
\partial_{\bar{z}}\zeta_2 (z,\bar{z})+\frac{1}{2}\left[\partial_{\varphi}^2
V\left(\varphi\!=\!\varphi_0(z,\bar{z})\right)\right]
=0\,.
\end{equation}
The ansatz
\begin{equation}
\zeta_2  =  \ln{\theta}(z,\bar{z})\quad\Rightarrow\quad 
A_2 =  \partial_{\bar{z}}\ln{\theta}\,,\,\,\, \bar{A}_2  =  
\partial_z \ln{\theta} 
\end{equation}
linearizes the HJE of second order in $y$:
\begin{equation}
\fbox{$\displaystyle\partial_{\bar{z}}\partial_z\theta(z,\bar{z})+\frac{1}{2}
\left[\partial_{\varphi}^2V\left(\varphi\!=\!\varphi_0(z,\bar{z})\right)
\right]\theta(z,\bar{z})=0\,.$}\label{variation}
\end{equation}
\\[1mm]
In order to obtain the expressions for $A_n$ and $\bar{A}_n$ one has to
substitute the power series (\ref{Reihe}) into the IC and the HJE and to 
compare the coefficients of the powers $y^{n-1}$ and $y^n$ respectively. 
Starting with the IC in $(n-1)$--th order:
\begin{eqnarray}
\partial_zA_n-A_{n+1}\partial_z\varphi_0+\sum_{p=0}^{n-1}\left(
\begin{array}{c}n-1\\p\end{array}\right)\bar{A}_{p+1}A_{n-p+1}\nonumber\\
=
\partial_{\bar{z}}\bar{A}_n-\bar{A}_{n+1}\partial_{\bar{z}}\varphi_0+\sum_{
p=0}^{n-1}\left(\begin{array}{c}n-1\\p\end{array}\right)A_{p+1}
\bar{A}_{n-p+1}
\end{eqnarray} 
we draw our attention to the coefficients $A_{n+1}$ and $\bar{A}_{n+1}$ of 
highest order. Fortunately they disappear from this equation as well as 
from the HJE due to the relations $A_1\! = \! \partial_{\bar{z}} \varphi_0$ 
and $\bar{A}_1\! = \! \partial_z \varphi_0$. Provided the coefficients 
$A_0,\ldots, A_{n-1}$ and $\bar{A}_0,\ldots,\bar{A}_{n-1}$ have
already been determined recursively, one gets an equation for the 
functions $A_{n}(z,\bar{z})$ and $\bar{A}_{n}(z,\bar{z})$:
\begin{eqnarray}
\partial_{z}\left(\theta^{n-2}A_n\right)-\partial_{\bar{z}}\left(\theta^{n-2}
\bar{A}_n\right) &=& \theta^{n-2}\sum_{p=2}^{n-2}\left[\left(\!\!
\begin{array}{c}n-1\\p\end{array}\!\!\right)-\left(\!\!\begin{array}{c}n-1
\label{intn}\nonumber\\
p-1\end{array}\!\!\right)\right]\!A_{p+1}\bar{A}_{n-p+1}\\&=:&
\mbox{Inh}\left(A_0,\ldots, A_{n-1};\bar{A}_0,\ldots,\bar{A}_{n-1}\right)
\,.\label{well}
\end{eqnarray}
The Inhomogeneity $\mbox{Inh}\left(A_0,\ldots, A_{n-1};\bar{A}_0,
\ldots,\bar{A}_{n-1}\right)$ vanishes for $n\le 4$. Similarly to 
(\ref{anull}) this hierarchy of equations is solved by splitting the solution 
in two parts: a general solution of the homogeneous part $\partial_z\left( 
\theta^{n-2}A_n^{hom}\right)\! - \!\partial_{\bar{z}}\left(\theta^{n-2} 
\bar{A}^{hom}_n\right)\! = \!0$, that is obtained --- at least locally --- by 
\begin{equation}
A^{hom}_n= \theta^{2-n}\partial_{\bar{z}}\zeta_n (z,\bar{z})\quad\mbox{and}
\quad 
\bar{A}^{hom}_n= \theta^{2-n}\partial_z\zeta_n (z,\bar{z}) 
\end{equation}
with an arbitrary smooth function $\zeta_n(z,\bar{z})$. The second part is a 
special solution of the inhomogeneous equation (\ref{intn}):
\begin{eqnarray}
A_n^{inh} =\theta^{2-n}\partial_{\bar{z}}\bar{\chi}_n(z,\bar{z})\quad
\mbox{and}\quad \bar{A}_n^{inh} = -\theta^{2-n}\partial_z\bar{\chi}_n(z,
\bar{z})\quad\mbox{with}\\
\bar{\chi}_n(z,\bar{z})=\frac{1}{2}\int^{z}\int^{\bar{z}}\mbox{Inh}\left(A_0,
\ldots, A_{n-1};\bar{A}_0,\ldots,\bar{A}_{n-1}\right)\,d{z}^{\prime}\, 
d{\bar{z}}^{\prime}\,.\label{inho}
\end{eqnarray}
The general solution 
\begin{equation}
A_n \! =\! \theta^{2-n}\partial_{\bar{z}}\left[\zeta_n (z,\bar{z})+
\bar{\chi}_n(z,\bar{z})\right]\quad\mbox{and}\quad \bar{A}_n \! =\! 
\theta^{2-n}\partial_{z}\left[\zeta_n (z,\bar{z})-\bar{\chi}_n(z,\bar{z})
\right]
\end{equation}
has to satisfy the HJE (\ref{HJE}) to n--th order in $y$, in which, 
remarkably, the coefficients of highest order $A_{n+1}$ and $\bar{A}_{n+1}$ 
vanish as well as in the IC, due to the relations $A_1\! = \! 
\partial_{\bar{z}} \varphi_0$ and $\bar{A}_1\! = \! \partial_z \varphi_0$.
Separating the functions of the remaining highest order $A_n$, $\bar{A}_n$
leads to the equation:
\begin{equation}
\partial_z\left(\theta^{n}A_n\right)+\partial_{\bar{z}}\left(\theta^n
\bar{A}_n\right) = -\theta^n\sum_{p=2}^{n-2}\left(\!\!
\begin{array}{c}n\\p\end{array}\!\!\right)\!\bar{A}_{p+1}A_{n-p+1}
-\theta^n\left.\frac{d^n}{d\varphi^n}V\left(\varphi\right)
\right|_{\varphi=\varphi_0}
\label{rr}
\end{equation}
It is convenient to set $\zeta_n(z,\bar{z})=\theta^{-1}
\chi_n$ before inserting into expression (\ref{rr}). Using the 
relation (\ref{variation}) the HJE of n--th order in $y$ yields an 
equation, which is nothing but the inhomogeneous extension of the linear PDE  
(\ref{variation}): 
\begin{equation}
\fbox{$\displaystyle\partial_z\partial_{\bar{z}}\chi_n+\frac{1}{2}\left[
\partial_{\varphi}^2V(\varphi_0)\right]\chi_n=\widetilde{\mbox{Inh}}\left(
\chi_0,\ldots, \chi_{n-1}\right)\,,$}\label{allgemein}
\end{equation}
with the inhomogeneity:
\begin{equation}
\widetilde{\mbox{Inh}}\left(\chi_0,\ldots, \chi_{n-1}\right)=
\partial_{\bar{z}}\theta\partial_z \bar{\chi}_n-\partial_z\theta
\partial_{\bar{z}}\bar{\chi}_n-\frac{1}{2}\theta^{n-1}\sum_{p=2}^{n-2}
\left(\!\!\begin{array}{c}n\\p\end{array}\!\!\right)\!\bar{A}_{p+1}A_{n-p+1}
-\frac{1}{2}\theta^{n-1}\partial^n_{\varphi}V(\varphi_0)\,.
\end{equation}
In general the coefficients $A_n$, $\bar{A}_n$ are determined by the 
$n$--th order of the HJE and the $n\!-\!1$ order of the IC for $n\ge 3$:
\begin{equation}
A_n= \theta^{2-n} \,\partial_{\bar{z}}
\left[\frac{\chi_n(z,\bar{z})}{\theta}-\bar{\chi}_n(z,\bar{z})\right]
\,,\quad\bar{A}_n=\theta^{2-n}\, \partial_z
\left[\frac{\chi_n(z,\bar{z})}{\theta}+\bar{\chi}_n(z,\bar{z})\right]\,.
\label{Koeff}
\end{equation}
Notably the infinite hierarchy of functions $\chi_n(z,\bar{z})$ has to fulfil 
{\em only} one PDE of second order: the equation (\ref{allgemein}). 
Its integral can be obtained by determining the general solution of the 
homogeneous PDE (\ref{variation}), {\em which is the same for all orders
$n \ge 2$} and {\em one} special solution of the inhomogeneous equation 
(\ref{allgemein}). It is given by using a Green function that can be 
chosen to be the same for all orders $n$ without any loss of generality.
The solutions of (\ref{variation}) can be obtained by employing B\"ack\-lund 
transformations. 
\section{B\"acklund transformations} 
B\"acklund transformations (BTs) are maps between the tangent bundles of 
integral submanifolds associated with PDEs. If we are able to find a BT
from the PDE which we wish to solve and to another one the general integral 
of which is known, we can obtain the general solution of the first one by 
integrating the BT. This treatment of a single PDE can be generalized to 
systems of partial differential equations \cite{Friede}.
\\[2mm]
By applying BTs we want to reduce the linear PDEs of second order with 
{\em nonconstant} coefficients of type (\ref{variation}) to linear PDEs with 
{\em constant} coefficients.
The inhomogeneous extension of equation (\ref{variation}) could be solved by
BTs, too, but for sake of simplicity we construct a special solution of the
inhomogeneous equation (\ref{variation}) by using a Green function and 
Fourier transformation. Like other authors, e.g.\ \cite{miura}, the general 
form of a BT we are starting from is given by two functions $F_1$, $F_2$:
\begin{eqnarray}
\partial_z\hat{\theta}(z,\bar{z})&=&F_1\left[z,\bar{z},\theta(z,\bar{z}),
\hat{\theta}(z,\bar{z}),\partial_z\theta(z,\bar{z})\right]\,,\nonumber\\
\partial_{\bar{z}}\hat{\theta}(z,\bar{z})& =& F_2\left[z,\bar{z},
\theta(z,\bar{z}),\hat{\theta}(z,\bar{z}),\partial_{\bar{z}}\theta(z,\bar{z})
\right]\,.
\end{eqnarray}
$\theta$ has to fulfil the relation (\ref{variation}), whereas $\hat{\theta}$
denotes the transformed function which is supposed to obey a linear PDE with 
a constant coefficient $m^2$: $\partial_z\partial_{\bar{z}}\hat{\theta}\!-\!
m^2\hat{\theta}\!=\!0$. Of course $m^2$ can be equal to zero which would 
yield the wave equation. Obviously  $F_1$ and $F_2$ have to fulfil 
the integrability condition $\partial_z\partial_{\bar{z}}
\hat{\theta}\!=\!\partial_{\bar{z}}\partial_z\hat{\theta}
\,\Rightarrow\,\mbox{d}F_1/\mbox{d}\bar{z}\!=\!\mbox{d}F_2/\mbox{d}z$.
This integrability condition does not lead to a restriction on the 
solutions of the PDE (\ref{variation}), if such a BT is found. Thus $F_1$ 
and $F_2$ have to fulfil the two equations
\begin{equation}
\mbox{d}F_1/\mbox{d}\bar{z}=\mbox{d}F_2/\mbox{d}z\,,\qquad
\mbox{d}F_1/\mbox{d}\bar{z}=m^2\hat{\theta}\,.\label{transfor}
\end{equation}
We would like to point out that the more general ansatz 
$\partial_z\partial_{\bar{z}}\hat{\theta}\!=\!K(z,\bar{z})\hat{\theta}$ 
leads to the same BTs (\ref{bt}) below at least for the models discussed here
\cite{wi}.
\\[2mm] 
Differentiating the eqs. (\ref{transfor}) with respect to $\partial_z\theta$ 
and $\partial_{\bar{z}}\theta$ leads to
\begin{eqnarray}
\partial_{\theta_z}^2 F_1=0\,,\quad 
&\Rightarrow&\quad F_1=f_1(\theta,\hat{\theta},z,\bar{z})\partial_z\theta
+m_1(\theta,\hat{\theta},z,\bar{z})\,, \\ 
\partial_{\theta_{\bar{z}}}^2 F_2=0\,,\quad&\Rightarrow&\quad
F_2=f_2(\theta,\hat{\theta},z,\bar{z})\partial_{\bar{z}}\theta+m_2(
\theta,\hat{\theta},z,\bar{z})\,.
\end{eqnarray}
Substituting these expressions into eqs. (\ref{transfor}), comparing
the coefficients of $\partial_{\bar{z}}\theta$, $\partial_z\theta$ and 
differentiating with respect to $\theta$, $\hat{\theta}$, we
conclude by lengthy but straightforward calculations that
\begin{eqnarray}
f_1&=&-\,f_2=c_0\,, \quad c_0\in {\cal R}\,,\quad c_0\not= 0\,,\nonumber\\ 
0&=&\partial_{\theta}m_1-f_1\partial_{\hat{\theta}}m_1\,,\quad \Rightarrow
\quad m_1=g_1(f_1\theta+\hat{\theta},z,\bar{z})\,,\\
0&=&\partial_{\theta}m_2+f_1\partial_{\hat{\theta}}m_2\,,\quad \Rightarrow
\quad m_2=g_2(f_1\theta-\hat{\theta},z,\bar{z})\,.
\end{eqnarray}
Inserting these results into relations (\ref{transfor}) and differentiating 
with respect to $f_1\theta\!+\!\hat{\theta}\!=\!
\eta_1$ and $f_1\theta\!-\!\hat{\theta}\!=\!\eta_2$ we obtain:
$\partial_{\eta_1}^2 m_1\!=\!0$, $\partial_{\eta_2}^2 m_2\!=\!0$.
Substituting $f_1\hat{\theta}\!\rightarrow\! \hat{\theta}$, 
$f_1\partial_z\hat{\theta}\!\rightarrow\! \partial_z\hat{\theta}$ and
$f_1\partial_{\bar{z}}\hat{\theta}\!\rightarrow\! \partial_{\bar{z}}
\hat{\theta}$ we get:
\begin{eqnarray}
\partial_z\hat{\theta}&=&\partial_z\theta+\alpha_1(z,\bar{z})\left[\theta+
\hat{\theta}\right]+\beta_1(z,\bar{z})\,,\label{vor1} \\
\partial_{\bar{z}}\hat{\theta}&=&-\partial_{\bar{z}}\theta+\alpha_2(z,
\bar{z})
\left[\hat{\theta}-\theta\right]+\beta_2(z,\bar{z})\,.\label{vor2}
\end{eqnarray}
Considering the two functions $\beta_1$, $\beta_2$ is only necessary for 
transformations between {\em inhomogeneous} PDEs. Thus, we here can choose 
$\beta_1\!=\!0\!=\!\beta_2$. Inserting our results (\ref{vor1}) and 
(\ref{vor2}) into equations (\ref{transfor}) and 
comparing the coefficients of $\theta$, $\hat{\theta}$ we finally obtain the 
linear BT
\begin{equation}
\fbox{$ \partial_z\hat{\theta}=\partial_z\theta+\left\{\theta+\hat{\theta}
\right\}\partial_z\psi\,,$}\quad
\fbox{$ \partial_{\bar{z}}\hat{\theta}=-\,\partial_{\bar{z}}\theta+\left\{
\hat{\theta}-\theta\right\}\partial_{\bar{z}}\psi\,,$}\label{bt}
\end{equation}
with the BT generating function $\psi\!=\!\psi(z,\bar{z})$ which is a 
{\em special} solution of
\begin{equation}
\partial_z\partial_{\bar{z}}\psi-(\partial_z\psi)(\partial_{\bar{z}}\psi)-
\frac{1}{2}\partial_{\varphi}^2V(\varphi_0)=0\,,\quad \partial_z\partial_{
\bar{z}}\psi+
(\partial_z\psi)(\partial_{\bar{z}}\psi)-m^2=0\,.
\label{psi}
\end{equation}
Substituting $\psi\!=\!-\ln\left(\tilde{\psi}\right)$ in these equations we 
conclude
\begin{equation}
\partial_z\partial_{\bar{z}}\tilde{\psi}+\frac{1}{2}\partial_{\varphi}^2V
(\varphi_0)\tilde{\psi}=0\,,\quad\,\,
\partial_z\partial_{\bar{z}}\left(\frac{1}{\tilde{\psi}}\right)-m^2
\left(\frac{1}{\tilde{\psi}}\right)
=0\,.\label{tildepsi}
\end{equation}
We therefore have to solve the following problem:
\\{2mm}
{\em We have to find a solution of the equation {\em (\ref{variation})} the 
inverse of which has to fulfil a Klein--Gordon or a wave equation, then
we can integrate the linear BT and obtain the general solution of 
{\em (\ref{variation})}.}
\\[2mm]
Another method of solving the eqs.\ (\ref{psi}) is based on their linear 
combinations:
\begin{equation}
\partial_z\partial_{\bar{z}}\psi=\frac{1}{4}\left(2m^2+
\partial_{\varphi}^2V(\varphi_0)\right)\,,\quad
(\partial_z\psi)(\partial_{\bar{z}}\psi)=\frac{1}{4}\left(2m^2-
\partial_{\varphi}^2V(\varphi_0)\right)\,.\label{psi2}
\end{equation}
Integrating the first equation and inserting this solution into the second 
expression imposes a restriction on $\partial_{\varphi}^2V(\varphi_0)$. We 
make use of eqs. (\ref{psi2}) when we study the $\phi^4$--model.

\section{Applications}
After solving the homogeneous equations (\ref{variation}), we calculate a 
special solution of their inhomogeneous extension (\ref{allgemein}) by 
determining a Green function --- without need of specifying the 
inhomogeneity {\em $\widetilde{\mbox{Inh}}$}. To obtain a solution of 
equation (\ref{allgemein}) specific for the models under consideration we 
have to fold the Green function with the inhomogeneity in 
every order $y^n$.
\\[2mm]
The Hamilton--Jacobi theory for the non--selfinteracting scalar field 
theories ${\cal L}_0\!=\!\partial_z\varphi\partial_{\bar{z}}\varphi$ and 
${\cal L}_1\!=\!\partial_z\varphi\partial_{\bar{z}}\varphi\!-
\!1/2m^2\varphi^2$ with the light cone variables $z\!=\!(x\!+\!t)/2$ and 
$\bar{z}\!=\!(x\!-\!t)/2$ leads to the wave or the Klein--Gordon equation 
(\ref{variation}) without need for a B\"acklund transformation or
specifying an extremal $\varphi_0(z,\bar{z})$:
\begin{equation}
{\cal L}_0:\,\, \partial_z\partial_{\bar{z}}\chi_n=\widetilde{\mbox{Inh}}
\left(\chi_0,\ldots,\chi_{n-1}\right)\,,\quad 
{\cal L}_1:\,\, \partial_z\partial_{\bar{z}}\chi_n+m^2\chi_n=
\widetilde{\mbox{Inh}}\left(\chi_0,\ldots,\chi_{n-1}\right)\,.
\end{equation}
The general solutions of these relations are known. Therefore we draw our 
attention to the more interesting case of selfinteracting theories:

\subsection{The homogeneous equations}
\subsubsection{Liouville model}
Applying our formalism to Liouville's theory ${\cal L}\!=\!\partial_z\varphi
\partial_{\bar{z}}\varphi\!+\!2\exp(\varphi)$, using an arbitrary solution 
of the equation of motion for which the general expression is known 
\cite{miura}: 
\begin{equation}
\varphi_0\!=\!\ln\left\{2\frac{(\partial_z s(z))(\partial_{\bar{z}} 
\bar{s}(\bar{z}))}{(s\!+\!\bar{s})^2}\right\}\,,
\end{equation}
with arbitrary smooth functions $s(z)$ and $\bar{s}(\bar{z})$, the relation 
(\ref{variation}) yields:
\begin{equation}
\partial_z\partial_{\bar{z}}\theta^L-2\frac{(\partial_z s)(\partial_{\bar{z}} 
\bar{s})}{(s+\bar{s})^2}\theta^L=0\,,\quad\Rightarrow\quad\partial_s
\partial_{\bar{s}}\theta^L(s,\bar{s})-2\frac{1}{(s+\bar{s})^2}
\theta^L(s,\bar{s})=0\label{liou}
\end{equation}
by employing a transformation of variables $z\rightarrow s(z), \bar{z}
\rightarrow\bar{s}(\bar{z})$. Obviously {\em one} special solution of this 
equation is $\theta^L_0=1/(s\!+\!\bar{s})$. Its inverse fulfils the wave 
equation $\partial_s\partial_{\bar{s}}\hat{\theta}^L\!=\!0$. Thus we know 
that there exists at least one BT which connects the integral submanifolds of
(\ref{liou}) and of the wave equation. Returning 
to the equations (\ref{tildepsi}) we conclude that $\psi^L\!=\!\ln(s\!+\!
\bar{s})$. So we can determine the BT by integrating the relations (\ref{bt}) 
and obtain the general solution of the equation (\ref{liou}):
\begin{equation}
\theta^L(s,\bar{s})=\partial_s C(s)-\partial_{\bar{s}}\bar{C}(\bar{s})+
\frac{2}{(s+\bar{s})^2}\{\bar{C}(\bar{s})-C(s)\}\label{end1}
\end{equation}
with two arbitrary smooth functions $C(s)$ and $\bar{C}(\bar{s})$.

\subsubsection{Sine \& Sinh--Gordon model}
In two dimensional space--time two types of solitons exist: the ``bell'' with 
the same asymptotic value at $x\!=\!-\infty$ and $x\!=\!\infty$ and a 
``kink'' soliton with different asymptotic values. Moreover there exist a 
topological conserved quantum number associated with the asymptotic behaviour 
of these solitons. The corresponding conserved current is given by: 
$J^{\mu}\!=\!\epsilon^{\mu\nu}\partial_{\nu}\varphi$ with the antisymmetric 
tensor $\epsilon^{\mu\nu}\!=\!-\epsilon^{\nu\mu},\,\mu,\nu\!=\!0,1$.
Thus the charge associated with this current is: $\displaystyle 
N\!=\!\int^{\infty}_{-\infty}J^0\mbox{d}x\!=\!\varphi|_{x=\infty}\!-\!
\varphi|_{x=-\infty}$, which vanishes obviously in the case of the bell 
solitons. For kinks it is a non--trivial quantum number.
\\[2mm]
The Sine--Gordon theory possesses a infinite hierarchy of
multikink solutions, which can be constructed by Auto--B\"acklund 
transformations. In this model the conserved quantity is associated with a 
particle number. For details as to solitons see e.g.\ \cite{ja}, \cite{sut}, 
\cite{mak} and \cite{Dodd}.
\\[2mm]
We want to embed for the Sine--Gordon model ${\cal L}_2\!=\!\partial_z\varphi
\partial_{\bar{z}}\varphi\!+\!2[1\!-\!\cos(\varphi)]$ the 
(anti-) kink solution $\varphi_0\!=\!\pm 4\arctan[\exp(z\!+\!\bar{z})]$ and 
in the case of the Sinh--Gordon model ${\cal L}_3\!=\!\partial_z\varphi
\partial_{\bar{z}}\varphi\!-\!2[1\!-\!\cosh(\varphi)]$ the bell solution 
$\varphi_0\!=\!\pm4\,\mbox{arctanh}[\exp(z\!+\!\bar{z})]$, which is only 
defined for $z\!+\!\bar{z}\!<\!0$. We then obtain for equation 
(\ref{variation}):
\begin{eqnarray}
\mbox{Sine--Gordon:}\qquad &&\partial_z\partial_{\bar{z}}\theta^{SG}-\{2
\tanh^2(z+\bar{z})-1\}\theta^{SG}=0\,,\label{sg}\\
\mbox{Sinh--Gordon:}\qquad &&\partial_z\partial_{\bar{z}}\theta^{Sh}-\{2
\coth^2(z+\bar{z})-1\}\theta^{Sh}=0\,,\label{sh}
\end{eqnarray}
respectively. Following the discussion of 
Liouville's theory we are able to solve these two equations by one BT. For 
this it is sufficient to realize that the inverse of the two 
solutions $\theta^{SG}_0\!=\!1/\cosh(z\!+\!\bar{z})$ and $\theta^{Sh}_0\!=
\!1/\sinh(z\!+\!\bar{z})$ solve the Klein--Gordon equation $\partial_z
\partial_{\bar{z}}\hat{\theta}\!=\!\hat{\theta}$. The functions $\theta^{
SG}_0\!=\!\tilde{\psi}^{SG}$ and $\theta^{Sh}_0\!=\!\tilde{\psi}^{Sh}$ can be 
calculated by using the relations $\tilde{\psi}\!=\!\exp(-\psi)$ and 
(\ref{xi}), (\ref{A}) shown in the Appendix. Thus the two generating 
functions are $\psi^{SG}\!=\!\ln(\cosh(z\!+\!\bar{z}))$ and $\psi^{Sh}\!=\!
\ln(\sinh(z\!+\!\bar{z}))$ which determine the BTs 
(\ref{bt}) between the Klein--Gordon eq. and the eqs. (\ref{sg}), (\ref{sh}).
Their general solutions can be calculated by integration of the linear BTs 
(\ref{bt}):
\begin{eqnarray}
\theta^{SG}\!\!&\!\!=\!\!&\!\!\int^{\infty}_{-\infty}\!\int^{\infty}_{-\infty}
\!\!\!\exp[-\dot{\imath}(qz+\bar{q}\bar{z})]\delta(q\bar{q}+1)\Upsilon^{SG}
(q,\bar{q})\left\{\!1-\!\frac{2}{q^2+1}
-\frac{2\dot{\imath}q}{q^2+1}\tanh(z\!
+\!\bar{z})\!\right\}\mbox{d}q\,\mbox{d}\bar{q} \nonumber\\ 
&&+c_0\cosh^{-1}(z+\bar{z})\,,\label{Erg 1}\\
\theta^{Sh}\!\!&\!\!=\!\!&\!\!\int^{\infty}_{-\infty}\!\int^{\infty}_{-\infty}
\!\!\exp[-\dot{\imath}(kz+\bar{k}\bar{z})]\delta(k\bar{k}+1)\Upsilon^{Sh}
(k,\bar{k})\left\{\!\frac{\dot{\imath}(k+\bar{k})}{2}-\coth(z+\bar{z})\!
\right\}\mbox{d}k\,\mbox{d}\bar{k}\nonumber\\
&&+c_1\sinh^{-1}(z+\bar{z})\,,\label{Erg sg}
\end{eqnarray}
with arbitrary constants $c_0,c_1 \in {\cal R}$ and two arbitrary 
functions $\Upsilon^{SG}(q,\bar{q}),\Upsilon^{Sh}(k,\bar{k})$ which have to 
be chosen in such a way that the integrals exist. A property of this 
BTs is that the solution which was used for the transformation, is 
multiplied with a constant and added to the modified solution of the 
Klein--Gordon or the wave equation.
\\[2mm]
This static kink $\varphi_0\!=\!\pm 4\arctan[\exp(x)]$
can be transformed by a Lorentz boost into time dependent solutions of the
relativistic invariant Euler--Lagrange equation $\varphi_0\!=\!
\pm 4\arctan[\exp(\gamma (x\!-\!vt)+\delta)],\gamma^2\!=\!1/(1\!-\!v^2)$ 
parametrized by the velocity $v$ and the phase shift $\delta$. We are able to 
include the embedding of these solutions in our discussion making use of the 
transformation of variables 
\begin{equation}
z\rightarrow w=z\gamma(1-v)+\delta/2\,,\qquad
\bar{z}\rightarrow\bar{w}=\bar{z}\gamma(1+v)+\delta/2\,.
\label{lorentz}
\end{equation}
The same holds for the static solitons of the relativistic covariant Sinh--,  
and $\phi^4$--models.

\subsubsection{$\phi^4$--model}
Contrary to the three previous models the following ones can only be solved 
by at least two BTs: the $\phi^4$--theories 
\begin{equation}
\mbox{I)}\quad {\cal L}=\partial_z
\varphi\partial_{\bar{z}}\varphi-4\varphi^2+2\varphi^4\quad\mbox{and} \quad
\mbox{II)}\quad {\cal L}=\partial_z\varphi\partial_{\bar{z}}\varphi+2
\varphi^2-2\varphi^4\,. 
\end{equation}
Both theories have soliton solutions. We choose the (anti-) kink 
$\varphi_0^{I}\!=\!\pm\tanh(z\!+\!\bar{z})$ and in case II the bell solution 
$\varphi_0^{II}\!=\!\pm\cosh^{-1}(z\!+\!\bar{z})$. Then we get the two 
relations (\ref{variation}) for $\theta^I$ and $\theta^{II}$:
\begin{eqnarray}
\mbox{I)}&:&\,\,
\partial_z\partial_{\bar{z}}\theta^{I}-\{6\tanh^2(z+\bar{z})-2\}
\theta^{I}=0\,,\label{phi}\\
\mbox{II)}&:&\,\,\partial_z\partial_{\bar{z}}
\theta^{II}-\{6\tanh^2(z+\bar{z})-5\}\theta^{II}=0\,.\label{KG3}
\end{eqnarray}
Except for the special values of same of the constants they are the same 
PDEs as that of the Sine--Gordon model, but they
cannot be solved by {\em one} BT only (see Appendix). Therefore we 
employ two BTs for each model: the first BTs leads to two eqs.\ in which 
the coefficient in front of $\tanh^2(z\!+\!\bar{z})$ is reduced to 2, the 
same as the one in the Sine--Gordon theory. This allows us to obtain two 
Klein--Gordon eqs.\ which differ by the choice of $m^2$ after a second BT 
for each model.
\\[2mm]
Inserting the function $\partial_{\varphi}^2V(\varphi_0)\!=\!v(z\!+\!\bar{z}
)$ in the results (\ref{K}) and (\ref{A}) of the Appendix we are able 
to calculate the functions $\psi^{I}$, $\psi^{II}$ and the 
two solutions of the equations (\ref{phi}), (\ref{KG3}) which are necessary 
for the transformations:
\begin{equation}
\mbox{I):}\,\,\theta^{I}_0=\cosh^{-2}(z+\bar{z})\,,\quad \mbox{II)}\,\,\,
\theta^{II}_0=\exp[d_0(z-\bar{z})]\cosh^{-2}(z+\bar{z})\,,\,\,d_0^2=3\,.
\end{equation}
Thus, with $\psi^{I}\!=\!2\ln\{\cosh(z\!+\!\bar{z})\}$ and 
$\psi^{II}\!=\!d_0(\bar{z}\!-\!z)\!+\!2\ln\{\cosh(z\!+\!\bar{z})\}$
we obtain after one BT and denoting the function $\hat{\theta}$ of eq. 
(\ref{bt}) by $\tilde{\theta}$:
\begin{eqnarray}
\mbox{I)}&:&\partial_z\partial_{\bar{z}}\tilde{\theta}^{I}-\left\{
2\tanh^2(z+\bar{z})+2\right\}\tilde{\theta}^{I}=0\,,\\
\mbox{II)}&:&\partial_z\partial_{\bar{z}}\tilde{\theta}^{II}-
\left\{2\tanh^2(z+\bar{z})-1\right\}\tilde{\theta}^{II}=0\,.
\end{eqnarray}
The second relation is the same as in the Sine--Gordon theory. Thus we 
only have to treat the first case here. This PDE has the special solution 
$\tilde{\theta}^{I}_0\!=\!1/\cosh(z\!+\!\bar{z})\exp[\dot{\imath}d_1(z\!-\!
\bar{z})],\,d_1^2\!=\!3$ the inverse of which fulfils a Klein--Gordon 
equation, namely: $\partial_z\partial_{\bar{z}}(1/\tilde{\theta}^I_0)\!=\!4/
\tilde{\theta}^I_0$. The solution $\tilde{\theta}^I_0$ was found according to 
the method discussed in the Appendix. The generating function for this BT is 
$\psi^{I}_2\!=\!\ln\{\cosh(z\!+\!\bar{z})\}\!+\!\dot{\imath}d_1(\bar{z}\!-\!
z)\!=\!-\ln(\tilde{\theta}^I_0)$. Hence the resulting solutions of the 
eqs. (\ref{phi}) and (\ref{KG3}) are:
\begin{eqnarray}
\theta^{I}&=&\int^{\infty}_{-\infty}\int^{\infty}_{-\infty}
\Upsilon^{I}(q,\bar{q})
\delta(q\bar{q}+4)\exp[-\dot{\imath}(qz+\bar{q}\bar{z})]\left\{q^4+16-4q^2
\right. \nonumber\\
&&\left. -\,(6\dot{\imath}q^3-24\dot{\imath}q)\tanh(z+\bar{z})-
12q^2\tanh^2(z+\bar{z})\right\}\,\mbox{d}q\,\mbox{d}\bar{q}\nonumber\\
&&+c_0\cosh^{-2}(z+\bar{z})
+c_1\cosh^{-1}(z+\bar{z})\tanh(z+\bar{z})\exp[\dot{\imath}d_1(z-\bar{z})]\,,
\label{Erg 2}\\ 
\theta^{II}&=&\int^{\infty}_{-\infty}\int^{\infty}_{-\infty}
\Upsilon^{II}(k,\bar{k})\delta(k\bar{k}+1)\exp[-\dot{\imath}(kz+\bar{k}
\bar{z})]\left\{-(k^2+1)^2\right.\nonumber\\
&&\left.-\,6\dot{\imath}k(1-k^2)\tanh(z+\bar{z})+12k^2\tanh^2(z+\bar{z})
\right\}\,\mbox{d}k\,\mbox{d}\bar{k}\nonumber\\
&&+c_2\cosh^{-2}(z+\bar{z})\exp[d_0(z-\bar{z})]+c_3\cosh^{-1}(z+\bar{z})
\tanh(z+\bar{z})\label{Erg K}
\end{eqnarray}
with $d_0^2\!=\!d_1^2\!=\!3$; $c_0,c_1,c_2,c_3\in{\cal R}$ and two arbitrary
functions $\Upsilon^{I}(q,\bar{q}),\Upsilon^{II}(k,\bar{k})$.
\\[3mm]
Expressions like (\ref{sg}) and (\ref{phi}) also occur e.g.\ in 
stability investigations or in discussion small fluctuations around the 
known soliton solutions $\varphi_0$ of these theories 
\cite{ja}\hspace{0.5mm}\cite{mak}. 
Carrying out the second variation of the action functional in the case of 
the Sine--Gordon and the $\phi^4$--theory employing the soliton solutions 
given above  --- or equivalently inserting $\varphi\!=\!\varphi_0\!+\!
\varphi_c,\,|\varphi_c|\ll |\varphi_0|$ into the Euler--Lagrange equation 
of the Sine--Gordon or the $\phi^4$--theory --- yields:
\begin{equation}
\left[-\partial_x^2+n(n+1)\tanh^2(x)+c_0\right]u(x)=\omega^2u(x)
\,,\label{cnum}
\end{equation}
where $\varphi_c(x,t)\!=\!\exp(\dot{\imath}\omega t)u(x)$. We have $n\!=\!1$, 
$c_0\!=\!2$ for the Sine--Gordon and $n\!=\!2$, $c_0\!=\!-2$ for the $\phi^4
$--model. Stability of the soliton solutions requires that all eigenvalues 
$\omega^2$ of this Schr\"odinger--like equation should be non--negative, so 
that small perturbations about $\varphi_0$ do not grow 
exponentially in time. The lowest eigenvalue is $\omega^2\!=\!0$ and the 
corresponding solution $\hat{\varphi}_c$ is the translation mode. It must be 
present, because of the translation invariance of our models under 
consideration. Remarkably we have used it for both models to 
construct the B\"acklund transformations: $\hat{\varphi}_c^{SG}\!=\!
1/\cosh (x\!=\!z\!+\!\bar{z})$ and $\hat{\varphi}_c^{I}\!=\!
1/\cosh^{2}(x)$.
\\[2mm]
The eqs.\ (\ref{sg}) are (\ref{phi}) reduced to eqs.\ (\ref{cnum}) if one
if one sets $x\! = \!z \! + \! \bar{z}$, $t\! = \!z\! - \!\bar{z}$ and 
$\exp(\dot{\imath}\omega t)u(x)$. The eqs.\ (\ref{cnum}) are solvable by 
transforming them into hypergeometric differential equations the solutions of
which can be given by {\em finite} series in powers of $\tanh(x)$ 
functions \cite{morse}:
\begin{eqnarray}
\underline{\mbox{SG}}:\quad \omega_k^2=k^2+1&:&\,
u(x)=\exp(\dot{\imath}kx)\{k+\dot{\imath}\tanh(x)\}\,,\\
\underline{\phi^4_I}:\qquad\,\quad \omega^2=3&:&\, u(x)=\tanh(x)/
\cosh(x)\,,\\
\omega_k^2=k^2+4&:&\,
u(x)=\exp(\dot{\imath}kx)\left\{-1-k^2-3k\dot{\imath}\tanh(x)+
3\tanh^2(x)\right\}\,.
\end{eqnarray}
These solutions are contained in our more general results (\ref{Erg 1}) and 
(\ref{Erg 2}).

\subsubsection{A mathematical remark}
As discussed in the Appendix our results can be generalized in order to 
reduce the hierarchy of linear PDEs: 
\begin{equation}
\partial_z\partial_{\bar{z}}\theta=\{n(n+1)\eta^2+a\}\theta\,,\quad a\in
{\cal R}\,,\quad n=0,1,2,\ldots\label{hie}
\end{equation}
by $n$ BTs to a Klein--Gordon or a wave equation successively. One BT can 
raise or lower the coefficient $n$ to $n\!+\!1$ or $n\!-\!1$. We have to 
assume that the smooth function $\eta(l)$ fulfils the nonlinear differential 
equation: $\partial_l\eta\!=\!\bar{b}\eta^2\!+\!\bar{c}$ with 
$\bar{c}\in{\cal R}$, $\bar{b}\!=\!\pm1$ and $l\!=\!z\!+\!\bar{z}$:
\begin{eqnarray}
\frac{\bar{b}}{\bar{c}}>0&:&\,\,
\eta=\sqrt{\bar{|c|}}\tan\left[\sqrt{\bar{|c|}}(l+l_0)\right]\,,\,\,\bar{c}
\in {\cal R}\,,\,\,l_0\in {\cal C}\,,\\
\frac{\bar{b}}{\bar{c}}<0&:&\,\,
\eta=-\sqrt{\bar{|c|}}\tanh\left[\sqrt{\bar{|c|}}(l+l_0)\right]\,,\quad
\bar{c}=0:\,\,\eta=\frac{1}{l_0-\bar{b}l}\,.\label{eta}
\end{eqnarray} 
The PDEs
(\ref{hie}) and (\ref{eta}) are solvable without need of specifying the 
constants $a$ or $\bar{c}$. 
\\[2mm]
Obviously the PDEs (\ref{phi}), (\ref{KG3}), (\ref{sg}) and (\ref{sh}) 
correspond to special choices of $a,\bar{b},\bar{c}$ and $n$: $\bar{b}\!=\!
-1$, $\bar{c}\!=\!1$, $a\!=\!-1$ and $n\!=\!1$ give the Sinh \& the 
Sine--Gordon models, whereas $n\!=\!2$, $\bar{b}\!=\!-1$, $\bar{c}\!=\!1$ 
give the $\phi^4$--theory with $a\!=\!-2$ (case I) and $a\!=\!-1$ 
(case II).

\subsection{The inhomogeneous equation}
A special solution of the inhomogeneous equation (\ref{allgemein}) can be 
obtained for the Sinh \& Sine--Gordon and both cases of the $\phi^4$--theory 
by employing Fourier transformations.  We discuss these theories first and 
return to the Liouville model later.

\subsubsection{Sinh \& Sine--Gordon, $\phi^4$ equations}
We introduce the Green function $\mbox{G}(z,\bar{z},\hat{z},\hat{\bar{z}})$ 
for the inhomogeneous equations (\ref{allgemein}):
\begin{equation}
\partial_z\partial_{\bar{z}}\mbox{G}+\frac{1}{2}\left\{\partial_{\varphi}^2V
(\varphi_0)\right\}\mbox{G}=\delta(z-\hat{z})\delta(\bar{z}-\hat{\bar{z}})
\,.\label{green1}
\end{equation}
If a Green function is found, we can calculate the solutions $\chi_n$ of eq.\ 
(\ref{allgemein}) to all orders of $y^n$:
\begin{eqnarray}
\chi_n=\frac{1}{2\pi}\int_{-\infty}^{+\infty}\int_{-\infty}^{+\infty}
\widetilde{\mbox{Inh}}\left\{\chi_0(\hat{z},\hat{\bar{z}}),\ldots,
\chi_{n-1}(\hat{z},\hat{\bar{z}})\right\}
\mbox{G}(z,\bar{z},\hat{z},\hat{\bar{z}})\,\mbox{d}\hat{z}\,
\mbox{d}\hat{\bar{z}}\,.
\end{eqnarray}
Introducing the Fourier transform $\widetilde{\mbox{G}}$ by:
\begin{equation}
\mbox{G}(z,\bar{z},\hat{z},\hat{\bar{z}})=\frac{1}{4\pi^2}
\int_{-\infty}^{+\infty}\int_{-\infty}^{+\infty}\exp\{\dot{\imath}[q(\hat{z}-
z)+\bar{q}(\hat{\bar{z}}-\bar{z})]\}\widetilde{\mbox{G}}(z,\bar{z},q,\bar{q})
\,\mbox{d}q\,\mbox{d}\bar{q}\,.\label{FT}
\end{equation}
As we need only a special $G$ we try for $\tilde{G}$ the ansatz that depends 
on $q$, $\bar{q}$ and $l \! = \! z \! + \! \bar{z}$ only. We then obtain
\begin{equation}
1 = \partial_l^2 \widetilde{\mbox{G}} - \dot{\imath} (q+\bar{q}) \partial_l 
\tilde{\mbox{G}} - \frac{1}{2}\partial_{\varphi}^2V \widetilde{\mbox{G}} \,. 
\label{ups}
\end{equation}
In our cases the potential term is only a function of $l$: 
$\partial_{\varphi}^2V\!=\! 2a_0 \!+\! 2n[n \! + \! 1]f^2(l),$ $a_0\in 
{\cal R}$, $n=1,2$ and $f$ denotes a $\tanh(l)$ (see (\ref{sg}), (\ref{KG3}) 
and (\ref{phi})) or $\coth(l)$ (see (\ref{sh})). In order to obtain one
solution of the PDE (\ref{ups}) it is sufficient to assume any $f$ that 
solves  $\partial_lf\!=\!1\!-\!f^2$. Since (\ref{ups}) is only a linear 
differential equation we are able to calculate a special inhomogeneous 
solution, if we know a homogeneous one:
\begin{equation}
\widetilde{\mbox{G}}_{inh}(q,\bar{q},l)=\int^{l}\widetilde{\mbox{G}}_{hom}
(q,\bar{q},\hat{l})^{-2}\exp(\dot{\imath}[q+\bar{q}]\hat{l})\int^{\hat{l}}
\widetilde{\mbox{G}}_{hom}(q,\bar{q},\tilde{l})\exp(-\dot{\imath}[q+\bar{q}]
\tilde{l})\,\mbox{d}\tilde{l}\,\mbox{d}\hat{l}\,.
\end{equation}
Substituting $\widetilde{\mbox{G}}_{hom}(q,\bar{q},l)\!=\!\exp[\dot{\imath}(q
\!+\!\bar{q})l/2]\widehat{\mbox{G}}(q,\bar{q},l)$ leads to:
\begin{equation}
\partial_l^2\widehat{\mbox{G}}-\left\{a_0-\frac{(q-\bar{q})^2}{4}+
n(n+1)f^2\right\}\widehat{\mbox{G}}=0\,.\label{green}
\end{equation}
Choosing the special ansatz: $\widehat{\mbox{G}}\!=\!(c_0\!+\!c_1f\!)\exp(c_3
l)$ and comparing the coefficients of powers of $f$ we get for $n\!=\!1$ with 
$a \! = \! a_0 \! - \! (q-\bar{q})^2/4$:
\begin{eqnarray}
a\not=-2&:&\,\, c_3=\pm \sqrt{a+2}\,,\,\,c_0=-c_1c_3\,,\,\,c_1\in {\cal R}\,,
\,\,c_1\not=0\,,\label{in}
\\
a=-2&:&\,\, c_3=0=c_0\,,\,\,c_1\in {\cal R}\,,\,\,c_1\not=0\,.
\end{eqnarray}
This choice of parameters provides the Green function for the Sinh \& 
Sine--Gordon theories (\ref{sh}) and (\ref{sg}). 
The case $n=2$ is of interest within the $\phi^4$--models (\ref{phi}) and 
(\ref{KG3}). Here we have to add the term $c_2f^2\exp(c_3h)$ to $\widehat{
\mbox{G}}$. This yields finally:
\begin{eqnarray}
a\not=-6&:&\,\,3c_0=c_2(5+a)\,,\,\,c_2=-\frac{c_1}{c_3}\,,\,\,c_3=\pm
\sqrt{a+6}\,,\,\,c_1\in{\cal R}\,,\,\,c_1\not=0\,,\label{in2}\\
a=-6&:&\,\,c_1=c_3=0\,,\,\,3c_0=-2c_2\,,\,\,c_2\in{\cal R}\,,\,\,c_2\not=0\,.
\end{eqnarray}
The sign of $c_3$ has to be chosen in such a way that the Fourier integral 
for the Green function $\widetilde{\mbox{G}}(z,\bar{z},q,\bar{q})$ exists.
\\[2mm]
Inserting these results into the expression for $\widetilde{\mbox{G}}_{inh}$
and inverting the Fourier transformation we are able to obtain --- in
principle --- the solution of equation (\ref{allgemein}) for every order of
$y^n$ by integration.

\subsubsection{Liouville Theorie}

For the Liouville model we start with equation (\ref{green1}), too and obtain 
instead of eqs.\ with relation (\ref{green}), where $l \! = \! s(z) \! + \! 
\bar{s}(\bar{z})$:
\begin{equation}
\partial_l^2\widehat{\mbox{G}} - \left\{ -\frac{(q - \bar{q})^2}{4} +
\frac{2}{l^2}\right\} \widehat{\mbox{G}} = 0\,.
\end{equation}
The solution is
\begin{equation}
\widehat{\mbox{G}}=c_0\left\{\frac{1}{l}a\sin(al)-a^2\cos(al)\right\}
+c_1\left\{\frac{1}{l}a\cos(al)+a^2\sin(al)\right\}
\,,\quad a=\pm \frac{(q - \bar{q})}{2}\,.
\end{equation}
Obviously we can not integrate $\widetilde{\mbox{G}}_{inh}$ explicitly, 
if use of this solution. However, the Green function G can be obtained
for the Liouville model. We return to equation (\ref{green1}), choose the 
ansatz $\mbox{G}(s(z),\hat{s},\bar{s}(\bar{z}),\hat{\bar{s}})\!=\!\mbox{H}(
s(z)\!-\!\hat{s})\mbox{H}(\bar{s}(\bar{z})\!-\!\hat{\bar{s}})\bar{\mbox{G}}(
s(z),\hat{s},\bar{s}(\bar{z}),\hat{\bar{s}})$, where H is the usual 
Heaviside step function. We insert the ansatz into eq.\ (\ref{FT}) and obtain
\begin{eqnarray}
0&=&\mbox{H}(s-\hat{s})\mbox{H}(\bar{s}-\hat{\bar{s}})\left\{\partial_s
\partial_{\bar{s}}-\frac{2}{(s+\bar{s})^2}\right\}\bar{\mbox{G}}+
\delta(s-\hat{s})\delta(\bar{s}-\hat{\bar{s}})[\bar{\mbox{G}}-1]\nonumber\\
&&+\mbox{H}(s-\hat{s})\delta(\bar{s}-\hat{\bar{s}})\partial_{s}
\bar{\mbox{G}}+\mbox{H}(\bar{s}-\hat{\bar{s}})\delta(s-\hat{s})
\partial_{\bar{s}}\bar{\mbox{G}}\,.
\end{eqnarray}
In order to solve this equation we have to find one solution 
$\bar{\mbox{G}}$ of the homogeneous equation (\ref{liou}) with the following 
properties: $\bar{\mbox{G}}(s\!=\!\hat{s},\bar{s}\!=\!\hat{\bar{s}})\!=\!1$,
$\partial_{\bar{s}}\bar{\mbox{G}}|_{s=\hat{s}}\!=\!0$ and 
$\partial_s\bar{\mbox{G}}|_{\bar{s}=\hat{\bar{s}}}\!=\!0$. It can easily be
verified that a solution is given by
\begin{equation}
\bar{\mbox{G}}=\frac{1}{\hat{s}+\hat{\bar{s}}}\left\{2s(z)-
\hat{s}+\hat{\bar{s}}-\frac{2}{s(z)+\bar{s}(\bar{z})}
[s(z)+\hat{\bar{s}}][s(z)-\hat{s}]\right\}\,.\label{inL}
\end{equation}
Thus we have determined an explicit expression for the Green function 
$\mbox{G}\!=\!\mbox{H}(z\!-\!\hat{z})\mbox{H}(\bar{z}\!-\!\hat{\bar{z}})
\bar{\mbox{G}}$.

\section{Related extremals}
Having determined solutions of the HJE (\ref{HJE}) combined with the IC 
(\ref{ICC}) associated with a given extremal in terms of power series we now
want to indicate how new extremals can be generated from a given one.
\\[2mm]
In order to connect the functions $\chi_n$ from equation (\ref{allgemein}),
with a one parameter family of extremals $\tilde{\varphi}
(z,\bar{z},u)$ embedded, we expand $\tilde{y}\!=\!(\tilde{\varphi}\!-\!
\varphi_0(z,\bar{z}))$ in the parameter $u$ of the 
solutions of the equation of motion. Therefore we have to consider the two 
dimensional submanifold $\Sigma\!:=\!\{(z,\bar{z},\tilde{\varphi}(z,\bar{z}))
\}$ of the extended configuration space ${\cal M}_{2+1}\!=\!\{(z,\bar{z},
\varphi)\}$. The starting points of this calculation are the slope functions 
(\ref{qq}):
\begin{eqnarray}
\partial_z\tilde{\varphi}(z,\bar{z})=\partial_{\varphi}\bar{S}|_{\varphi=
\tilde{\varphi}}=\bar{A}_1+\sum^{\infty}_{i=2}\frac{1}{(i-1)!}\bar{A}_i
\tilde{y}^i\,,\quad&\Rightarrow&\quad \partial_z\tilde{y}=\sum^{\infty}_{i=1}
\frac{1}{i!}\bar{A}_{i+1}\tilde{y}^{i+1}\,,\label{v1}\\
\partial_{\bar{z}}\tilde{\varphi}(z,\bar{z})=\partial_{\varphi}
S|_{\varphi=\tilde{\varphi}}=A_1+\sum^{\infty}_{i=2}\frac{1}{(i-1)!}A_i
\tilde{y}^i\,,\quad&\Rightarrow&\quad \partial_{\bar{z}}\tilde{y}=
\sum^{\infty}_{i=1}\frac{1}{i!}A_{i+1}\tilde{y}^{i+1}\label{v2}
\end{eqnarray}
with $\tilde{y}\!=\!\tilde{\varphi}\!-\!\varphi_0$, $A_1=\partial_{\bar{z}}
\varphi_0$ and $\bar{A}_1=\partial_z\varphi_0$. Expanding $\tilde{y}$ in a
power series of $u$ 
\begin{equation}
\tilde{y}(z,\bar{z},u)=\tilde{\varphi}(z,\bar{z},u)-\varphi_0(z,\bar{z})=u
\sum^{\infty}_{k=0}\Lambda_k(z,\bar{z})\frac{u^k}{k!}\label{ex}
\end{equation}
yields:
\begin{eqnarray}
\partial_{\bar{z}}\sum^{\infty}_{k=0}\frac{1}{k!}\Lambda_ku^{k+1}
&=&\sum^{\infty}_{i=1}\frac{1}{i!}A_{i+1}\left\{\sum^{\infty}_{j=0}
\!\frac{1}{j!}\Lambda_ju^{j+1}\right\}^i\,,\label{reihe1}\\
\partial_z\!\sum^{\infty}_{k=0}\frac{1}{k!}\Lambda_ku^{k+1}
&=&\sum^{\infty}_{i=1}\frac{1}{i!}\bar{A}_{i+1}\left\{\sum^{\infty}_{j=0}
\frac{1}{j!}\Lambda_ju^{j+1}\right\}^i\,.\label{reihe2}
\end{eqnarray}
Comparing the coefficients in $l\!+\!1$--th order leads to:
\begin{eqnarray} 
\partial_{\bar{z}}\Lambda_l&=&A_2\Lambda_l+\mbox{Inh}(A_2,...,A_{l+1};\Lambda_0
,...,\Lambda_{l-1})\,,\\
\partial_z\Lambda_l&=&\bar{A}_2\Lambda_l+\mbox{Inh}(\bar{A}_2,...,\bar{A}_{l+1}
;\Lambda_0,...,\Lambda_{l-1})\,,
\end{eqnarray}
with the inhomogeneities Inh which depends on  $A_i$, $\bar{A}_i$
calculated above and the functions of lower order $\Lambda_j, j<l$.
Making use of $A_2\!=\!\partial_{\bar{z}}\ln(\theta)$ and $\bar{A}_2\!=\!
\partial_z\ln(\theta)$ we get:\vspace{-5mm}
\begin{center}
\fbox{\parbox{12.5 cm}{\begin{eqnarray}
\Lambda_l&=&\theta\left\{\int^z\frac{1}{\theta}\mbox{Inh}
(\bar{A}_2,...,\bar{A}_{l+1};\Lambda_0,...,\Lambda_{l-1})\,\mbox{d}\hat{z}+
\mbox{const}\right\}\nonumber\\
&=&\theta\left\{\int^{\bar{z}}\frac{1}{\theta}\mbox{Inh}(A_2,...,A_{l+1};
\Lambda_0,...,\Lambda_{l-1})\,\mbox{d}\hat{\bar{z}}+\mbox{const}\right\}\,.
\end{eqnarray}}}\vspace{-5mm}
\end{center}
So the solutions of the Euler--Lagrange 
equations $\tilde{\varphi}\!=\!\varphi_0\!+\!\sum^{\infty}_{k=0}\Lambda_k
u^{k+1}/k!$ can be obtained by successive integration of the 
coefficients $\Lambda_l$. Regarding the lowest order ($l\!=\!0$) $\partial_z
\Lambda_0\!=\!\bar{A}_2\Lambda_0$, $\partial_{\bar{z}}\Lambda_0\!=\!A_2
\Lambda_0$ leads to:
\begin{equation}
\Lambda_0=c_0\theta\,,\quad c_0\in{\cal R}\,,\quad\Rightarrow\quad
\tilde{y}=u c_0\theta+\ldots\,.
\end{equation}
Since $\theta$ obeys the linear PDE (\ref{variation}) $c_0$ can always be
absorbed into it. $\theta$ is discussed in the ch. Applications above see
(\ref{Erg 1}), (\ref{Erg sg}), (\ref{Erg 2}) and (\ref{Erg K}).
\\[2mm]
Thus the one parametric family of extremals in the vicinity (u $\ll 1$) of the

original solution of the equation of motion $\varphi_0(z,\bar{z})$ is 
determined by:
\begin{equation}
\fbox{$\tilde{\varphi}(z,\bar{z},u)=\varphi_0(z,\bar{z})+u 
\theta(z,\bar{z})\,.$}\label{klein}
\end{equation}
\\[2mm]
We compare our considerations with an expansion of the field variable 
$\tilde{\varphi}$ within the Euler--Lagrange equation. $\tilde{\varphi}$ is 
expanded in the parameter of the families of the extremals denoted as $u$:
\begin{equation}
\tilde{\varphi}(z,\bar{z},u)=\tilde{\varphi}_0(z,\bar{z})+
y(z,\bar{z},u)=\tilde{\varphi}_0(z,\bar{z})+u\sum^{\infty}_{n=0}
\frac{1}{n!}\Lambda_n(z,\bar{z})u^n\,.
\end{equation}
$\tilde{\varphi}_0(z,\bar{z})$ is an arbitrary extremal. The potential
is assumed to be a analytic function of $\tilde{\varphi}$:
\begin{equation}
V(\tilde{\varphi})=\!\sum^{\infty}_{m=0}\frac{1}{m!}\partial_{\tilde{\varphi}}
^mV(\tilde{\varphi})|_{\tilde{\varphi}=\tilde{\varphi}_0}(\tilde{\varphi}-
\tilde{\varphi}_0)^m=\!\sum^{\infty}_{m=0}\frac{1}{m!}\partial_{\tilde{
\varphi}}^mV(\tilde{\varphi})|_{\tilde{\varphi}=\tilde{\varphi}_0}u^m\left
\{\sum^{\infty}_{n=0}\!\frac{1}{n!}\Lambda_n(z,\bar{z})u^n\right\}^m.
\end{equation}
Similar to the results of chapter 3, where we study the Hamilton--Jacobi 
theory, we obtain by comparing the coefficients of equal order of $u^n$: 
\begin{equation}
u^0:\quad \partial_z\partial_{\bar{z}}\tilde{\varphi}_0(z,\bar{z})+
\frac{1}{2}\partial_{\tilde{\varphi}}V(\tilde{\varphi})|_{\tilde{\varphi}=
\tilde{\varphi}_0}=0\,,
\end{equation}
which is fulfilled due to assumption and:
\begin{eqnarray}
u^1&:&\quad \partial_z\partial_{\bar{z}}\Lambda_0+
\frac{1}{2}\partial_{\tilde{\varphi}}^2V(\tilde{\varphi}_0)\Lambda_0=0\,,\\
u^{n+1}&:& \partial_z\partial_{\bar{z}}\Lambda_n+\frac{1}{2}\partial_{\tilde{
\varphi}}^2V(\tilde{\varphi}_0)\Lambda_n=\mbox{Inh}(\Lambda_0...\Lambda_{n-1},
\partial_{\tilde{\varphi}}^2V(\tilde{\varphi}_0)...\partial_{\tilde{\varphi}}
^{n+1}V(\tilde{\varphi}_0))\,.\label{Lambda}
\end{eqnarray}
The first one of these PDEs yields:
\begin{equation}
\tilde{\varphi}=\tilde{\varphi}_0+u\Lambda_0\,.
\end{equation}
Because $\Lambda_0$ has to fulfil the same linear equation as $\theta$, 
which we introduced in the Hamilton--Jacobi theory (\ref{variation}), we are 
able to identify $\Lambda_0$ and $\theta$ with each other.
Thus our result (\ref{klein}) is {\em equivalent} to a 
second variation of the action functional, which is commonly employed e.g.\ 
with semiclassical considerations \cite{ja} and stability investigations
\cite{mak}.
\\[2mm]
The PDE (\ref{Lambda}) is analogous to the equation (\ref{allgemein}),
which we obtained in the Hamilton--Jacobi framework. Both can be used
to determine the fluctuations in a neighbourhood of a given extremal 
$\tilde{\varphi}_0$ in every order of $u$. 
\\[2mm]
If one is able to find the general integral of the Hamilton--Jacobi equation
and the integrability condition, the general solution of the Euler--Lagrange
equation can be obtained.

\subsection{An example}
Here we would like to calculate the Hamilton--Jacobi functions $S(z,\bar{z},
\varphi)$, $\bar{S}(z,\bar{z},\varphi)$ and a related family of extremals 
$\tilde{\varphi}(z,\bar{z},u)$ according to the formalism developed in the
chapters 4, 6 and 7. For this we choose the Sine--Gordon model with the
1--kink--solution $\varphi_0 \!=\! 4\arctan(\exp(z\!+\!\bar{z}))$.
Though the functions $S$, $\bar{S}$ and the related extremals are determined
perturbatively the corresponding formal series (\ref{Reihe}), (\ref{ex})
can be obtained explicitly.

\subsubsection{The Hamilton--Jacobi potentials $S$, $\bar{S}$}
The coefficients $A_n(z,\bar{z})$ and $A_n(z,\bar{z})$ (\ref{Reihe}) 
are determined by calculating the functions $\chi_n(z,\bar{z})$
and $\bar{\chi}_n(z,\bar{z})$ as solutions of the PDEs (\ref{allgemein}) in 
every order $n\!=\!0,1,2,\ldots$. 
\\[2mm] 
The coefficients $A_0(z,\bar{z})$, $\bar{A}_0(z,\bar{z})$ can be calculated 
from (\ref{A0}), (\ref{anull}) where the Lagrangian ${\cal L}_0$ is given by
\begin{equation}
{\cal L}_0:=\partial_z\varphi_0\partial_{\bar{z}}\varphi_0- V(\varphi_0)
=32\left(\frac{e^{z+\bar{z}}}{1+e^{2(z+\bar{z})}}\right)^2=8\frac{1}{(\cosh(
z+\bar{z}))^2}
\end{equation}
on the single extremal $\varphi_0$. Then we obtain 
\begin{eqnarray}
A_0(z,\bar{z})&=&-4\cos\left(\frac{\varphi_0}{2}\right)+\partial_{\bar{z}}
\chi_0(z,\bar{z})=4\tanh{(z+\bar{z})}+\partial_{\bar{z}}
\chi_0(z,\bar{z})\,,\label{bsp1}\\
\bar{A}_0(z,\bar{z})&=&-4\cos\left(\frac{\varphi_0}{2}\right)-\partial_{
z}\chi_0(z,\bar{z})=4\tanh{(z+\bar{z})}-\partial_{z}
\chi_0(z,\bar{z})\,.\label{bsp2}
\end{eqnarray}
The coefficients of first order are determined by the embedding conditions:
\begin{equation}
A_1(z,\bar{z})=\partial_{\bar{z}}\varphi_0=\frac{2}{\cosh{l}}\,,\quad
\bar{A}_1(z,\bar{z})=\partial_{z}\varphi_0=\frac{2}{\cosh{l}}\,,\label{bsp3}
\end{equation}
with the substitution $l\!=\!z\!+\!\bar{z}$. Obviously this expressions for 
the coefficients $A_0$, $\bar{A}_0$, $A_1$, $\bar{A}_1$ obtained are the 
general solutions of these equations (\ref{bsp1}), (\ref{bsp2}), 
(\ref{bsp3}) in contrast to the following coefficients of higher orders of 
(\ref{Reihe}).
\\[2mm]
For the coefficients of the second order $A_2(z,\bar{z})$, $\bar{A}_2(z,
\bar{z})$ we choose the translation mode $\theta \!=\! 1/\cosh(l)$ 
of (\ref{Erg 1}) 
\begin{equation} 
A_2(z,\bar{z})=\partial_{\bar{z}}\ln(\theta)=-\tanh(l)\,,\quad
\bar{A}_2(z,\bar{z})=\partial_{z}\ln(\theta)=-\tanh(l)\,.
\end{equation}
The inhomogeneity of the equation for $\bar{\chi}_3$ (see (\ref{inho}) and 
(\ref{Koeff})) always vanishes. Therefore, according to (\ref{well}), 
(\ref{inho}) and the discussion thereby we can choose 
$\bar{\chi}_3(z,\bar{z})\! \equiv \!0$ 
without any loss of generality. Hence the coefficients $A_3$ and $\bar{A}_3$ 
are only determined by the function ${\chi}_3(z,\bar{z})$, which fulfill 
the inhomogeneous PDE (\ref{allgemein}):
\begin{equation}
\partial_z\partial_{\bar{z}}\chi_3-\{2\tanh^2(z+\bar{z})-1\}\chi_3=
-\frac{1}{2}\partial_{\varphi}^3V(\varphi)|_{\varphi=\varphi_0}\theta^2=
2\frac{\tanh(l)}{\cosh^3(l)}\,.
\end{equation}
A special solution is given by $\chi_3\!=\!-\tanh(l)/(2\cosh(l))$, which 
yields:
\begin{equation}
A_3(z,\bar{z})=\frac{1}{\theta}\partial_{\bar{z}}\left(\frac{\chi_3}{
\theta}\right)=-\frac{1}{2\cosh(l)}\,,\quad\bar{A}_3(z,\bar{z})=\frac{1}{
\theta}\partial_{z}\left(\frac{\chi_3}{\theta}\right)=-\frac{1}{2\cosh(l)}
\,.
\end{equation}
For $n\!=\!4$ the inhomogeneity of the wave equation (\ref{well}) for the 
coefficient $\bar{\chi}_4$ (\ref{inho}) vanishes. Thus without any loss of 
generality we may set $\bar{\chi}_4\!=\!0$. The inhomogeneity of the equation 
(\ref{well}) for the function $\chi_4$ also vanishes. So we may choose the 
translation mode again:
\begin{equation}
A_4(z,\bar{z})=\frac{1}{\theta^2}\partial_{\bar{z}}\left(\frac{\chi_4}{
\theta}\right)=\frac{1}{4}\tanh(l)\,,\quad\bar{A}_4(z,\bar{z})=\frac{1}{
\theta^2}\partial_{z}\left(\frac{\chi_4}{\theta}\right)=\frac{1}{4}\tanh(l)
\,.
\end{equation}
Regarding the series of coefficients we see, that those for odd and even 
indices contain the same functions $1/ \cosh(l)$ and $\tanh(l)$, 
respectively. So we assume the same for the higher orders:
\begin{equation}
A_{2n}=\bar{A}_{2n}=\frac{(-1)^n}{2^{2n-4}}\tanh(l)\quad\mbox{and}\quad 
A_{2n+1}=\bar{A}_{2n+1}=\frac{(-1)^n}{2^{2n-3}}\frac{1}{\cosh(l)},\;
n=0,1,2,3,\ldots\,.
\end{equation}
Inserting these expressions into the formal expansion (\ref{Reihe}) we 
obtain:
\begin{eqnarray}
S(z,\bar{z},\varphi)&=&4\left(\sum_{n=0}^{\infty}\frac{(-1)^ny^{2n}}{n!2^{2n
}}\right)\tanh(l)+4\left(\sum_{n=0}^{\infty}\frac{(-1)^ny^{2n+1}}{n!2^{2n+1}
}\right)\frac{1}{\cosh(l)}+\partial_{\bar{z}}\chi_0(z,\bar{z})\nonumber\\
&=&4\cos\left(\frac{y}{2}\right)\tanh(l)+4\sin\left(\frac{y
}{2}\right)\frac{1}{\cosh(l)}+\partial_{\bar{z}}\chi_0(z,\bar{z})
\nonumber\\
&=&-4\cos\left(\frac{y+\varphi_0}{2}\right)+\partial_{z}\chi_0(z,\bar{z})=
-4\cos\left(\frac{\varphi}{2}\right)+\partial_{z}\chi_0(z,\bar{z})
\label{DWWF1}
\,,
\\
\bar{S}(z,\bar{z},y)&=&4\left(\sum_{n=0}^{\infty}\frac{(-1)^ny^{2n}}{n!
2^{2n}}\right)\tanh(l)+4\left(\sum_{n=0}^{\infty}\frac{(-1)^ny^{2n+1}}{n!
2^{2n+1}}\right)\frac{1}{\cosh(l)}-\partial_{\bar{z}}\chi_0(z,\bar{z})
\nonumber\\
&=&4\cos\left(\frac{y}{2}\right)\tanh(l)+4\sin\left(\frac{y}{2}\right)
\frac{1}{\cosh(l)}-\partial_{\bar{z}}\chi_0(z,\bar{z})
\nonumber\\
&=&-4\cos\left(\frac{y+\varphi_0}{2}\right)-\partial_{\bar{z}}\chi_0(z,
\bar{z})=-4\cos\left(\frac{\varphi}{2}\right)-\partial_{\bar{z}}\chi_0(z,
\bar{z})\label{DWWF2}\,.
\end{eqnarray}
These solutions satisfy the HJE (\ref{HJE}) and the IC (\ref{ICC}). The
embedding condition (\ref{qq}) is fulfilled on the single extremal 
$\varphi_0$ by construction. 
\subsubsection{Embedded extremals}
The embedded extremals $\tilde{\varphi}(z,\bar{z},u)$ can be determined by a 
straightforward integration of the eqs.:
\begin{eqnarray} 
\partial_z\tilde{\varphi}(z,\bar{z},u)&=&\Phi=\partial_{\varphi}\bar{S
}\left(z,\bar{z},\varphi=\tilde{\varphi}(z,\bar{z},u)\right)=2\sin\left(
\frac{\tilde{\varphi}}{2}\right)
\,, \,\label{au1}
\\ 
\partial_{\bar{z}}\tilde{\varphi}(z,\bar{z},u)&=&\bar{\Phi}=\partial_{
\varphi}S\left(z,\bar{z},\varphi=\tilde{\varphi}(z,\bar{z},u)\right)=2\sin
\left(\frac{\tilde{\varphi}}{2}\right)
\,\label{au2}
\end{eqnarray}
which leads to
\begin{eqnarray}
\int^{\tilde{\varphi}}\frac{\mbox{d}w}{2\sin\left(
w/2\right)}&=&\ln\arctan\left(\frac{\tilde{\varphi}}{4}\right)=
z+\bar{f}(\bar{z})
\,,\\
\int^{\tilde{\varphi}}\frac{\mbox{d}w}{2\sin\left(
w/2\right)}&=&\ln\arctan\left(\frac{\tilde{\varphi}}{4}\right)=
\bar{z}+f(z)\,.
\end{eqnarray}
Obviously this system of algebraic equations can only be satisfied by the 
functions $f(z)\!=\!z\!+u\!$ and $\bar{f}(\bar{z})\!=\!\bar{z}\!+u\!,$ $u\!=
\!\mbox{const.}$ which
gives the family of extremals:
\begin{equation}
\tilde{\varphi}(z,\bar{z},u)=4\arctan\left(\exp(z+\bar{z}+u)\right)\,,
\end{equation}
parametrized by one parameter $u$.
\\[2mm]
However, here we would like to show how the embedded extremals can be 
calculated by the recursive formalism developed above, which is necessary 
if we are not able to determine the DeDonder \& Weyl Hamilton--Jacobi 
functions or the corresponding familily of extremals explicitly.
\\[2mm]
First we calculate the first three orders ($\Lambda_0,\Lambda_1,\Lambda_2$) 
of the expansion (\ref{ex}) of $\tilde{y}(z,\bar{z},u)$ in $u$ which will 
turn out to be sufficient to guess the general result for the embedded 
extremals $\tilde{y}$. {}From the eqs.\ (\ref{v1}), (\ref{v2}) and (\ref{ex}) 
we get
\begin{eqnarray}
\partial_z\tilde{y}&=&\bar{A}_2\tilde{y}+\frac{1}{2!}\bar{A}_3\tilde{y}^2+
\frac{1}{3!}\bar{A}_4\tilde{y}^3+O(u^4)\,,\\
\partial_{\bar{z}}\tilde{y}&=&A_2\tilde{y}+\frac{1}{2!}A_3\tilde{y}^2+
\frac{1}{3!}A_4\tilde{y}^3+O(u^4)\,,
\end{eqnarray}
in which the formal expansion: 
\begin{equation}
\tilde{y}(z,\bar{z},u)=\tilde{\varphi}(z,\bar{z},u)-\varphi_0(z,\bar{z})=u
\sum^{\infty}_{k=0}\Lambda_k(z,\bar{z})\frac{u^k}{k!}
\end{equation}
has to be inserted. For the first orders we obtain
\begin{eqnarray}
u^1:\quad&&\partial_z\Lambda_0=\frac{1}{\theta}\partial_z(\theta)\Lambda_0\,,
\\&&\partial_{\bar{z}}\Lambda_0=\frac{1}{\theta}\partial_{\bar{z}}(\theta)
\Lambda_0\,,\\
u^2:\quad&&\partial_z\Lambda_1=\frac{1}{\theta}\partial_z(\theta)\Lambda_1+
\frac{1}{2!\theta}\partial_z\left(\frac{\chi_3}{\theta}\right)\Lambda_0^2\,,\\
&&\partial_{\bar{z}}\Lambda_1=\frac{1}{\theta}\partial_{\bar{z}}(\theta)
\Lambda_1+\frac{1}{2!\theta}\partial_{\bar{z}}\left(\frac{\chi_3}{\theta}
\right)\Lambda_0^2\,,\\
u^3:\quad&&\partial_z\Lambda_2=2\frac{1}{2!\theta}\partial_z(\theta)\Lambda_2+
2\frac{1}{3!\theta^2}\partial_z\left(
\frac{\chi_4}{\theta}\right)\Lambda_0^3+
4\frac{1}{2!\theta}\partial_z\left(\frac{\chi_3}{\theta}\right)\Lambda_0
\Lambda_1\,,\\
&&\partial_{\bar{z}}\Lambda_2=2\frac{1}{2!\theta}\partial_{\bar{z}}(\theta)
\Lambda_2+2\frac{1}{3!\theta^2}\partial_{\bar{z}}\left(
\frac{\chi_4}{\theta}\right)\Lambda_0^3+
4\frac{1}{2!\theta}\partial_{\bar{z}}\left(\frac{\chi_3}{\theta}\right)
\Lambda_0\Lambda_1\,.
\end{eqnarray}
The solutions of these eqs.\ are determined up to a constant which may be 
absorbed by a redefinition of the parameter $u$.
\begin{eqnarray}
\Lambda_0&=&2\theta=\frac{2}{\cosh(l)}\,,\quad\Lambda_1=2\chi_3=-\frac{
\tanh(l)}{\cosh(l)}=\frac{\mbox{d}}{\mbox{d}l}\left(\frac{1}{\cosh(l)}\right)
\,,\\
\Lambda_2&=&\frac{8}{3}\chi_4+4\frac{\chi_3^2}{\theta}-\frac{1}{3}\theta=
\frac{4}{3\cosh^2(l)}\left(\sinh^2(l)-\frac{1}{2}\cosh^2(l)\right)\nonumber\\
&=&\frac{\mbox{d}^2}{\mbox{d}l^2}\left(\frac{2}{3\cosh(l)}\right)\,.
\end{eqnarray}
{}From these coefficients we can can already guess the form of the 
coefficients of the higher orders:
\begin{equation} 
\Lambda_{k}=\frac{1}{k+1}\frac{\mbox{d}^k}{\mbox{d}l^k}\left( 
\frac{2}{\cosh(l)}\right)
\end{equation}
{}from which we obtain the embedded extremals:
\begin{eqnarray}
\tilde{\varphi}&=&\varphi_0+u\sum_{k=0}^{\infty}\frac{u^k}{(k+1)!}\frac{
\mbox{d
}^k}{\mbox{d}l^k}\left(\frac{2}{\cosh(l)}\right)=\varphi_0+\int^l\sum_{k=0
}^{\infty}\frac{u^{k+1}}{(k+1)!}\frac{\mbox{d}^{k+1}}{\mbox{d}l^{\prime k+1
}}\left(\frac{2}{\cosh(l^{\prime})}\right)\,\mbox{d}l^{\prime}\nonumber\\
&=&\varphi_0+\int^l\frac{2}{\cosh(l^{\prime}+u)}\,\mbox{d}l^{\prime}-\int^l
\frac{2}{\cosh(l^{\prime})}\,\mbox{d}l^{\prime}=4\arctan\left(\exp(l+u)
\right)
\,.\label{next}
\end{eqnarray}
Obviously we get a one parametric family of extremals satisfying the 
E.L.--equations, as well as the equations (\ref{au1}) and (\ref{au2}). 
It covers a strip of the extended configuration space: $0<\varphi<2
\pi$, $z,\bar{z}\in{\cal R}$. By translations $\varphi \!\rightarrow 
\! \varphi \!+\! 2\pi$ the whole ${\cal R}^3$ parametrized by $z,\bar{z},
\varphi\in {\cal R}$ can be covered by families of these extremals with the 
exception of the parallel planes $\varphi \!=\! 2k\pi$, $k\!=\! 0, \pm 1,
\pm 2,\ldots$, which are solutions of the equations of motion, too. These 
are the so--called ``vacuum" solutions in the Sine--Gordon theory. So this 
set of families of extremals, counted by the integer number $k$ can be 
completed by these planes $\varphi \!=\! 2k\pi$, $k\!=\! 0, \pm 1,
\pm 2,\ldots$, so that the whole space ${\cal R}^3$ is covered by extremals.

\section{Wave Fronts}
In order to determine the wave fronts we have to turn to Carath\'eodory's 
framework, i.e.\ it is necessary to transform the DeDonder \& Weyl 
Hamilton--Jacobi functions $S(z,\bar{z
},\varphi)$ and $\bar{S}(z,\bar{z},\varphi)$ appearing in the expansion 
(\ref{Reihe}) and by the series (\ref{Koeff}) into those of Carath\'eodory 
$S^z(z,\bar{z},\varphi)$, $S^{\bar{z}}(z,\bar{z},\varphi)$, namely $S^z$ and 
$S^{\bar{z}}$ of eq.\ (\ref{carat}).

\subsection{Carath\'eodory's Hamilton--Jacobi functions}
As stated in chapter 3 the two Hamiltonian formulations are algebraically 
equivalent for fields with only one field component. Thus the Hamiltonian 
density ${\cal H}$ and the momenta $p$, $\bar{p}$ are the same in both 
formalisms. Therefore we can use the equality of the momenta in order to 
determine $S^z$, $S^{\bar{z}}$ from the functions $S$ and $\bar{S}$ 
calculated above. 
\\[2mm]
We therefore return to the transversality conditions in (\ref{trans}), 
(\ref{trans2}) and obtain:
\begin{eqnarray}
\partial_{\bar{ z}} S^{\bar{z}}\partial_{\varphi}S^z-
\partial_{\bar{z}} S^{z}\partial_{\varphi}S^{\bar{z}}&=&p=
\partial_{\varphi}S=\sum^{\infty}_{i=0}\frac{1}{i!}A_{i+1}y^i
\,,\label{trans3}\\
\partial_{z} S^{z}
\partial_{\varphi}S^{\bar{z}}-\partial_{z} S^{\bar{z}}\partial_{\varphi}S^{z}
&=&\bar{p}=
\partial_{\varphi}\bar{S}=\sum^{\infty}_{i=0}\frac{1}{i!}\bar{A}_{i+1}y^i
\label{trans4}\,.
\end{eqnarray}
Inserting the functions $S^z$ and $S^{\bar{z}}$ expanded in powers of $y\!=
\!\varphi\!-\!\varphi_0(z,\bar{z})$, like $S$ and $\bar{S}$,
\begin{equation}
S^z(z,\bar{z},\varphi)=\sum_{i=0}^{\infty}\frac{1}{i!}A^z_i(z,\bar{z}) y^i\,,
\quad 
S^{\bar{z}}(z,\bar{z},\varphi)=\sum_{i=0}^{\infty}\frac{1}{i!}A^{\bar{z}}_i
(z,\bar{z}) y^i
\label{Reihecar}\end{equation}
and comparing powers of $y^n$ yields: 
\begin{eqnarray*}
\bar{A}_{n+1}-\sum^{n}_{i=0}\left(\begin{array}{c}n\\i\end{array}\right)
\left\{\left[
\partial_z A^z_i-A^z_{i+1}\partial_z\varphi_0
\right]A^{\bar{z}}_{n-i+1}-
\left[
\partial_z A^{\bar{z}}_i - A^{\bar{z}}_{i+1}\partial_z
\varphi_0
\right]A^z_{n-i+1}
\right\}&=&0\,,
\\
A_{n+1}+\sum^{n}_{i=0}\left(\begin{array}{c}n\\i\end{array}\right)
\left\{\left[\partial_{\bar{z}} A^z_i-A^z_{i+1}\partial_{\bar{z}}\varphi_0
\right]A^{\bar{z}}_{n-i+1}-
\left[
\partial_{\bar{z}} A^{\bar{z}}_i - A^{\bar{z}}_{i+1}\partial_{\bar{z}}
\varphi_0\right]A^z_{n-i+1}
\right\}&=&0\,.
\end{eqnarray*}
These equations determine the coefficients $A^z_{n+1}(z,\bar{z})$, $
A^{\bar{z}}_{n+1}(z,\bar{z})$ recursively: 
\begin{eqnarray}
(\partial_{z}A_0^{z})A^{\bar{z}}_{n+1}-
(\partial_{z}A_0^{\bar{z}})A^{z}_{n+1}&=&\bar{A}_{n+1}+\mbox{Inh}^{n+1}_1(
A^z_0,A^{\bar{z}}_0,\ldots,A^z_n,A^{\bar{z}}_n)\,,
\label{glei1}\\
(\partial_{\bar{z}}A_0^{\bar{z}})A^{z}_{n+1}-
(\partial_{\bar{z}}A_0^{z})A^{\bar{z}}_{n+1}
&=&A_{n+1}+\mbox{Inh}^{n+1}_2(A^z_0,A^{\bar{z}}_0,\ldots,A^z_n,
A^{\bar{z}}_n)\,.\label{glei2}
\end{eqnarray}
The functions $\mbox{Inh}^{n+1}_1$ and $\mbox{Inh}^{n+1}_2$ depend only on 
the coefficients of lower powers of $y$, namely $A^z_0,A^{\bar{z}}_0,\ldots,
A^z_n,A^{\bar{z}}_n$. They vanish for $n\! = \!0$: $\mbox{Inh}^{1}_1\! = \!
\mbox{Inh}^{1}_2\! = \! 0$. For $n\! = \!1$ we get
\begin{equation}
\mbox{Inh}^{2}_1=A^{\bar{z}}_1\partial_z A^{z}_1-
A^{z}_1\partial_z A^{\bar{z}}_1\quad\mbox{and}\quad
\mbox{Inh}^{2}_2=-A^{\bar{z}}_1\partial_{\bar{z}} A^{z}_1+
A^{z}_1\partial_{\bar{z}} A^{\bar{z}}_1
\end{equation}
and for arbitrary orders $n>1$:
\begin{eqnarray}
\mbox{Inh}^{n+1}_1 &=&
-\sum^{n-1}_{i=1}\left(\begin{array}{c}n\\i\end{array}\right)
\left\{\left[
\partial_z A^z_i-A^z_{i+1}\partial_z\varphi_0
\right]A^{\bar{z}}_{n-i+1}-
\left[
\partial_z A^{\bar{z}}_i - A^{\bar{z}}_{i+1}\partial_z
\varphi_0
\right]A^z_{n-i+1}
\right\}
\nonumber\\
&&+A^{\bar{z}}_1\partial_z A^{z}_n-
A^{z}_1\partial_z A^{\bar{z}}_n
\label{Inh1}\,,\\
\mbox{Inh}^{n+1}_2 &=&
+\sum^{n-1}_{i=1}\left(\begin{array}{c}n\\i\end{array}\right)
\left\{\left[\partial_{\bar{z}} A^z_i-A^z_{i+1}\partial_{\bar{z}}\varphi_0
\right]A^{\bar{z}}_{n-i+1}-
\left[
\partial_{\bar{z}} A^{\bar{z}}_i - A^{\bar{z}}_{i+1}\partial_{\bar{z}}
\varphi_0\right]A^z_{n-i+1}
\right\}
\nonumber\\
&&-A^{\bar{z}}_1\partial_{\bar{z}} A^{z}_n+
A^{z}_1\partial_{\bar{z}} A^{\bar{z}}_n
\,.\label{Inh2}
\end{eqnarray}
The determinant  
\begin{equation}
\Delta=(\partial_{\bar{z}}A_0^{\bar{z}})(\partial_zA_0^z)-(\partial_zA_0^{
\bar{z}})(\partial_{\bar{z}}A_0^z)={\cal L}|_{\varphi=\varphi_0}={\cal L}_0
=\partial_z\varphi_0\partial_{\bar{z}}\varphi_0-V(\varphi_0)
\end{equation}
of this linear system of {\em algebraic} equations is assumed to be not 
zero. We consider only those extremals 
$\varphi_0$ and regions in the Minkowski space where the Lagrange density
${\cal L}|_{\varphi=\varphi_0}$ does not vanish. It makes sense to exclude 
those focal points and caustics, where ${\cal L}|_{\varphi=\varphi_0} \!=\! 
0$, because the transversality relations of the wave fronts and extremals 
are violated otherwise \cite{ka}.
\\[2mm]
To obtain the functions $A_0^{\bar{z}}$ and $A_0^z$ we have to
consider the zeroth order of the Hamilton--Jacobi equation in (\ref{trans2}):
\begin{equation}
(\partial_{\bar{z}}A_0^{\bar{z}})(\partial_zA_0^z)-(\partial_zA_0^{
\bar{z}})(\partial_{\bar{z}}A_0^z)=
\partial_z\varphi_0\partial_{\bar{z}}\varphi_0-V(\varphi_0)={\cal L}_0\,.
\label{A_0}
\end{equation}
If ${\cal L}_0\!\neq\!0$ 
then one of the functions $A^z_0$, $A^{\bar{z}}_0$ can be chosen arbitrarily 
and the other one has to be calculated according to this PDE. For example we 
may choose $A^{\bar{z}}_0\!=\!\bar{z}$, $A^z_0\!=\!\int {\cal L}_0\,
\mbox{d}z$. In the case of the kink solution $\varphi_0\!=\!\pm 4\arctan (
\exp(z\!+\!\bar{z}))$ for the Sine--Gordon theory we get $A_0^z\!=\!8\tanh(z
\!+\!\bar{z})$.
\\[2mm]
The Hamilton--Jacobi equation of Carath\'eodory is satisfied automatically in 
any order of $y^n$, $n\ge 1$, because the corresponding equation of 
DeDonder \& Weyl is fulfilled:
\begin{equation}
\partial_{\varphi}\left\{ \partial_zS^{z}\partial_{\bar{z}} S^{\bar{z}} - 
\partial_{z}S^{\bar{z}} \partial_{\bar{z}} S^z+{\cal{H}}\right\}=
\partial_z p+\partial_{\bar{z}}\bar{p}+\partial_{\varphi}{\cal H}=
\partial_{\varphi}\left\{\partial_z S+\partial_{\bar{z}} \bar{S}+{\cal H}
\right\}=0\,.
\end{equation}
The integrability condition holds too, because it imposes a constraint 
on the slope functions which are independent of the special Hamiltonian
description for one component field theories.
\\[2mm]
The linear system of equations for $A^z(z,\bar{z})$, $A^{\bar{z}}(z,\bar{z})$ 
can be solved easily:
\begin{eqnarray}
A^z_{n+1}&=&\frac{1}{{\cal L}_0}\left[(\partial_zA_0^z)A_{n+1}+(
\partial_{\bar{z}}A_0^z)\bar{A}_{n+1}\right]+\widetilde{\mbox{Inh}}_1^{n+1}
(A^z_0,A^{\bar{z}}_0,\ldots,A^z_n,A^{\bar{z}}_n)\,,\label{A^z1}\\
A^{\bar{z}}_{n+1}&=&\frac{1}{{\cal L}_0}\left[(\partial_zA_0^{\bar{z}})
A_{n+1}+(\partial_{\bar{z}}A_0^{\bar{z}})\bar{A}_{n+1}\right]+
\widetilde{\mbox{Inh}}_2^{n+1}(A^z_0,A^{\bar{z}}_0,\ldots,A^z_n,A^{\bar{z}}_n
)\label{A^z}\,.
\end{eqnarray}
The functions $\widetilde{\mbox{Inh}}_1^{n+1}$, $\widetilde{\mbox{Inh}}_2^{
n+1}$ are linear combinations of the inhomogeneities (\ref{Inh1}) and 
(\ref{Inh2}) may easily be determined. So we get e.g.\ for the coefficients  
$A^z_1,A^{\bar{z}}_1$ and $A^z_2,A^{\bar{z}}_2$ the expressions:
\begin{eqnarray}
A^z_{1}&=&\frac{1}{{\cal L}_0}\left[(\partial_zA_0^z)A_{1}+(
\partial_{\bar{z}}A_0^z)\bar{A}_{1}\right]=
\frac{\left[(\partial_zA_0^z)\partial_{\bar{z}}\varphi_0+(
\partial_{\bar{z}}A_0^z)\partial_{z}\varphi_0\right]}{\partial_z\varphi_0
\partial_{\bar{z}}\varphi_0-V(\varphi_0)}
\label{A1}\,,\\
A^{\bar{z}}_{1}&=&\frac{1}{{\cal L}_0}\left[(\partial_zA_0^{\bar{z}})A_{1}+
(\partial_{\bar{z}}A_0^{\bar{z}})\bar{A}_{1}\right]=
\frac{\left[(\partial_zA_0^{\bar{z}})\partial_{\bar{z}}\varphi_0+(
\partial_{\bar{z}}A_0^{\bar{z}})\partial_{z}\varphi_0\right]}{\partial_z
\varphi_0\partial_{\bar{z}}\varphi_0-V(\varphi_0)}
\label{AQ1}
\end{eqnarray}
and 
\begin{eqnarray}
A^z_{2}\!&\!=\!&\!\frac{1}{{\cal L}_0}\!\left[(\partial_zA_0^z)\left(A_{2}
\!-\!A^{\bar{z}}_1\partial_{\bar{z}} A^{z}_1\!+\!A^{z}_1\partial_{\bar{z}} 
A^{\bar{z}}_1\right)+(\partial_{\bar{z}}A_0^z)\left(\bar{A}_{2}\!+\!A^{\bar{z
}}_1\partial_z A^{z}_1\!-\!A^{z}_1\partial_z A^{\bar{z}}_1\right)
\right] \label{A^z0}\!,
\\
A^{\bar{z}}_{2}\!&\!=\!&\!\frac{1}{{\cal L}_0}\!\left[(\partial_zA_0^{\bar{z}
})\left(A_{2}\!-\!A^{\bar{z}}_1\partial_{\bar{z}} A^{z}_1\!+\!A^{z}_1
\partial_{\bar{z}} A^{\bar{z}}_1\right)
+
(\partial_{\bar{z}}A_0^{\bar{z}})\left(\bar{A}_{2}\!+\!A^{\bar{z}}_1
\partial_z A^{z}_1\!-\!A^{z}_1\partial_z A^{\bar{z}}_1\right)\right]\!.
\label{A^z00}
\end{eqnarray}
The two coefficients $A^z_0$ and $\bar{A}^{\bar{z}}_0$ have to fulfil only
one equation (\ref{A_0}), because the closed 
2--form $\Omega\!=\!\mbox{d}S^z\wedge\,\mbox{d}S^{\bar{z}}$ is
invariant under transformations $S^z\rightarrow \hat{S}^z(S^z,S^{\bar{z}})$,
$S^{\bar{z}}\rightarrow \hat{S}^{\bar{z}}(S^z,S^{\bar{z}})$ with a Jacobi 
determinant that equals one. A family of wave fronts given by 
\begin{eqnarray}
S^z(z,\bar{z},\varphi)&=&\sum_{i=0}^{\infty}\frac{1}{i!}A^z_i(z,\bar{z})
\left(\varphi-\varphi_0(z,\bar{z})\right)^i
=\sigma=\mbox{const.}\,,\label{welle1}\\
S^{\bar{z}}(z,\bar{z},\varphi)&=&\sum_{i=0}^{\infty}\frac{1}{i!}A^{\bar{z}}_i
(z,\bar{z})\left(\varphi-\varphi_0(z,\bar{z})\right)^i=\bar{\sigma}=
\mbox{const.}\label{welle}
\end{eqnarray}
is not changed by this transformation, but only reparametrized \cite{wulf} 
$\sigma \rightarrow \hat{\sigma} \! = \!\hat{S}^z(\sigma,\bar{\sigma})$, 
$\bar{\sigma} \rightarrow \hat{\bar{\sigma}} \! = \!
\hat{S}^{\bar{z}}(\sigma,\bar{\sigma})$.

\subsection{An explicit representation for the wave fronts}
In order to obtain an explicit expression for the {\em one dimensional} wave 
fronts $z(\sigma,\bar{\sigma},\varphi)$ and $\bar{z}(\sigma,\bar{\sigma},
\varphi)$ --- $\sigma,\bar{\sigma}$ fixed --- we have to invert the 
relations (\ref{welle}) which can be regarded as the defining equations for 
these functions $z(\varphi)$, $\bar{z}(\varphi)$. Here they are assumed to 
be analytic functions of $\varphi$:
\begin{equation}
z=\sum^{\infty}_{i=0}\frac{1}{i!}\alpha_i(\sigma,\bar{\sigma})\left(
\hat{y}\!=\!\varphi\!-\!\hat{\varphi}_0(\sigma,\bar{\sigma})\right)^i
\,,\quad
\bar{z}=\sum^{\infty}_{i=0}\frac{1}{i!}\bar{\alpha}_i(\sigma,\bar{\sigma})
\left(\hat{y}\!=\!\varphi\!-\!\hat{\varphi}_0(\sigma,\bar{\sigma})\right)^i
\,,
\end{equation}
i.e.\ they can be expanded in powers of the difference $\hat{y}\!=\!\varphi\!
-\!\hat{\varphi}_0(\sigma,\bar{\sigma})$, where $\hat{\varphi}_0(\sigma,\bar{
\sigma}) \! = \! \varphi_0\left(z(\sigma,\bar{\sigma}),\bar{z}(\sigma,\bar{
\sigma})\right)$ denotes the extremal in terms of the variables $\sigma, 
\bar{\sigma}$. Obviously there is a difference  
between the quantities $y \!=\! \varphi\! - \!\varphi_0(z,\bar{z})$ and 
$\hat{y}\! = \!\varphi\!-\!\hat{\varphi}_0(\sigma,\bar{\sigma})$. Thus 
in order to determine the coefficients $\alpha_n,\bar{\alpha}_n$ we have to 
insert this series into the defining equations for the wave fronts 
(\ref{welle1}) and (\ref{welle}):
\begin{eqnarray}
\sigma\!&\!=\!&\!\sum_{i=0}^{\infty}\frac{y^i}{i!}A^z_i\!\left(\!z=\alpha_0(
\sigma,\bar{\sigma}) +\!\! \sum^{\infty}_{n=1}\frac{{\hat{y}}^n}{n!}\alpha_n
(\sigma,\bar{\sigma}),\,\bar{z}=\bar{\alpha}_0(\sigma,\bar{\sigma})+\!\!
\sum^{\infty}_{m=1}\frac{{\hat{y}}^m}{m!}\bar{\alpha}_m(\sigma,\bar{\sigma})
\right)\!,
\label{repara}
\\
\bar{\sigma}\!&\!=\!&\!\sum_{i=0}^{\infty}\frac{y^i}{i!}A^{\bar{z}}_i\!
\left(\! z=\alpha_0(\sigma,\bar{\sigma}) +\!\! \sum^{\infty}_{n=1}\frac{{
\hat{y}}^n}{n!}\alpha_n(\sigma
,\bar{\sigma}),\,\bar{z}=\bar{\alpha}_0(\sigma,\bar{\sigma})+\!\!
\sum^{\infty}_{m=1}\frac{{\hat{y}}^m}{m!}\bar{\alpha}_m(\sigma,\bar{\sigma})
\right)\label{repara1}
\end{eqnarray}
where the variables $y$ have to be expanded in powers of $\hat{y}$, too:
\begin{equation}
y=\varphi-\varphi_0\left(\!
z=\alpha_0(\sigma,\bar{\sigma})
+\!\!\sum^{\infty}_{n=1}\frac{{\hat{y}}^n}{n!}\alpha_n(\sigma,\bar{\sigma}),
\,\bar{z}=\bar{\alpha}_0(\sigma,\bar{\sigma})+\!\!\sum^{\infty}_{m=1}
\frac{{\hat{y}}^m}{m!}\bar{\alpha}_m(\sigma,\bar{\sigma})\right)
\,.\label{ups1}\end{equation}
The wave fronts on the extremals $\alpha_0(\sigma,\bar{\sigma}), \bar{\alpha
}_0(\sigma,\bar{\sigma})$ are determined by the zeroth order of the eqs.\
(\ref{repara}), (\ref{repara1}) and (\ref{ups1}):
\begin{equation}
\sigma = A_0^z\left(z=\alpha_0(\sigma,\bar{\sigma}),\bar{z}=\bar{\alpha}_0(
\sigma,\bar{\sigma})\right)\quad\mbox{and}\quad\bar{\sigma} = A_0^{\bar{z}}
\left(z=\alpha_0(\sigma,\bar{\sigma}),\bar{z}=\bar{\alpha}_0(\sigma,\bar{
\sigma})\right)\,.\label{nullte}
\end{equation}
For the Sine--Gordon theory we obtain from the choice we have made for 
the functions $A_0^z, A_0^{\bar{z}}$: $\bar{z}\!=\!\bar{\alpha}_0(\sigma,
\bar{\sigma})\!=\!\sigma$, $z\!=\!\alpha_0(\sigma,\bar{\sigma}) \!=\! 
\mbox{artanh}(\bar{\sigma})-\sigma$.
\\[2mm]
Locally the functions $\alpha_0(\sigma,\bar{\sigma})$ and $\bar{\alpha}_0(
\sigma,\bar{\sigma})$ are determined uniquely, because the Jacobi 
determinant $\Delta \!=\!(\partial_{\bar{z}}A_0^{\bar{z}})(\partial_zA_0^z) 
\!-\!(\partial_zA_0^{\bar{z}})(\partial_{\bar{z}}A_0^z) \!=\! {\cal L}|_{
\varphi=\hat{\varphi}_0}$ of this transformation $z,\bar{z} \rightarrow 
\sigma, \bar{\sigma}$ does not vanish, as assumed.
\\[2mm] 
Expanding the expressions (\ref{repara}) in  $\hat{y}^n$, inserting the 
series (\ref{ups1}) and separating the coefficients $\alpha_n(\sigma,\bar{
\sigma})$, $\bar{\alpha}_n(\sigma,\bar{\sigma})$ we obtain a pair of linear 
algebraic equations for these coefficients, that can be solved like  
the system (\ref{glei1}), (\ref{glei2}). After inserting the coefficients 
$A_1,\bar{A}_1,\ldots,A_{n},\bar{A}_{n}$ of the DeDonder and Weyl 
Hamilton--Jacobi theory we get the final result for $n\ge 2$:
\\[3mm]
\fbox{\parbox{15.5cm}{
\begin{eqnarray}\hspace*{-1cm}
\alpha_n \!\!&\!\!=\!\!&\!\!\frac{(-{\cal L}_0)^{n-1}}{{
\cal H}_0^{n+1}}\!\left(V(\varphi_0)A_n\!-\!(\partial_{\bar{z}}\varphi
_0)^2\bar{A}_n\right)\!
+\!\mbox{Inh}(A_1,
\bar{A}_1,\ldots,A_{n-1},\bar{A}_{n-1},\mbox{deriv.}),\label{fast ende}
\\\hspace*{-1cm}
\bar{\alpha}_n\!\!&\!\!=\!\!&\!\!\frac{(-{\cal L}_0)^{n-1}}{{
\cal H}_0^{n+1}}\!\left(V(\varphi_0)\bar{A}_n\!-\!(\partial_{z}\varphi_0)^2 
A_n\right)\!+\!\mbox{Inh}(A_1,\bar{A}_1,\ldots,A_{n-1},\bar{A}_{n-1},
\mbox{deriv.})
.
\label{ende}
\end{eqnarray}}}\\[3mm]
The term ``deriv.'' in the inhomogeneities denotes the derivatives of $A^{z
}_i$, $A^{\bar{z}}_i$, $i\!=\! 1,\ldots,n-1$ 
with respect to $z$ and $\bar{z}$. The functions $A^{z}_n$, 
$A^{\bar{z}}_n$ and their derivatives are short--cuts of e.g. $\partial_z
A^z_n\!=\!\partial_zA^z_n|_{z=\alpha_0(\sigma,\bar{\sigma}),\bar{z}=\bar{
\alpha}_0(\sigma,\bar{\sigma})}$ and thus depending on $\sigma$, $\bar{
\sigma}$ only.
\\[1mm]
Obviously a not degenerated transformation $z,\bar{z} 
\rightarrow \sigma\!=\! S^z, \bar{\sigma}\!=\! S^{\bar{z}}$ only exists if
the Lagrangian ${\cal L}_0$ and the Hamiltonian densities ${\cal H}_0$ on the 
extremals do not vanish.
\\[1mm]
The coefficients of zeroth order are given in (\ref{nullte}), those of 
the next two powers in $\hat{y}$ are:
\begin{equation}
\alpha_1(\sigma,\bar{\sigma})=\frac{\partial_{\bar{z}}\varphi_0}{{
\cal H}_0}=\frac{\partial_{\bar{z}}\varphi_0}{\partial_z
\varphi_0\partial_{\bar{z}}\varphi_0+V(\varphi_0)}
\;,\quad
\bar{\alpha}_1(\sigma,\bar{\sigma})=\frac{\partial_{z}\varphi_0}{{\cal H}_0}
=\frac{\partial_{z}\varphi_0}{\partial_z\varphi_0\partial_{\bar{z}}\varphi_0+
V(\varphi_0)}\,,\label{erste}\end{equation}
and 
\begin{eqnarray}
\alpha_2(\sigma,\bar{\sigma}) &=& \frac{V(\varphi_0)}{{\cal H}_0^{3}}\left[
-{\cal L}_0\partial_{\bar{z}}\ln(\theta)+\partial_{\bar{z}}\left(\partial_z
\varphi_0\partial_{\bar{z}}\varphi_0\right)\right]\nonumber\\&-&
\frac{\partial_{\bar{z}}
\varphi_0\partial_{\bar{z}}\varphi_0}{{\cal H}_0^{3}}\left[
-{\cal L}_0\partial_{z}\ln(\theta)+\partial_{z}\left(\partial_z
\varphi_0\partial_{\bar{z}}\varphi_0\right)\right]+\frac{\partial_{\bar{z}}
\varphi_0}{{\cal H}_0^{2}}\partial_{\varphi}
V(\varphi_0)\,,\label{zweite1}
\\
\bar{\alpha}_2(\sigma,\bar{\sigma}) &=& \frac{V(\varphi_0)}{{\cal H}_0^{3}}
\left[-{\cal L}_0\partial_{z}\ln(\theta)+\partial_{z}\left(\partial_z
\varphi_0\partial_{\bar{z}}\varphi_0\right)\right]
\nonumber
\\&-&
\frac{\partial_z
\varphi_0\partial_{z}\varphi_0}{{\cal H}_0^{3}}\left[
-{\cal L}_0\partial_{\bar{z}}\ln(\theta)+\partial_{\bar{z}}\left(\partial_{
z}\varphi_0\partial_{\bar{z}}\varphi_0\right)\right]+\frac{\partial_{z}
\varphi_0}{{\cal H}_0^{2}}\partial_{\varphi}V(\varphi_0)\,.
\label{zweite2}
\end{eqnarray}
Thus the wave fronts $z(\sigma,\bar{\sigma},\varphi)$ and $\bar{z}(\sigma,
\bar{\sigma},\varphi)$ can be calculated from the coefficients $A_n$, 
$\bar{A}_n$ of DeDonder and Weyl's Hamilton--Jacobi framework and the 
coefficients $A^z_0$, $A^{\bar{z}}_0$ of Carath\'eodory's one. 
Carath\'eodory's coefficients $A^z_n$ and $A^{\bar{z}}_n$, $n\ge 1$ are not 
necessarily be involved in the final expressions (\ref{fast ende}), 
(\ref{ende}), since they can be substituted by those of DeDonder and 
Weyl --- determined by B\"acklund transformations.

\subsection{An alternative way to determine the wave fronts}
If one is not interested to get an explicit representation of 
Carath\'eodory's Hamilton--Jacobi functions $S^z(z,\bar{z},\varphi)$, $S^{
\bar{z}}(z,\bar{z},\varphi)$ it is possible to get the coefficients 
$\alpha_n$, $\hat{\alpha}_n$ of the last paragraph much more easily. However, 
this method can only be applied, if the Hamiltonian density ${\cal H}$ does 
not vanish! If ${\cal H}\!=\!0$ we have to calculate Carath\'eodory's 
Hamilton--Jacobi functions $S^z(z,\bar{z},\varphi)$, $S^{\bar{z}}(z,\bar{z},
\varphi)$ as described above.
\\[2mm]
For nonvanishing Hamiltonian densities ${\cal H}$ we may change the 
independent variables $z$, $\bar{z}$ to $\sigma \!=\! S^z(z,\bar{z},\varphi)
$ and $\bar{\sigma} \!=\!S^{\bar{z}}(z,\bar{z},\varphi)$, while the field 
$\varphi$ remains unchanged. The functional determinant of this 
transformation is just ${\cal H}$ according to Carath\'eodory's 
Hamilton--Jacobi equation (\ref{CHJ}). Then the wavefronts $z\!=\!z(\varphi,
\sigma,\bar{\sigma}),\,\bar{z}\!=\!\bar{z}(\varphi,\sigma,\bar{\sigma})$ can 
be determined explicitly from the equations
\begin{equation}
\partial_{\varphi}z(\varphi,\sigma,\bar{\sigma})=\frac{p}{\cal H}\;,\quad
\partial_{\varphi}\bar{z}(\varphi,\sigma,\bar{\sigma})=\frac{\bar{p}}{\cal H}
\,.
\end{equation}
These equations are obtained by comparing the coefficients of the wedge 
products $\mbox{d}\varphi\wedge \mbox{d}\sigma$ and $\mbox{d}\varphi\wedge 
\mbox{d}\sigma$ in the equation (\ref{CHJ}), if the variables $\sigma \!=\! 
S^z$ and $\bar{\sigma} \!=\!S^{\bar{z}}$ and $\varphi$ are regarded as 
the independent ones \cite{Teil2}. So we can e.g.\ immediately determine the 
coefficients of the first order $\alpha_1$
and $\bar{\alpha}_1$ (\ref{erste}). The remaining coefficients $\alpha_n$
and $\bar{\alpha}_n$ in (\ref{ende}) and (\ref{fast ende})
can be determined by expanding the DeDonder and Weyl momenta $p$, $\bar{p}$ 
and the Hamiltonian density ${\cal H}$ in powers of $\hat{y}$.

\subsection{An example: the Sine--Gordon theory}
We return to our example in chapter 7 in order to illustrate the formalism
discussed above.

\subsubsection{Carath\'eodory's Hamilton--Jacobi functions}
To obtain Carath\'eodory's Hamilton--Jacobi functions $S^z(z,\bar{z},
\varphi)$, $S^{\bar{z}}(z,\bar{z},\varphi)$ from those of DeDonder \& Weyl, 
(\ref{DWWF1}) and (\ref{DWWF2}), we first have to solve the equation 
(\ref{A_0}) in order to get the coefficients $A^z_0(z,\bar{z})$, $A_0^{
\bar{z}}(z,\bar{z})$ of order $y^0$. Because in our example the
Lagrangian density depends only on $l \!=\! z \!+\! \bar{z}$ on the extremals
it is useful to transform the independent variables $z, \bar{z}$ into 
$l \!=\! z \!+\! \bar{z}$ and $\bar{l} \!=\! z \!-\! \bar{z}$
\begin{equation}
(\partial_{\bar{l}}A_0^{z})(\partial_{l}A_0^{\bar{z}})-(\partial_{\bar{l}
}A_0^{\bar{z}})(\partial_{l}A_0^z)=\frac{1}{2}{\cal L}_0=\frac{4}{\cosh^2(l)
}\,.
\end{equation}
Due to the invariance of the 2--form $\Omega\!=\!\mbox{d}S^z\wedge\,\mbox{d}
S^{\bar{z}}$ with respect to the transformations $S^z\rightarrow \hat{S}^z(
S^z,S^{\bar{z}})$, $S^{\bar{z}}\rightarrow \hat{S}^{\bar{z}}(S^z,
S^{\bar{z}})$ 
with a Jacobi determinant equal to one we can choose any single solution of 
this equation without any loss of generality, e.g.
\begin{equation}
A_0^z=A^z_0(l)=4\tanh(z+\bar{z})\;,\quad A_0^{\bar{z}}=A_0^{\bar{z}}=-\bar{l}
=\bar{z}-z\,.
\end{equation}
Inserting this result into the equations (\ref{A1}) and (\ref{AQ1}) we get 
the coefficients of first order
\begin{equation}
A_1^z=A^z_1(l)=\frac{2}{\cosh(l)}\;,\quad A_1^{\bar{z}}=0\,,
\end{equation}
which leads, using the eqs.\ (\ref{A^z1}), (\ref{A^z}), (\ref{A^z0}) and 
(\ref{A^z00}), to the coefficients of the second and the third order in $y$ 
\begin{equation}
A_2^z=A^z_2(l)=-\tanh(l)\;,\quad A_3^z=A^z_3(l)=-\frac{1}{2\cosh(l)}
\;,\quad A_2^{\bar{z}}=A_3^{\bar{z}}=0\,.\label{two}
\end{equation}
Inspecting the series of these coefficients we get an ansatz for those of all
orders $y^n$, $n\ge 1$
\begin{equation}
A_{2n}^z=A^z_{2n}(l)=4\frac{(-1)^{n}}{2^{2n}}\tanh(l)\,,\quad
A_{2n-1}^z=A^z_{2n-1}(l)=4\frac{(-1)^{n}}{2^{2n-1}}\frac{1}{\cosh(l)}\,,\quad
A_n^{\bar{z}}=0\,,
\end{equation}
{}from which we obtain Carath\'eodory's Hamilton--Jacobi functions $S^z(z,
\bar{z},\varphi)$, $S^{\bar{z}}(z,\bar{z},\varphi)$ by determining the sum 
(\ref{Reihecar}):
\begin{eqnarray}
S(z,\bar{z},\varphi)&=&4\left(\sum_{n=0}^{\infty}\frac{(-1)^ny^{2n}}{n!2^{2n
}}\right)\tanh(l)+4\left(\sum_{n=0}^{\infty}\frac{(-1)^ny^{2n+1}}{n!2^{2n+1}
}\right)\frac{1}{\cosh(l)}
\,,
\nonumber\\
&=&4\cos\left(\frac{y}{2}\right)\tanh(l)+4\sin\left(\frac{y
}{2}\right)\frac{1}{\cosh(l)}=-4\cos\left(\frac{\varphi}{2}\right)
\label{HJC1}\,,\\
S^{\bar{z}}(z,\bar{z},\varphi)&=&\bar{z}-z\,.
\label{HJC2}
\end{eqnarray}
These functions satisfy Carath\'eodory's Hamilton--Jacobi equation 
(\ref{HJE}) and the integrability condition (\ref{ICC}) as well as the 
embedding conditions (\ref{qq1}), (\ref{qq}). The solutions (\ref{HJC1}), 
(\ref{HJC2}) for $S^z$, $S^{\bar{z}}$ are the only ones provided that we have 
fixed the coefficients $A_0^z$ and $A_0^{\bar{z}}$, because the eqs.\ 
(\ref{A^z1}), (\ref{A^z}), (\ref{A^z0}) and (\ref{A^z00}) lead to unique 
solutions . The general solution of the system of PDEs (\ref{CHJ}), 
(\ref{ICC}) and (\ref{qq}) is obtained by applying arbitrary smooth 
transformations $S^z\rightarrow \Sigma(S^z,S^{\bar{z}})$, $S^{\bar{z}}
\rightarrow \bar{\Sigma}(S^z,S^{\bar{z}})$ to the functions (\ref{HJC1}), 
(\ref{HJC2}) with a Jacobi determinant $\partial_1 \Sigma\partial_2\bar{
\Sigma} \!-\!  \partial_2 \Sigma\partial_1\bar{\Sigma} \! = \! 1$
\begin{equation}
S^z(z,\bar{z},\varphi)=\Sigma\left(\bar{z}-z,-4\cos\left(\varphi/2
\right)\right)
\quad
\mbox{and}
\quad
S^{\bar{z}}(z,\bar{z},\varphi)=\bar{\Sigma}\left(\bar{z}-z,-4\cos\left(
\varphi/2\right)\right)
\,.\label{expl}
\end{equation}

\subsubsection{An explicit representation of the wave fronts -- the singular 
case}
Carath\'eodory's Hamilton--Jacobi functions $S^z$, $S^{\bar{z}}$ are given 
now by the equations (\ref{expl}). If we would like to 
have them in the explicit form $z(\sigma,\bar{\sigma},\varphi),\bar{z}(
\sigma,\bar{\sigma},\varphi)$, we have to invert the equations 
(\ref{welle1}), (\ref{welle}).
Doing this we get only the zeroth order coefficients 
\begin{equation}
\alpha_0(\sigma,\bar{\sigma})=\frac{1}{2}\left(\mbox{artanh}\frac{\sigma}{4}+
\bar{\sigma}\right)
\,\,
\mbox{and}
\,\,\,
\bar{\alpha}_0(\sigma,\bar{\sigma})=\frac{1}{2}\left(\mbox{artanh}\frac{
\sigma}{4}-\bar{\sigma}\right)
\end{equation}
without any difficulties. All the other coefficients $\alpha_n$, 
$\bar{\alpha}_n$, $n\ge 1$ {\em do not exist}, because the Hamiltonian 
density ${\cal H}_0=p_0\bar{p}_0+V_0(\varphi)$ with $V_0\!= \!2(\cos(\varphi)
-1)$ vanishes on the given extremals $\varphi_0$ (see (\ref{fast ende})
and (\ref{ende}))!
\\[2mm]
{}From the results (\ref{HJC1}), (\ref{HJC2}) it becomes obvious why this 
happens. The function (\ref{HJC2}) depends only on the difference 
$z\!-\!\bar{z}$ whereas (\ref{HJC1}) does not depend on the 
variables $z,\bar{z}$ at all! Therefore it is impossible to invert the 
equations (\ref{welle1}), (\ref{welle}) to get the functions $z(\sigma,\bar{
\sigma},\varphi)$ and $\bar{z}(\sigma,\bar{\sigma},\varphi)$ at every point 
$(\sigma,\bar{\sigma})$ of the parameter space $\Upsilon
\!=\!\{(\sigma,\bar{\sigma})\}$. Hence the wave fronts are only defined at 
points $(\sigma,\bar{\sigma},\varphi)$ for which  
\begin{equation}
\varphi\stackrel{!}{=}g(\sigma)=2\,\mbox{arccos}\left(\sigma/4\right)\,\,,
-4\le\sigma\le 4
\end{equation}
holds.
The real parameter $\bar{\sigma}$ remains arbitrary. Nevertheless 
the wave fronts are one dimensional straight lines parallel to each other: 
$z\!-\!\bar{z}\!=\!\bar{\sigma}\!=\!const.$.  They cover the extended 
configuration space ${\cal M}_{2+1}\!=\!\{z,\bar{z},\varphi\}$ as required, 
but the wave fronts cannot be given as functions $z(\sigma,\bar{\sigma},
\varphi)$ and $\bar{z}(\sigma,\bar{\sigma},\varphi)$ because they 
are parallel to the surfaces $\varphi\! = \! const.$. 
\\[2mm]
Due to the transversality condition (\ref{transvers}) the 2--parametric 
family of 1--d wave fronts (see Fig.\ 1)
\begin{equation}
z - \bar{z} = \bar{\sigma} = const.\,,\qquad\, \varphi = g(\sigma)  
= const. \label{wsing}
\end{equation} 
and the 1--parametric set of the 2--d extremals $\tilde{\varphi}=4\arctan
\left(\exp(l+u)\right)$ intersect each other transversally everywhere in 
${\cal M}_{2+1}\!=\!\{z,\bar{z},\varphi\}$, because the Lagrangian density 
${\cal L}$ differs from zero in ${\cal M}_{2+1}$ contrary to the Hamiltonian 
density which vanishes everywhere in ${\cal M}_{2+1}$. The angle $\angle{(
w,e)}$ between the basis vector $W$ of the tangent space ${\cal T_PW}$
of the wave front 
\begin{equation}
w=2\sin(\varphi/2)(\partial_z+\partial_{\bar{z}})
\end{equation}
and an arbitrary nonvanishing vector $e$ in the tangent space ${\cal T_PE}$
of the extremal pa\-ra\-me\-tri\-zed by the variables $\lambda, \bar{
\lambda}$ 
\begin{equation}
e=\lambda\partial_z+\bar{\lambda}\partial_{\bar{z}}+2(\lambda+\bar{\lambda})
\sin(\varphi/2)\partial_{\varphi}\,,\,\mbox{with}\,\,|\lambda|+|\bar{\lambda}
|>0
\end{equation}
at the point $P\!=\!(z,\bar{z},\varphi)$ is given by
\begin{equation}
\angle{(w,e)}=\arccos\left(\frac{(w,e)}{|w||e|}\right)=
\arccos\left(\frac{\lambda+\bar{\lambda}}{\sqrt{2}
\sqrt{\lambda^2+\bar{\lambda}^2+4(\lambda+\bar{\lambda})^2\sin^2(\varphi/2)}}
\right)\,,
\end{equation}
the minimum of which with respect to a variation of the parameters $\lambda,
\bar{\lambda}$ gives the angle under which the wavefront intersects the 
extremal at $P$:
\begin{equation}
\angle({\cal W},{\cal E})=\mbox{min}\angle{(w,e)}=\arccos\left(\frac{1}{
\sqrt{1+8\sin^2(\varphi/2)}}\right)=\arccos\left(\frac{1}{\sqrt{9-
\sigma^2/2}}\right)\,,
\end{equation}
$\angle({\cal W},{\cal E})$ is constant on a given wave front  $\sigma \!=\! 
\mbox{const.}$,
$\bar{\sigma}\!=\!\mbox{const.}$. This angle $\angle({\cal W},{\cal E})$
takes its minimum $\angle({\cal W},{\cal E})\! = \!0$ for the values 
$\varphi \!=\! \varphi_k=2k\pi,\, k \! = \! 0, \pm 1,\pm 2,\ldots$, namely 
$\angle({\cal W},{\cal E}) \!=\! 0$. So in the limit $z\!+\!\bar{z}
\!=\!l \rightarrow \pm\infty$ the transversality condition is violated, 
since the wave fronts and extremals (\ref{next}) are asymptotically parallel 
at all points $P$ in the planes $\{z, \bar{z} \in {\cal R}, \varphi \! = \! 
2k\pi\} \in {\cal M}_{2+1}$ parametrized by the integer number $k\! = \!0,
\pm 1,\pm 2,\ldots$. Any of these planes separates the two one--parameter 
families $\tilde{\varphi}\! = \!\arctan(\exp(l \!+\! u))\! + \!2(k \!-\! 1)
\pi$ and $\tilde{\varphi}\! = \!\arctan(\exp(l \!+\! u))\! + \!2k\pi$. Every 
extremal of these families ``touches'' the plane $\varphi\!=\!2k\pi$ in the 
limits $l\rightarrow -\infty$ or $l\rightarrow\infty$. The maximum value for 
$\angle({\cal W},{\cal E})\! = \!\arccos(1/3)$ indicates that the wave fronts 
and the extremals are never perpendicular to each other.
\\[3mm]
Notice, that the singular situation discussed above is essentially a 
coordinate singularity resulting from the singular change of variables $z$, 
$\bar{z}$ to $\sigma \!=\! S^z(z,\bar{z},\varphi)$ and $\bar{\sigma} \!=\!
S^{\bar{z}}(z,\bar{z},\varphi)$. It shows that we cannot choose the 
parameters $\sigma$, $\bar{\sigma}$, $\varphi$ as independent variables to 
represent the wave fronts. The wave fronts are not degenerated at all. 
\\[2mm]
However the transversality between the wave fronts and extremals is  
violated in this case on the boundaries of the regions ${\cal M}_k \!=\! \{2k
\pi \!<\!\varphi \!<\! 2(k\! +\! 1)\pi, z, \bar{z} \!\in \!{\cal R}\}$, where 
the families of extremals (\ref{next}) are defined --- a singularity that 
cannot be circumvented by a coordinate transformation. 
\\[3mm]
These results can be obtained in a straightforward manner by using 
the DeDonder--\-Weyl Hamilton--Jacobi functions $S(z,\bar{z},\varphi)$ and 
$\bar{S}(z,\bar{z},\varphi)$ from (\ref{DWWF1}), (\ref{DWWF2}). 
\\[2mm]
The Hamiltonian density ${\cal H}$ vanishes on the extremals 
$\tilde{\varphi}(z,\bar{z},u)$ of (\ref{next}):
\begin{equation}
{\cal H}=\partial_{\varphi}S\partial_{\varphi}\bar{S}+V(\varphi)=\left(2\sin(
\varphi/2)\right)^2-2(1-\cos(\varphi))=0\,.
\end{equation}
Notice that the ``usual" canonical Hamiltonian density:
\begin{equation}
{\cal H}_{can.}(x,t)=\frac{1}{2}\pi^2+\frac{1}{2}(\partial_x\varphi)^2+
(1-\cos(\varphi))\quad\mbox{with}\quad\pi=\frac{\partial{\cal L}}{
\partial(\partial_t\varphi)}
\end{equation}
does not vanish on the kink solution $\varphi_0$, because the 
Legendre--Transformation is applied only with respect to the time derivatives
$\partial_t\varphi$ of the field $\varphi$ and does not affect its 
spatial derivative $\partial_x\varphi$. This underlines again that the 
energy density ${\cal H}_{can.}$ is different from the ``covariant'' 
Hamiltonian density ${\cal H}$ (\ref{ham}), that we use in this article.
\\[2mm]
Carath\'eodory's Hamilton--Jacobi equation (\ref{CHJ}) on the family of 
extremals (\ref{next}):
\begin{equation} 
\partial_zS^{z}\partial_{\bar{z}} S^{\bar{z}} - \partial_{z}S^{\bar{z}} 
\partial_{\bar{z}} S^z=-{\cal H}= 0\,,\label{HJex}
\end{equation}
the solutions of which determine the wave fronts transversal to the extremals 
we are interested in, shows that the change of the variables $z$, $\bar{z}$ 
to $\sigma \!=\! S^z(z,\bar{z},\varphi)$ and $\bar{\sigma} \!=\!S^{\bar{z}}(z,
\bar{z},\varphi)$ is {\em not} a regular one, because ${\cal H}$ is nothing 
but the functional determinant of this transformation.
\\
The general solution of equation (\ref{HJex}) is:
\begin{equation}
S^z=\Sigma^z(f(z,\bar{z}),\varphi)\,,\quad S^{\bar{z}}=\Sigma^{\bar{z}}(f(z,
\bar{z}),\varphi)\,,
\end{equation}
with an arbitrary smooth function $f(z,\bar{z})$.
Inserting this result into the conditions on the momenta (\ref{carat}): 
\begin{eqnarray}
p\!&\!=\!&\!2\sin(\varphi/2)=\partial_{\bar{ z}} S^{\bar{z}}\partial_{
\varphi}S^z-\partial_{\bar{z}} S^{z}\partial_{\varphi}S^{\bar{z}}=+(
\partial_zf)\left(\partial_f \Sigma^z\partial_{\varphi}\Sigma^{\bar{z}}-
\partial_{\varphi}\Sigma^z\partial_f \Sigma^{\bar{z}}\right)\!,
\\
\bar{p}\!&\!=\!&\!2\sin(\varphi/2)=\partial_{z} S^{z}\partial_{\varphi}S^{
\bar{z}}-\partial_{z} S^{\bar{z}}\partial_{\varphi}S^{z}=
-(\partial_{\bar{z}}f)\left(\partial_f \Sigma^z\partial_{\varphi}\Sigma^{
\bar{z}}-\partial_{\varphi}\Sigma^z\partial_f \Sigma^{\bar{z}}\right)
\end{eqnarray}
shows, after subtracting these PDEs from each other, that the function $f$ 
depends on the difference $\bar{l} \!=\! z\! - \! \bar{z}$ only: $f \!=\! 
f(z\! - \! \bar{z})$. Due to the invariance of the basic 2--form 
(\ref{basca}) with respect to symplectic transformations we may choose
$f \!=\! z\! - \! \bar{z}$, $\Sigma^z\!=\! -4\cos(\varphi/2)$ and $\Sigma^{
\bar{z}} \!=\! \bar{z}\! - \!z$ without any loss of generality. This result 
coincides with (\ref{HJC1}) and (\ref{HJC2}).

\subsubsection{An explicit representation of the wave fronts -- the regular 
case}
If we would like to circumvent the spurious singularity discussed above
we have to ensure that the Hamiltonian density does not vanish in ${\cal M}_{
2+1}$, e.g.\ by adding a global constant $c_0$ to the Lagrangian 
density:
\begin{equation}
{\cal L}\,\Rightarrow\,{\cal L}=\partial_z\varphi\partial_{\bar{z}}\varphi+
2(1-\cos(\varphi))+c_0\,,\label{shift}
\end{equation}
which does not influence the equation of motion. More generally, we can add 
any exact two form $\Gamma \! = \! \mbox{d}(f(z,\bar{z})\mbox{d}z \!+\! 
\bar{f}(z,\bar{z})\mbox{d}\bar{z})$ to the basic form (\ref{basic}), $\Omega
\rightarrow \Omega\!+\!\Gamma$, without affecting the momenta $p,\bar{p}$, 
the slope functions $v,\bar{v}$ and the equation of motion, but modifying the 
Hamiltonian and the Lagrangian densities ${\cal H}$, ${\cal L}$: 
\begin{equation}
{\cal H}\,\,\rightarrow\,\,{\cal H}+\partial_{\bar{z}}f-\partial_z\bar{f}\,,
\qquad{\cal L}\,\,\rightarrow\,\,{\cal L}-\partial_{\bar{z}}f+\partial_z
\bar{f}\,.
\end{equation}
We see that 
the form of the wave fronts is influenced by adding such a term $\Gamma$ 
while the family of extremals (\ref{next}) remains unchanged --- a property 
which is known in mechanics, too. For the sake of simplicity we choose $f\!=
\!0$, $\bar{f}\!=\!c_0z$ which shifts ${\cal L}$ and ${\cal H}$ merely by 
a constant $c_0$.
\\[2mm]
With the exception of the coefficients $A_0$, $\bar{A}_0$ of zeroth order, 
which are of no interest in determing the embedded extremals and the 
corresponding wave fronts, the DeDonder and Weyl Hamilton--Jacobi functions 
$S(z,\bar{z},\varphi)$ and $\bar{S}(z,\bar{z},\varphi)$ and the resulting 
momenta $p\!=\!\partial_{\varphi}S$, $\bar{p}\!=\!\partial_{\varphi}\bar{S}$
are not affected by this shift contrary to the wave fronts. 
\\[2mm]
Using Carath\'eodory's Hamilton--Jacobi equation (\ref{CHJ}):
\begin{equation} 
\partial_zS^{z}\partial_{\bar{z}} S^{\bar{z}} - \partial_{z}S^{\bar{z}} 
\partial_{\bar{z}} S^z=-{\cal H}\equiv c_0
\end{equation}  
which represents nothing but the determinant of the linear system 
of equations (see (\ref{trans2}))
\begin{eqnarray}
p&=&2\sin(\varphi/2)=\partial_{\bar{ z}} S^{\bar{z}}\partial_{\varphi}S^z-  
\partial_{\bar{z}} S^{z}\partial_{\varphi}S^{\bar{z}}
\label{pe}\,,\\
\bar{p}&=&2\sin(\varphi/2)=\partial_{z} S^{z}\partial_{\varphi}S^{\bar{z}}- 
\partial_{z} S^{\bar{z}}\partial_{\varphi}S^{z}
\label{pebar}\,.
\end{eqnarray}
For the functions $\partial_{\varphi}S^z$ and $\partial_{\varphi}S^{\bar{z}}
$ we get a system of decoupled linear PDEs for the functions $S^z$, $S^{
\bar{z}}$
\begin{eqnarray}
\left({\cal H}\partial_{\varphi}+p\partial_z+\bar{p}\partial_{\bar{z}}
\right)S^{z}&=&\left(-c_0\partial_{\varphi}+2\sin(\varphi/2)\left[
\partial_z+\partial_{\bar{z}}\right]\right)S^{z}=0\,,
\\
\left({\cal H}\partial_{\varphi}+p\partial_z+\bar{p}\partial_{\bar{z}}
\right)S^{\bar{z}}&=&\left(-c_0\partial_{\varphi}+2\sin(\varphi/2)\left[
\partial_z+\partial_{\bar{z}}\right]\right)S^{\bar{z}}=0\,.
\end{eqnarray}
The general solution: 
\begin{eqnarray}
S^z&=&\Sigma\left(-c_0z/2+2\cos(\varphi/2),-c_0\bar{z}/2+2\cos(\varphi/2)
\right)\,,
\\
S^{\bar{z}}&=&\bar{\Sigma}\left(-c_0z/2+2\cos(\varphi/2),-c_0\bar{z}/2+2\cos(
\varphi/2)\right)
\end{eqnarray}
of this system PDEs is obtained by the method of 
characteristics. The functions $\Sigma, \bar{\Sigma}$ have only to satisfy 
Carath\'eodory's Hamilton--Jacobi equation (\ref{CHJ}). Taking the 
invariance with respect to symplectic transformations into account we may 
choose the functions
\begin{equation}
S^z=\sqrt{2}\left[\bar{z}-(4/c_0)\cos(\varphi/2)\right]
\,\,\quad
\mbox{and}
\quad\,\,
S^{\bar{z}}=\sqrt{2}\left[-c_0z/2+2\cos(\varphi/2)\right]
\label{reg}
\end{equation}
without any loss of generality.
\\[2mm]
Obviously the transformation 
\begin{equation}
(z,\bar{z}) \rightarrow \left(\sigma \!=\! S^z(z,\bar{z},\varphi),\,\bar{
\sigma} \!=\!S^{\bar{z}}(z,\bar{z},\varphi)\right)
\end{equation}
exists now, leading to an explicit representation of the wave fronts:
\begin{equation}
z(\sigma,\bar{\sigma},\varphi)=-\frac{\sqrt{2}\bar{\sigma}}{c_0}+\frac{4}{
c_0}\cos(\varphi/2)\,\,,\quad\bar{z}(\sigma,\bar{\sigma},\varphi)=\frac{
\sigma}{\sqrt{2}}+\frac{4}{c_0}\cos(\varphi/2)
\label{exx}\end{equation}
with some properties different from those obtained in the ``singular'' case: 
the 1--d wave fronts are not straight lines in the extended configuration 
space. Subtracting the eqs.\ (\ref{exx}) from each other shows that they lie 
in planes parallel to the l-axis like in the singular case.
\\[2mm]
Similar to the ``singular" case the angle $\angle{(w,e)}$ between the basis 
vector of the tangent space ${\cal T_PW}$ of the wave front 
\begin{equation}
w=2\sin(\varphi/2)(\partial_z+\partial_{\bar{z}})-c_0\partial_{\varphi}
\end{equation}
and an arbitrary nonvanishing vector in the tangent space ${\cal T_PE}$
of the extremal 
\begin{equation}
e=\lambda\partial_z+\bar{\lambda}\partial_{\bar{z}}+2(\lambda+\bar{\lambda})
\sin(\varphi/2)\partial_{\varphi}\,,\,\mbox{with}\,\,|\lambda|+|\bar{\lambda}
|>0
\end{equation}
at the point $(z,\bar{z},\varphi)$ is given by
$$
\angle{(w,e)}\!=\!\arccos\left(\frac{(w,e)}{|w||e|}\right)\!=\!\arccos\left(
\frac{2(1-c_0)(\lambda+\bar{\lambda})\sin(\varphi/2)}{\sqrt{c_0^2+8\sin^2(
\varphi/2)}\sqrt{\lambda^2\!+\!\bar{\lambda}^2\!+\!4(\lambda+\bar{\lambda}
)^2\sin^2(\varphi/2)}}\right),
$$
the minimum of which with respect to a variation of the parameters $\lambda,
\bar{\lambda}$ gives the angle $\angle({\cal W},{\cal E})$ in which the 
1--dimensional wavefront intersects the 2--dimensional extremal at $P$:
\begin{equation}
\angle({\cal W},{\cal E})=\mbox{min}\angle{(w,x)}=\arccos\left(\frac{
2\sqrt{2}|1-c_0||\sin(\varphi/2)|}{\sqrt{c_0^2+8\sin^2(\varphi/2)}\sqrt{1+8
\sin^2(\varphi/2)}}\right)\,. 
\end{equation} 
Obviously in the case $c_0 \! = \! 1$ the wave fronts cross the extremals 
always {\em perpendicularly}, because $\angle({\cal W},{\cal E}) \! = \! \pi
/2$ holds {\em everywhere} in ${\cal M}_{2+1}$ (see Fig.\ 3). 
\\[2mm]
For $c_0 \! \neq \! 0,1$ the maximum value $\angle({\cal W},{\cal E})\!=\!
\pi/2$ is taken only for $\varphi_k\!=\!2k\pi,\, k \! = \! 0, \pm 1,\pm 2,
\ldots$, {\em contrary} to the singular case $c_0\!=\!0$, where for these 
values of the field variable the minimum of the angle $\angle({\cal W},{\cal 
E})$ vanishes, indicating that 
the transversality relation between the wave fronts and extremals  
is violated there. The angle $\angle({\cal W},{\cal E})$ has a local minimum 
at those points where $|\cos(\angle({\cal W},{\cal E}))|$ becomes maximal:
\begin{equation} 
|\cos\left(\angle({\cal W},{\cal E})_{min 1}\right)|=\frac{
2\sqrt{2}|1-c_0|}{3\sqrt{c_0^2+8}}\,,\,\,\,\,
|\cos\left(\angle({\cal W},{\cal E})_{min 2}\right)|=\frac{
|1-c_0|}{1+|c_0|}\,.\label{min}
\end{equation}
We find the first minima $\angle({\cal W},{\cal E})_{min 1}$  
on planes $\varphi \!=\! (2k+1)\pi,\, k \! = \! 0, \pm 1,\pm 2, \ldots$
where the cosine of the field variable vanishes. They do exist for all 
values of the parameter $c_0 \!\in\! {\cal R}\setminus\{0\}$. For $c_0
\!\neq\!- 8$ the angle $\angle({\cal W},{\cal E})_{min 1}$ differs from 
zero, i.e.\ the transversality relation is fulfilled. If $c_0\!=\!- 8$ both 
minima (\ref{min}) coincide, therefore this case is discussed below.
\\[2mm]
The second minima $\angle({\cal W},{\cal E})_{min 2}$ exist only in 
the range $|c_0| \!\le\! 8$, $c_0\!\neq\!0$. These minima lie at values of 
the field variable: $\varphi \!=\!2k\pi\pm 2 \arcsin(\sqrt{|c_0|/8}), k \!=
\! 0, \pm 1, \pm 2, \ldots$. If $c_0 \!<\!0$, this angle vanishes, 
whereas it differs from zero for all $c_0 \!>\!0$. 
\\[2mm]
It results that in the case $c_0\!>\!0$ or $c_0\!<
\!-8$ the wave fronts and extremals are never parallel to each other 
($\angle({\cal W},{\cal E})_{min 1,2} \!\neq\! 0$). This guarantees the 
transversality relations everywhere in the extended configuration space 
${\cal M}_{2+1}$ --- even on the planes $\varphi_k \!=\! 2k\pi$, 
$k \!=\! 0, \pm 1,\pm 2,\ldots$ that separate the one parameter families of 
extremals --- contrary to the singular case. 
This result coincides with the fact that both the Lagrangian and the 
Hamiltonian densities do not vanish in ${\cal M}_{2+1}$ for $c_0\!>\!0$ or 
$c_0\!<\!-8$.
\\[2mm]
For $-8\!\le\!c_0\!<\!0$ the minimum $\angle({\cal W},{\cal E})_{min 2}$ is 
equal to zero. This happens just for the points on the plane 
\begin{equation}
\varphi=2k\pi\pm 2\arcsin(\sqrt{-c_0/8})\,,\quad k = 0, \pm 1,\pm 2, \ldots
\label{plane}
\end{equation}
where the shifted Lagrangian density (\ref{shift}) vanishes, as expected 
from the transversality relation (\ref{transvers}) (see Fig.\ 2). The case 
$c_0\!=\!- 8$ is a special one: here the two planes (\ref{plane}) that exist 
in every region 
$2k\pi \!<\! \varphi\! <\! 2(k \!+\! 1)\pi$, $z,\bar{z}\in{\cal R}$,
$k \!=\! 0, \pm 1,\pm 2,\ldots$ coincide. Hence we get only one plane 
$\varphi\!=\!2k\pi+\pi$ in the range $2k\pi \!<\! \varphi\! <\! 2(k \!+
\! 1)\pi$, where the transversality relations are violated due to a 
vanishing shifted Lagrangian density.
\\[2mm]
The dependence of the angle $\angle({\cal W},{\cal E})$ on the constant 
$c_0$ shows, that the geometrical properties of the wave fronts and even 
the transversality relation may be affected by changes of the Lagrangian 
density that do not influence the equation of motions and the corresponding 
extremals at all.
\\[3mm]
Now we would like to show how these results can be obtained using the 
recursion formulas given in sections (8.1) and (8.2). We first calculate 
Carath\'eodory's Hamilton--Jacobi functions $S^z(z,\bar{z},\varphi)$, $S^{
\bar{z}}(z,\bar{z},\varphi)$ from those of DeDonder \& Weyl, (\ref{DWWF1}) 
and 
(\ref{DWWF2}). For the sake of simplicity we choose the shift $c_0\!=\!-2$
in order to eliminate the constant $2$ in the Lagrangian of the Sine--Gordon
model ${\cal L}\!=\!\partial_z\varphi\partial_{\bar{z}}\varphi \!+\! 2(1-
\cos(\varphi))$.
\\[2mm]
The coefficients $A^{z}_0(z,\bar{z})$ and $A^{\bar{z}}_0(z,\bar{z})$ can be 
chosen as special solutions:
\begin{equation}
A_0^z=A^z_0(l)=4\tanh(l)-l\;,\quad A_0^{\bar{z}}=A_0^{\bar{z}}=
-\bar{l}
\label{anu}
\end{equation}
of the equation (\ref{A_0}) written in terms of the variables $l \!=\! z \!+
\! \bar{z}$ and $\bar{l} \!=\! z \!-\! \bar{z}$. 
\begin{equation}
(\partial_{\bar{l}}A_0^{z})(\partial_{l}A_0^{\bar{z}})-(\partial_{\bar{l}
}A_0^{\bar{z}})(\partial_{l}A_0^z)=\frac{{\cal L}_0}{2}=\frac{4}{\cosh(l
)^2}-1\,.
\end{equation}
The equations (\ref{A1}) and (\ref{AQ1}) yields the coefficients of first 
order:
\begin{equation}
A_1^z=A^z_1(l)=\frac{2}{\cosh(l)}\;,\quad A_1^{\bar{z}}=0\,,
\end{equation}
and the relations (\ref{A^z1}), (\ref{A^z}), (\ref{A^z0}) and 
(\ref{A^z00}) lead to those of the second and the third order in $y$. They 
coincide with those of the singular case (\ref{two}) since the ratios
$(\partial_zA^z_0)/{\cal L}$ and $(\partial_{\bar{z}}A^z_0)/{\cal L}$ are the 
same in the regular and in the singular case. The same holds for the 
coefficients $A^{z}_n(z,\bar{z})$ and $A^{\bar{z}}_n(z,\bar{z})$ of higher 
orders in $y$. So we get for $n \! > \! 0$: 
\begin{equation}
A_{2n}^z=A^z_{2n}(l)=4\frac{(-1)^{n}}{2^{2n}}\tanh(l)\,,\quad
A_{2n-1}^z=A^z_{2n-1}(l)=4\frac{(-1)^{n}}{2^{2n-1}}\frac{1}{\cosh(l)}\,,\quad
A_n^{\bar{z}}=0\,,
\end{equation}
which combine the expressions (\ref{anu}) yield the Hamilton--Jacobi 
functions $S^z(z,\bar{z},\varphi)$, $S^{\bar{z}}(z,\bar{z},\varphi)$:
\begin{eqnarray}
S(z,\bar{z},\varphi)\!&\!=\!&\!4\left(\sum_{n=0}^{\infty}
\frac{(-1)^ny^{2n}}{n!2^{2n
}}\right)\tanh(l)+4\left(\sum_{n=0}^{\infty}\frac{(-1)^ny^{2n+1}}{n!2^{2n+1}
}\right)\frac{1}{\cosh(l)}-l,
\nonumber\\
\!&\!=\!&\!4\cos\left(\frac{y}{2}\right)\tanh(l)+4\sin\left(\frac{y
}{2}\right)\frac{1}{\cosh(l)}\!-\!l=-4\cos\!\left(\frac{\varphi}{2}\right)
\!-\!z\!-\!\bar{z}\label{HJC21},\\
S^{\bar{z}}(z,\bar{z},\varphi)\!&\!=\!&\!\bar{z}-z\label{HJC22}\,.
\end{eqnarray}
Both functions $S^z$ and $S^{\bar{z}}$ satisfy Carath\'eodory's 
Hamilton--Jacobi equation (\ref{HJE}), the integrability criterion 
(\ref{ICC}) and the embedding conditions (\ref{qq}). They are related to 
the solutions (\ref{reg}) by the transformation
\begin{equation}
S^z\,\rightarrow\,\tilde{S}^z=-\frac{1}{\sqrt{2}}(S^z+S^{\bar{z}})\,,\quad 
S^{\bar{z}}\,\rightarrow\,\tilde{S}^{\bar{z}}=\frac{1}{\sqrt{2}}(S^{
\bar{z}}-S^z)
\,.\vspace{3mm}\label{con}
\end{equation}
The wave fronts $z(\sigma,\bar{\sigma},\varphi)$ and $\bar{z}(\sigma,\bar{
\sigma},\varphi)$ can be determined recursively following the ideas discussed 
in the section (8.2):
\\[2mm]
The coefficients $\alpha_0(\sigma,\bar{\sigma}), \bar{\alpha}_0(\sigma,\bar{
\sigma})$ of zeroth order in $\hat{y}$ are given by inverting the eqs.\ 
(\ref{nullte}):
\begin{eqnarray}
\sigma &=& A_0^z\left(z=\alpha_0(\sigma,\bar{\sigma}),\bar{z}=\bar{\alpha}_0(
\sigma,\bar{\sigma})\right)=4\tanh(\alpha_0+\bar{\alpha}_0)-(\alpha_0+\bar{
\alpha}_0)\,,\\
\bar{\sigma} &=& A_0^{\bar{z}}\left(z=\alpha_0(\sigma,\bar{\sigma}),\bar{z}=
\bar{\alpha}_0(\sigma,\bar{\sigma})\right)=-\alpha_0+\bar{\alpha}_0\,,
\end{eqnarray}
which yields:
\begin{equation}
\alpha_0=\frac{1}{2}(f^{-1}(\sigma)-\bar{\sigma})\quad\mbox{and}\quad
\bar{\alpha}_0=\frac{1}{2}(f^{-1}(\sigma)+\bar{\sigma})\,,
\end{equation}
where the symbol $f^{-1}$ denotes the inverse of the function 
$f(x)\!:=\!\tanh(x)-x$. The coefficients of $\hat{y}$, $\hat{y}^2$ and 
$\hat{y}^3$ are given by (\ref{ende}), (\ref{erste}), (\ref{zweite1}) 
and (\ref{zweite2}):
\begin{eqnarray}
\alpha_1(\sigma)=\frac{\partial_{\bar{z}}\varphi_0}{{
\cal H}_0}=\frac{1}{\cosh(f^{-1}(\sigma))}
\,,&&
\bar{\alpha}_1(\sigma)=\frac{\partial_{z}\varphi_0}{{\cal H}_0}
=\frac{1}{\cosh(f^{-1}(\sigma))}\,,
\\
\alpha_2(\sigma)=-\frac{1}{2}\tanh(f^{-1}(\sigma))\,,&&
\bar{\alpha}_2(\sigma)=-\frac{1}{2}\tanh(f^{-1}(\sigma))\,,
\\
\alpha_3(\sigma)=-\frac{1}{4\cosh(f^{-1}(\sigma))}\,,&&
\bar{\alpha}_3(\sigma)=-\frac{1}{4\cosh(f^{-1}(\sigma))}\,.
\end{eqnarray}
This leads to a general ansatz for the coefficients $\alpha_n$ and $\bar{
\alpha}_n$, $n\!\ge\!1$:
\begin{equation}
\alpha_{2n}=\bar{\alpha}_{2n}=2\frac{(-1)^{n}}{2^{2n}}\tanh(f^{-1}(\sigma
))\,,\quad \alpha_{2n-1}=\bar{\alpha}_{2n-1}=2\frac{(-1)^{n}}{2^{2n-1}}
\frac{1}{\cosh(f^{-1}(\sigma))}\,,
\end{equation}
which give the wave fronts
\begin{eqnarray}
z(\sigma,\bar{\sigma},\varphi) &=& 2\left(\sum_{n=0}^{\infty}\frac{(-1)^ny^{
2n}}{n!2^{2n}}\right)\tanh(f^{-1}(\sigma))+2\left(\sum_{n=0}^{\infty}
\frac{(-1)^ny^{2n+1}}{n!2^{2n+1}}\right)\frac{1}{\cosh(f^{-1}(\sigma))}
\nonumber\\
&&-2\tanh(f^{-1}(\sigma))+\frac{1}{2}(f^{-1}(\sigma)-\bar{\sigma})=
-2\cos\left(\frac{\varphi}{2}\right)-\frac{1}{2}(\sigma+\bar{\sigma}),\\
{\bar{z}}(\sigma,\bar{\sigma},\varphi)
&=&-2\cos\left(\frac{\varphi}{2}\right)
-\frac{1}{2}(\sigma-\bar{\sigma})\,.
\end{eqnarray}
which are related to the representation (\ref{exx}) by the 
transformation (\ref{con}) applied to $S^z\!=\!\sigma$ and $S^{\bar{z}} \!=\! 
\bar{\sigma}$, which reparametrizes the family of wave fronts only.

\section{Conclusions}
Within the manifest covariant Hamilton--Jacobi canonical frameworks of 
De\-Don\-der \& Weyl and of Ca\-ra\-th\'eo\-do\-ry we have investigated 
relations between families of extremals and Hamilton--Jacobi wave fronts for 
2--dimensional one component field theories. This is of interest, since the 
dynamics of fields can be described either by the Euler--Lagrange  
or the Hamilton--Jacobi equations supplemented by the integrability 
conditions.
\\[2mm]
We developed a formalism to solve the DeDonder \& Weyl Hamilton--Jacobi 
equation and the integrability condition perturbatively by expanding the 
Hamilton--Jacobi functions in powers of the field variable. Starting from a 
single given extremal it is then possible to calculate new ones from it by 
using the two DeDonder \& Weyl Hamilton--Jacobi functions.
\\[2mm]
This formalism is useful especially for investigating extremals in the 
neighbourhood of known extremals in several 2--dimensional field theories: 
the massless and massive Klein--Gordon, the Sine-- and Sinh--Gordon, the 
Liouville as well as the $\phi^4$ theory. In the Sine--, Sinh--Gordon 
and in the $\phi^4$ theory we have studied the embedding of topologically 
nontrivial soliton solutions. This approach is related to the usual 
stability investigations of solitons where perturbations are considered
which are a product of functions depending on the time and the space 
variables separately. We determined the {\em general} solutions of the 
equations of the second variation by using B\"acklund transformations. 
In this manner we have obtained all the extremals in the vicinity of a given 
one.
\\[2mm]
Calculating proper Hamilton--Jacobi wave fronts makes the use of 
Carath\'eodory's Ha\-mil\-ton--\-Ja\-co\-bi functions $S^z(z,\bar{z},
\varphi), \,S^{
\bar{z}}(z,\bar{z},\varphi)$ necessary. Solving Ca\-ra\-th\'eo\do\ry's 
Ha\-mil\-ton--\-Ja\-co\-bi equation is considerably simplified by using the 
corresponding DeDonder--Weyl Hamilton--Jacobi functions. One first obtains
the wave fronts as equipotential surfaces $S^z\!=\!\sigma\!=\!const.,\,S^{
\bar{z}}\!=\!\bar{\sigma}\!=\!const.$ in an implicit form. If the 
transformation of variables $(z,\bar{z},\varphi) \rightarrow (\sigma,\bar{
\sigma},\varphi)$ 
is a regular one, i.e.\ if Carath\'eodory's Hamilton density does not vanish 
on the family of extremals under consideration, we get an explicit 
representation $z\!=\!z(\varphi,\sigma,\bar{\sigma}),\,\bar{z}\!=\!\bar{z}(
\varphi,\sigma,\bar{\sigma})$ for the wave fronts.
\\[2mm]
These general results have been applied in detail to a special single kink 
solution of the Sine--Gordon equation. After calculating the DeDonder \& 
Weyl Hamilton--Jacobi potentials $S$, $\bar{S}$ we obtained a corresponding 
one--parameter family of embedded extremals: a field of kink solutions of 
constant energy covering the extended configuration space. {}From the 
functions $S$, $\bar{S}$ we have determined Carath\'eodory's 
Hamilton--Jacobi functions $S^z$, $S^{\bar{z}}$ explicitly. The wave fronts 
have been determined for the singular (${\cal H}\!=\!0$) as well as for the 
regular case (${\cal H}\!\neq\!0$). In addition the transversality 
conditions between the wave fronts and the embedded extremals have been 
analyzed.

\begin{appendix}
\section{Appendix}
B\"acklund transformations are employed to map the integral submanifolds of
the PDE (\ref{variation}) to those of the Klein--Gordon or wave equation, 
the {\em general} solutions of which are known. In 
this chapter we determine all the functions $\partial_l^2V(\varphi_0(l))$ 
with $l \!=\! z \!+\! \bar{z}$
that allow this map by {\em one} BT. But unfortunately it turns out that the 
nonconstant coefficients $\partial_l^2V(\varphi_0(l))$ in (\ref{phi}), 
(\ref{KG3}) for the $\phi^4$ models do not belong to this class. 
Thus, we need at least two BTs to connect these two equations to one PDE 
with constant coefficients. Moreover we have to discuss which relations of 
type (\ref{variation}) can be reduced to free field equations by one or 
a {\em finite} number of BTs at all.
\\[2mm]
We show that there is no BT which relate the PDEs (\ref{phi}), (\ref{KG3}) 
and $\partial_z\partial_{\bar{z}}\hat{\theta}\!=\!m^2\hat{\theta}$, 
$m^2\in {\cal R}$. Even if we make the more general 
assumption that the transformed function $\hat{\theta}$ obeys the eq.\
$\partial_z\partial_{\bar{z}}\hat{\theta}\!=\!m^2(z,\bar{z})\hat{\theta}$ 
with a nonconstant but separable coefficient $m^2$: $m^2\!=\!B(z)\bar{B}(
\bar{z})$ the equations (\ref{phi}) and (\ref{KG3}) for the $\phi^4$ theories 
cannot be 
reduced by {\em one} BT to such an equation for $\hat{\theta}$ and thus not 
solved like the PDEs (\ref{liou}), (\ref{sg}), (\ref{sh}) for the Liouville, 
the Sine--  and the Sinh--Gordon models. The reason is that equation 
$\partial_z\partial_{\bar{z}}\hat{\theta}\!=\!m^2(z,\bar{z})\hat{\theta}$ 
can be reduced to a Klein--Gordon one by a suitable transformation of 
variables $z\!\rightarrow\! B(z)$ and $\bar{z}\!\rightarrow\!\bar{B}(\bar{z}
)$ \cite{wi}. The equations (\ref{phi}) and (\ref{KG3}) {\em can} 
be transformed by {\em two} BTs to Klein--Gordon equations. This result can 
be generalized to PDEs of the following type:
\begin{equation}
\partial_z\partial_{\bar{z}}\theta=\{n[n+1]\eta^2(l=z+\bar{z})+a\}\theta\,
\quad n=1,2,\ldots\,,\quad a\in {\cal R}\,,
\end{equation}
if $\eta$ is a solution of $\partial_l\eta\!=\!\bar{b}\eta^2\!+\!\bar{c}$ 
with $\bar{b}^2\!=\!\pm1$ and $\bar{c}\in {\cal R}$ by $h$ BTs. Especially
the PDEs (\ref{phi}), (\ref{KG3}) can be obtained by choosing $\bar{b}\!=\!
-1$, $\bar{c}\!=\!1$ and $n\!=\!2$. 
We start from eqs.\ (\ref{psi2}), assuming $m^2 \! \rightarrow m^2(l)$ to be 
a function of $l\!=\!z\!+\!\bar{z}$ and rename $\partial_{\varphi}^2V(
\varphi_0(l))$ as $2v(l)$. {}From the equations (\ref{psi2}) we infer
\begin{eqnarray}
\psi&=&\psi_0+A(z)+\bar{A}(\bar{z})\,,\quad\psi_0(l)=\frac{1}{2}\int^l\,
\mbox{d}l^{\prime}\int^{l^{\prime}}\,\mbox{d}l^{\prime\prime}(m^2(l^{\prime
\prime})+v(l^{\prime\prime}))\,,
\label{psi_0}\label{A}\\
\Rightarrow\, 0&=&\partial_l^2\psi_0+(\partial_l\psi_0+\partial_zA)
(\partial_l\psi_0+\partial_{\bar{z}}\bar{A})-m^2(l)\label{K und V}
\end{eqnarray}
with two arbitrary functions $A$ and $\bar{A}$. We are interested in the 
relation between $m^2(l)$ and $v(l)$. First we have to determine the 
functions $A$ and $\bar{A}$, which depend only on $z$ and $\bar{z}$ 
respectively. Differentiating the PDE (\ref{K und V}) with respect to 
$\bar{l}\!=\!z\!-\!\bar{z}$ leads to:
\begin{equation}
\partial_l\psi_0[\partial_z^2A-\partial_{\bar{z}}^2\bar{A}]=
\partial_{\bar{z}}^2\bar{A}\partial_zA-\partial_z^2A
\partial_{\bar{z}}\bar{A}\,,\label{psi0}
\end{equation}
because $\psi_0$ depends on $l$ only. Now, if $\partial_{\bar{z}}^2\bar{A}\!
=\!0$ or $\partial_z^2A \!=\! 0$, we obtain from eq.\ (\ref{psi0}) that 
either $\partial_{\bar{z}}^2\bar{A} \!=\! 0$ and $\partial_z^2A \!=\! 0$ 
(case I) or $\partial_l\psi_0 \!=\! -a_0$ (case II). The case II leads to 
$\partial_l^2\psi_0 \! = \!0$ and using relation (\ref{K und V}) gives  
$m^2(l) \!=\! 0$. The definition of $\psi_0$ 
finally yields $v\!=\!0$, which is of no interest for us. Without any loss 
of generality we have chosen: $\partial_z^2A\!=\!0\Rightarrow\partial_zA \!
= \! a_0,\, a_0 \in {\cal R}$.  
\\[2mm]
If $0\!=\!\partial_z^2A\!-\!\partial_{\bar{z}}^2\bar{A}$ (case III), then 
it follows that either $\partial_z^2A\!=\!\partial_{\bar{z}}^2\bar{A} \! = 
\!0$ or $\partial_zA\!=\!\partial_{\bar{z}}\bar{A} \! = 
\!a \!=\! const.$, the second case being a special case of the first one. 
So we have $\partial_zA\! = \!a_1$ and $\partial_{\bar{z}}\bar{A}\!=\!\bar{
a}_1$ with $a_1,\bar{a}_1\in {\cal R}$. Thus case III is contained in the 
first one.
\\[2mm] 
If $\partial_z^2A\!-\!\partial_{\bar{z}}^2\bar{A}\!\neq\! 0$ (case IV) we 
are able to divide the relation (\ref{K und V}) by it and to apply the 
operator $\partial_{\bar{l}}$ once more:
\begin{eqnarray}
0&=&(\partial_z^3A\partial_{\bar{z}}^2\bar{A}+\partial_z^2A
\partial_{\bar{z}}^3\bar{A})(\partial_zA-\partial_{\bar{z}}\bar{A})
-2\partial_z^2A\partial_{\bar{z}}^2\bar{A}(\partial_z^2A-
\partial_{\bar{z}}^2\bar{A})\,,\label{II}
\\
\Rightarrow\,0&=&\partial_z\left\{\frac{\partial_z^3A}{\partial_z^2A}
\right\}-2\partial_z\left\{\frac{\partial_z^2A-\partial_{\bar{z}}^2\bar{A}}{
\partial_zA-\partial_{\bar{z}}\bar{A}}\right\}\,.
\end{eqnarray}
Differentiating this equation with respect to $\bar{z}$, dividing the 
result by $\partial_{\bar{z}}^2\bar{A}$ and applying $\partial_{\bar{z}}$
leads to:
\begin{equation}
\partial_{\bar{z}}^3\bar{A}=\partial_{\bar{z}}^2\bar{A}[e_0\partial_{\bar{z}}
\bar{A}+e_1]\,,\quad\partial_z^3A=\partial_z^2A[e_2\partial_zA+e_3]\,,\quad 
e_0,e_1,e_2,e_3\in {\cal R}\,.\label{sonne}
\end{equation}
Substituting these results into the equation (\ref{II}) yields for $A(z)$ and 
$\bar{A}(\bar{z})$ the differential equations:
\begin{equation}
\partial_{\bar{z}}^2\bar{A}=\frac{1}{2}(\partial_{\bar{z}}\bar{A})^2e_0
+e_4\partial_{\bar{z}}\bar{A}+e_5\,,\quad
\partial_z^2A=\frac{1}{2}(\partial_zA)^2e_0+e_4\partial_zA+e_5\,,
\end{equation}
with the same constants $e_0,e_4,e_5$ for both functions. Inserting these 
expressions into relation (\ref{psi0}) we get for $\partial_l^2\psi_0\!=\!(
m^2\!+\!v)/2$:
\begin{equation}
\partial_l^2\psi_0=-\frac{e_0\partial_z^2A\partial_{\bar{z}}^2
\bar{A}}{2[\frac{e_0}{2}(\partial_zA+\partial_{\bar{z}}\bar{A})+e_4]^2}\,.
\end{equation}
{}From equation (\ref{K und V}) it follows that 
\begin{eqnarray}
m^2(l)&=&\,\,\left(1-\frac{e_0}{2}\right)\partial_z^2A\partial_{\bar{z}}^2
\bar{A}\left[\frac{e_0}{2}(\partial_zA+\partial_{\bar{z}}\bar{A})+e_4\right]
^{-2},\\
v(l)&=\!&\!-\,\left(1+\frac{e_0}{2}\right)\partial_z^2A
\partial_{\bar{z}}^2\bar{A}\left[\frac{e_0}{2}(\partial_zA+\partial_{\bar{z}}
\bar{A})+e_4\right]^{-2}.
\end{eqnarray}
Thus we have to solve the differential equations (\ref{sonne}) for 
$\partial_z A$ and $\partial_{\bar{z}} \bar{A}$ with different choices of the 
constants $e_0,e_4$ and $e_5$. Depending on the sign of $\Delta\!=\!2e_0e_5
\!-\! e_4^2$ there exist three different cases:
\begin{eqnarray}
\mbox{IV}.1)\,\Delta=0&:&\partial_zA=-\frac{e_4}{e_0}-\frac{2}{c(z-z_0)}\,,\,
\,\partial_{\bar{z}}\bar{A}=-\frac{e_4}{e_0}-\frac{2}{c(\bar{z}-\bar{z}_0)}\,
,\,\,z_0,\bar{z}_0\in{\cal C}\\
\mbox{IV}.2)\,\Delta<0&:&\partial_zA=\frac{\sqrt{\Delta}}{e_0}\tan\left(
\frac{\sqrt{\Delta}}{2}(z-z_0)\right)-\frac{e_4}{e_0}\,,\\
&&\partial_{\bar{z}}\bar{A}=\frac{\sqrt{\Delta}}{e_0}\tan\left(\!\frac{
\sqrt{\Delta}}{2}(\bar{z}-\bar{z}_0)\right)-\frac{e_4}{e_0}\,,\\
\mbox{IV}.3)\,\Delta>0&:&\partial_zA=-\,\frac{\sqrt{-\Delta}}{e_0}\tanh
\left(\frac{\sqrt{-\Delta}}{2}(z-z_0)\right)-\frac{e_4}{e_0}\,,\\
&&\partial_{\bar{z}}\bar{A}=-\,\frac{\sqrt{-\Delta}}{e_0}\tanh\left(\frac{
\sqrt{\Delta}}{2}(\bar{z}-\bar{z}_0)\right)-\frac{e_4}{e_0}\,.
\end{eqnarray}
For the case IV.1) we obtain:
\begin{equation}
m^2(l)=\frac{4}{e_0^2}\left(1-\frac{e_0}{2}\right)[l-l_0]^{-2}
\,,\quad v=-\,\frac{4}{e_0^2}\left(1+\frac{e_0}{2}\right)[l-l_0]
^{-2}\,,\quad l_0=z_0+\bar{z}_0\,.
\end{equation}
If we choose $e_0\!=\!2$, then $m^2(l)$ vanishes and $v\!=\!-2[l\!
-\!l_0]^{-2}$ is for $l_0\!=\!0$ identical with 
$\partial_{\varphi}^2V(\varphi_0)$ in Liouville's model and on the other 
hand if one would like to obtain the PDE (\ref{liou}) only (this means $m^2(l
)\! = \!-2[l\!-\!l_0]^{-2}$) one has to calculate $e_0$ from the eq.\ $-2\!=
\!4(1\!-\!e_0/2)/e_0^2$ yielding $e_{0_{1,2}}\!=\!-1,2$. Therefore one 
obtains the wave equation or once again Liouville's  model. The last case 
represents only a map of (\ref{liou}) onto itself (for details of Auto--BTs 
see e.g.\ \cite{Laughlin}). 
\\[2mm]
The case IV.2) yields:
\begin{equation}
m^2(l)=\frac{\Delta}{e_0^2}\left(1-\frac{e_0}{2}\right)\sin^{-2}
\left[\frac{\sqrt{\Delta}}{2}(l-l_0)\right]\,,\quad
v=-\,\frac{\Delta}{e_0^2}\left(1+\frac{e_0}{2}\right)\sin^{-2}
\left[\frac{\sqrt{\Delta}}{2}(l-l_0)\right]\,.
\end{equation}
This is of no interest for our special models. Notice, however, that the 
special choice $v \!=\! -\Delta 2^{-1}\sin^{-2}(\sqrt{\Delta}[l\!-\!l_0]/2)$ 
leads to the wave equation.
\\[2mm]
The case IV.3) is obviously similar to IV.2). The resulting functions $m^2(l
)$ and $v(l)$ are:
\begin{equation}
m^2(l)=\frac{\Delta}{e_0^2}\left(1-\frac{e_0}{2}\right)\sinh
^{-2}\left[\frac{\sqrt{-\Delta}}{2}(l-l_0)\right],\,\,
v=\frac{\Delta}{e_0^2}\left(1+\frac{e_0}{2}\right)\sinh^{-2}
\left[\frac{\sqrt{-\Delta}}{2}(l-l_0)\right]\,.\vspace{2mm}
\end{equation}
As the equations (\ref{phi}) and (\ref{KG3}) for the $\phi^4$ models are not 
contained in the cases discussed up to now we return to case I: $\partial_z A 
\!=\! a_1$ and $\partial_{\bar{z}}\bar{A} \!=\! \bar{a}_1$. Assuming $m^2$ to 
be independent of $l$ equation (\ref{K und V}) then gives:
\begin{equation}
\partial_l^2\psi_0+(\partial_l\psi_0)^2+(a_1+\bar{a}_1)\partial_l\psi_0
+a_1\bar{a}_1-m^2=0\,.
\end{equation}
Substituting $\xi\!=\!\partial_l\psi_0\!+\!(a_1\!+\!\bar{a}_1)/2$ and 
$\tilde{\Delta}\!=\!m^2\!+\!(a_1\!-\!\bar{a}_1)^2/4$ leads to:
\begin{eqnarray}
\mbox{I.1})\,\tilde{\Delta}>0\!&\!:\!&\! \xi=\tilde{\Delta}\tanh(\tilde{
\Delta}(l-l_0))\,,\,\,\Rightarrow\,\,v=2\tilde{\Delta}^2(1-\tanh^2[
\tilde{\Delta}(l-l_0)])-m^2.\label{xi}\\
\mbox{I.2})\,\tilde{\Delta}<0\!&\!:\!&\! \xi=-\,\tilde{\Delta}\tan(\tilde{
\Delta}(l-l_0))\,,\,\,\Rightarrow\,\,v=-\,2\tilde{\Delta}^2(1\!+\!\tan^2[
\tilde{\Delta}(l-l_0)])-m^2.\\
\mbox{I.3})\,\tilde{\Delta}=0\!&\!:\!&\! \xi=\frac{1}{l-l_0}\,,\,\,
\Rightarrow\,\,v=-\,2\frac{1}{(l-l_0)^2}-m^2\,.
\end{eqnarray}
The case I.1 is the essential one for us. Because of the special type of
the equations (\ref{phi}) and (\ref{KG3}), we are able to eliminate $\Delta$ 
by a transformation of variables $z\rightarrow\Delta z$ and 
$\bar{z}\rightarrow\Delta\bar{z}$. Thus only the class of PDEs with a 
coefficient 2 in front of their $\tanh^2(l)$:
\begin{equation}
\partial_z\partial_{\bar{z}}\theta=\left\{n_0\tanh^2(l)+\bar{n}_0
\right\}\theta\,,\quad n_0=2\,,\quad \bar{n}_0\in {\cal R} \label{b=2}
\end{equation}
can be reduced by {\em one} BT to a 
Klein--Gordon or wave equation, like the relations we obtain in the 
Sine--Gordon (\ref{sg}) and the Sinh--Gordon (\ref{sh}) theory. 
For the $\phi^4$ models with $n_0\!=\!6$ we need at least two BTs.
\\[2mm]
Substituting $m^2(l)\!=\!2\psi_0(l)\!-\!v(l)$ in equation 
(\ref{K und V}) leads to:
\begin{equation}
\partial_l^2\psi_0-(\partial_l\psi_0+a_1)(\partial_l\psi_0+\bar{a}_1)-
v=0\,.\label{00}
\end{equation}
Choosing $v$ to be equal to $-b\eta^2(l)\! - \!c$ where the function 
$\eta$ has to obey the equation $\partial_l\eta\! = \!\bar{b}\eta^2\!+\!
\bar{c}$ with $\bar{b},\bar{c}\in {\cal R}$, making the special
ansatz $\partial_l\psi_0\!=\!d_0\!+\!d_1\eta$ with two constants $d_0,d_1$,
inserting all this into equation (\ref{00}) and comparing the coefficients 
in front of the powers of $\eta$ yields:
\begin{eqnarray}
\underline{\eta^2}&:&d_1\bar{b}+b-d_1^2=0\,,\,\,\Rightarrow\,\,d_{1_{1,2}}=
\frac{\bar{b}}{2}\pm\sqrt{\left(\frac{\bar{b}}{2}\right)^2+b}\,,\\
\underline{\eta^1}&:&2d_0+a_1+\bar{a}_1=0\,,\,\,\Rightarrow\,\,2d_0=-\,a_1
-\bar{a}_1\,,\\
\underline{\eta^0}&:&d_1\bar{c}+c-(d_0+a_1)(d_0+\bar{a}_1)=0\,,\,\,
\Rightarrow\,\,\bar{a}_1=\pm2\sqrt{-d_1\bar{c}-c}+a_1.
\end{eqnarray}
For $m^2(l)$ we obtain:
\begin{equation}
m^2(l)=2\partial_l^2\psi_0-v=c+2d_1\bar{c}+(2d_1\bar{b}+
b)\eta^2\,.\label{K}
\end{equation}
Of special interest is the coefficient in front of $\eta^2$. Inserting 
$d_1$ into eq.\ (\ref{K}) we get:
\begin{equation}
2d_1\bar{b}+b=\bar{b}^2+b\pm\bar{b}\sqrt{\bar{b}^2+4b}\,,
\end{equation}
with $b\!=\!n(n\!+\!1)$ and choosing $\bar{b}$ to be equal to $\pm1$ we 
obtain:
\begin{equation}
2d_1\bar{b}+b=n(n+1)+1\pm\sqrt{1+4n(n+1)}=\left\{\begin{array}{l}
n(n-1)\\(n+1)(n+2)\end{array}\right.\,.
\end{equation}
Thus starting with a coefficient $n(n\!+\!1)$ one BT can raise or lower 
$n$ by $1$. Choosing $n\!=\!1,2,\ldots$ we are able to calculate the  
functions $\psi_i$, which are essential in order to determine the $i$--th BT 
of the hierarchy of $n$ BTs, using the equation (\ref{psi_0}).
Moreover the solutions 
$\tilde{\psi}_i\!=\!\exp(-\psi_i)$ of the $i$--th PDE 
$\partial_z\partial_{\bar{z}}\theta\!+\!v_i\theta\!=\!0$ which reciprocal 
fulfils $\partial_z\partial_{\bar{z}}\hat{\theta}\!+\!m^2_i(l)\hat{\theta}\!
=\!0$ (see (\ref{bt})) can be calculated. Obviously the special case 
$\bar{b}\!=\!-1$ and $\bar{c}=1$ leads to:
\begin{equation}
\partial_z\partial_{\bar{z}}\theta=[n(n+1)\tanh^2(z+\bar{z})+a]\theta\,.
\end{equation}
For $n\!=\!2$ and special choices of $a$ this relation yields the PDEs 
(\ref{phi}), (\ref{KG3}) for the $\phi^4$--theories. So they can be solved 
by $n \!=\! 2$ BTs.
\end{appendix}

\newpage
\subsection*{Figures}
Here we display the 2--dimensional projections of the 2--dimensional 
extremals ((\ref{next})) on the plane $\bar{l} \!=\! \gamma (t-vx)\! = \! 
const.$ and those 1--dimensional wave fronts (\ref{wsing}) that lie entirely
in this plane. The ``singular" case (\ref{wsing}) is shown in Fig.\ 1, the 
``regular" ones (\ref{exx}) for $c_0\!=\! -2$ and $c_0\!=\! 1$ in Figs.\ 2 
and 3, respectively. The extremals are plotted in solid lines, whereas the 
wave fronts are given in dotted ones. The axes are the field variable and 
the independent variable $l \!=\! \gamma (x-vt)$ which parametrizes the kink 
solution, obtained from the static one by a Lo\-rentz transformation (see 
(\ref{lorentz})). 
\\[3mm]
{\bf Fig.\ 1: The singular case} 
\begin{center}
\epsfig{file=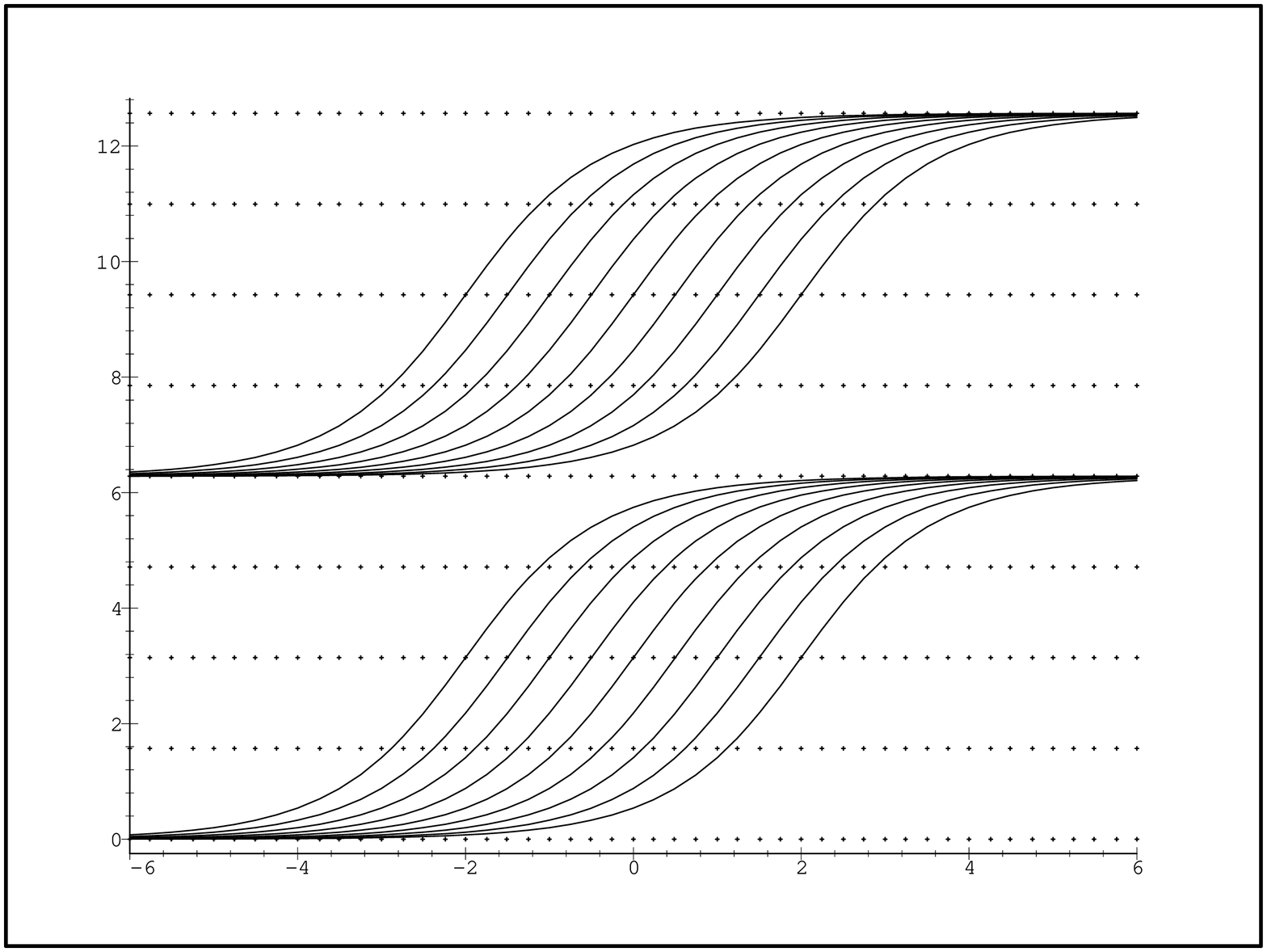,width=12cm, angle=-90}
\\[-85mm]\hspace*{-9.2cm}\small$\varphi$\normalsize\\[29mm]\hspace*{-96mm}
\scriptsize$2\pi$\normalsize\\[32mm]\hspace*{93mm}\small$l$\normalsize
\\[12mm]
\end{center}
In this case the transversality relations are fulfilled in the extended 
configuration space ${\cal M}_{2+1}$ except on the planes $\varphi \!=\! 2k
\pi$, $k \!=\! 0,\pm 1,\pm 2,\ldots$, where the extremals and wave fronts are 
parallel. 
\newpage
{\bf Fig.\ 2: The regular case for $c_0\!=\!-2$}
\begin{center}
\epsfig{file=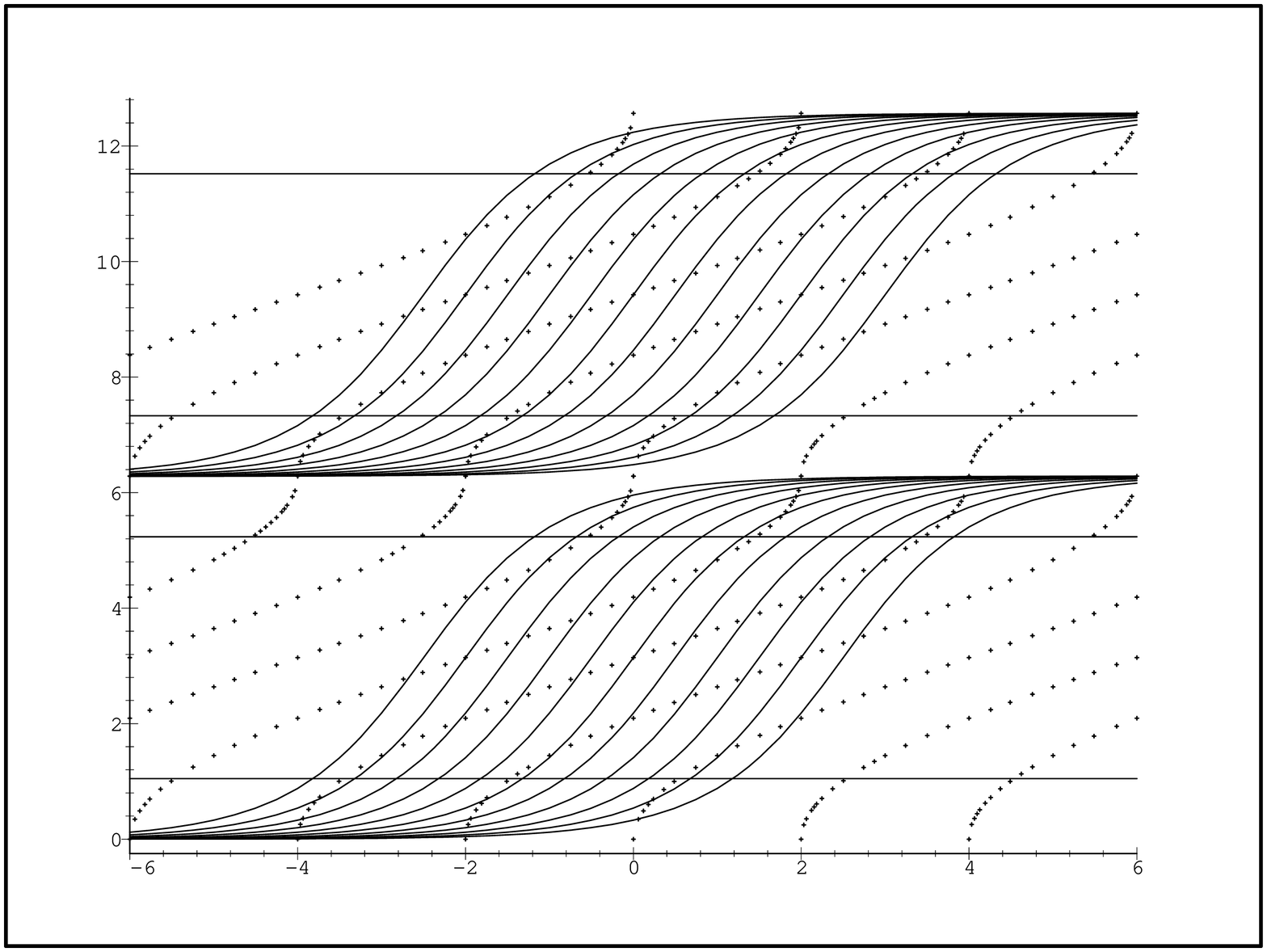,width=12cm, angle=-90}
\\[-85mm]\hspace*{-9.2cm}\small$\varphi$\normalsize\\[29mm]\hspace*{-94mm}
\scriptsize$2\pi$\normalsize\\[32mm]\hspace*{93mm}\small$l$\normalsize
\\[-16mm]\hspace*{-93mm}\scriptsize$\frac{1}{3}\pi$\normalsize
\\[-25.5mm]\hspace*{-93mm}\scriptsize$\frac{5}{3}\pi$\normalsize
\\[-15.75mm]\hspace*{-93mm}\scriptsize$\frac{7}{3}\pi$\normalsize
\\[-26mm]\hspace*{-94mm}\scriptsize$\frac{11}{3}\pi$\normalsize
\\[72mm]
\end{center}
Here the transversality relations are fulfilled outside the planes $\varphi 
\!=\! 2k\pi\pm \pi/3$, $k \!=\! 0,\pm 1,\pm 2,\ldots$, where the extremals 
and wave fronts are parallel since the Lagrangian density vanishes there.
Notice, that the transversality relations are fulfilled on the boundaries  
of the regions ${\cal M}_k \!=\! \{2k\pi \!<\!\varphi \!<\! 2(k\! +\! 1)\pi,
z, \bar{z} \!\in \!{\cal R}\}$, where the families of extremals (\ref{next}) 
are defined. Therefore the wave fronts can be continued from one of these 
regions to the next smoothly, contrary to the extremals.
\newpage
{\bf Fig.\ 3: The regular case for $c_0 \!=\! 1$}
\begin{center}
\epsfig{file=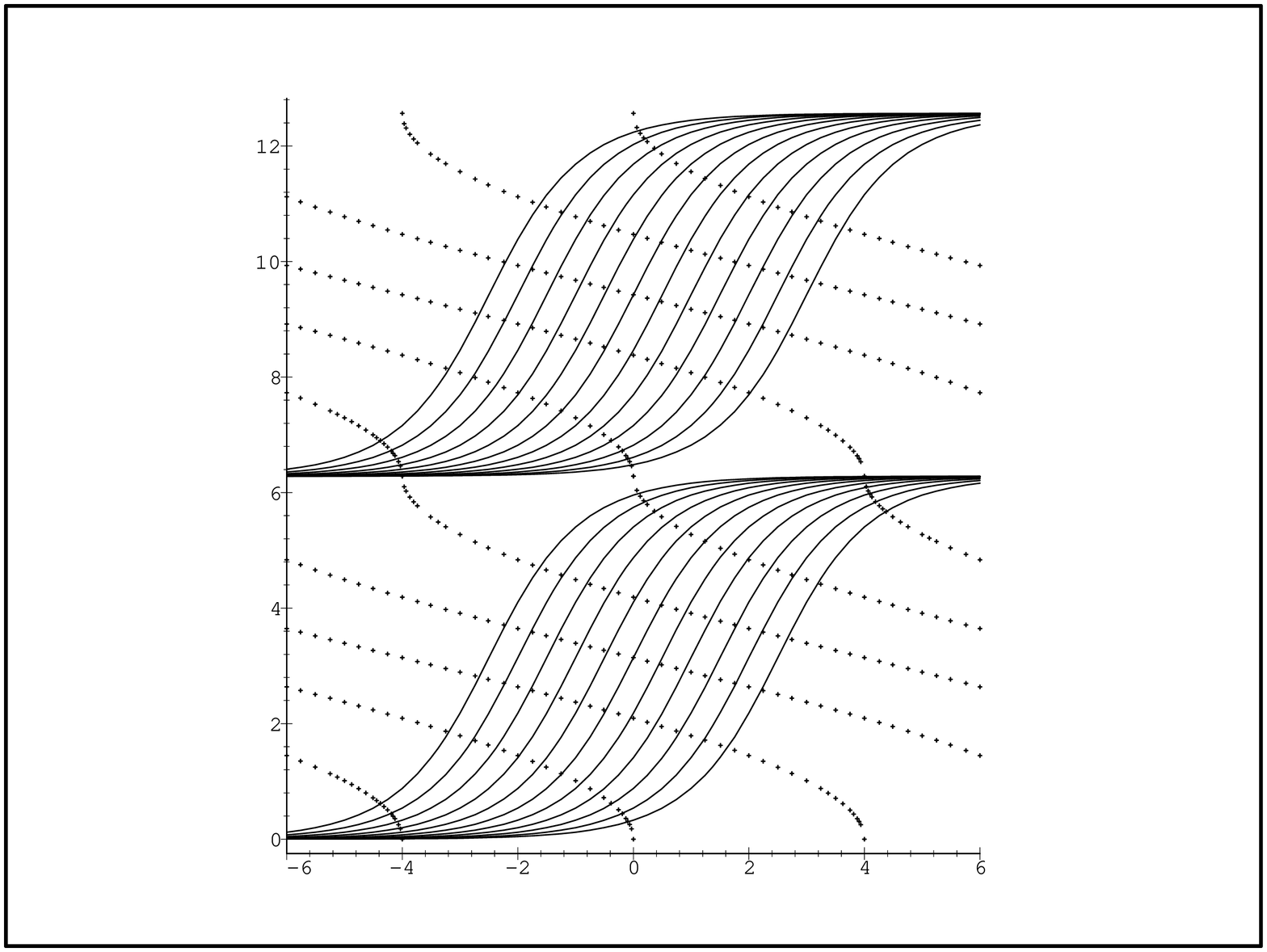,width=12cm, angle=-90}
\\[-83mm]\hspace*{-66mm}\small$\varphi$\normalsize\\[27mm]\hspace*{-68mm}
\scriptsize$2\pi$\normalsize\\[31mm]\hspace*{64mm}\small$l$\normalsize
\\[12mm]
\end{center}
Notably, not only the transversality relation is satisfied {\em everywhere} 
in the extended configuration space, but in addition the wave fronts 
intersect the extremals {\em everywhere} perpendicularly.
\end{document}